\documentclass[prd,nofootinbib,shownopacs,preprint]{revtex4}
\usepackage{graphicx}
\usepackage{epsfig}
\usepackage{amsmath}
\usepackage{amsfonts}
\usepackage{amssymb}
\usepackage{url}
\usepackage{hyperref}
\usepackage{subfigure}
\usepackage{hhline}
\usepackage{color}
\usepackage{bm}
\usepackage{cancel}
\usepackage{xspace}
\newcommand{\nn}{\nonumber}
\newcommand{\be}{\begin{equation}}
\newcommand{\ee}{\end{equation}}
\newcommand{\bea}{\begin{eqnarray}}
\newcommand{\eea}{\end{eqnarray}}

\newcommand{\dss}{D^{**}}

\newcommand{\dSs}{D^*_0}
\newcommand{\dVs}{D^*_1}
\newcommand{\dV}{D_1}
\newcommand{\dTs}{D^*_2}
\newcommand{\bmt}{\begin{pmatrix}}
\newcommand{\emt}{\end{pmatrix}}

\def\d{{\rm d}}

\newcommand{\lqcd}{\ensuremath{\Lambda_{\rm QCD}}\xspace}

\newcommand{\Bbar}{\,\overline{\!B}{}}
\newcommand{\Dbar}{\,\overline{\!D}{}}
\newcommand{\Kbar}{\,\overline{\!K}{}}
\def\B0bar{\Bbar{}^0}
\def\D0bar{\Dbar{}^0}
\def\K0bar{\Kbar{}^0}

\def\rt{\rho_\tau}
\def\rl{\rho_\ell}
\usepackage[toc,page]{appendix}
\usepackage{comment}
\begin{document}
\title{Delving into new physics in semileptonic $b \to c \tau \bar \nu$ transitions}
\author{Aishwarya Bhatta}
\email{aish.bhatta@gmail.com}

\author{Rukmani Mohanta}
\email{rmsp@uohyd.ac.in}
\affiliation{School of Physics,  University of Hyderabad, Hyderabad-500046,  India
}
\begin{abstract}
In recent times, there have been several observations suggesting deviations from lepton universality in semileptonic decays of $B$ mesons. These deviations have been noticed both in the neutral-current transitions ($b \to s ll $) and charged-current transitions ($b \to c l \bar \nu_l$). Motivated by these intriguing findings, we investigate the semileptonic decays involving the quark-level transitions of $b \to c l \bar \nu_l$. To explore the presence of new physics, we adopt a model-independent approach and analyze the processes $\Lambda_b \to \Lambda_c \tau \bar \nu_\tau$, $B_c^+ \to \eta_c \tau^+ \nu_\tau$, and $B \to D^{**} \tau \bar \nu_\tau$, where $D^{**} = {{D_0}^*, {D_1}^*, D_1, {D_2}^*}$ represent the four lightest excited charm mesons. Considering a comprehensive effective Hamiltonian, we perform a global fit using various sets of new coefficients and incorporate measurements of $R_{D}$, $R_{D^{*}}$, $R_{J/\psi}$, $P_{\tau}^{D^*}$, as well as the upper limit on the branching ratio of $B_c^+ \to \tau^+ \nu_\tau$. Subsequently, we examine the impact of constrained new couplings on the branching ratios, forward-backward asymmetry parameters, lepton non-universality ratios (LNU), as well as lepton and hadron polarization asymmetries in these decay modes.
\end{abstract}
\maketitle
\section{Introduction}
Considering the Standard Model (SM) as the theory of elementary particles narrating their interactions at the most fundamental level, nearly all the observed data from colliders can be explained to an unprecedented level, but still there are some unanswered questions such as  baryon asymmetry of the universe, dark matter and dark energy, origin of neutrino mass, etc., which  necessitate to explore the physics beyond it. In this regard, the rare $B$ meson decays mediated by the flavor changing neutral current (FCNC) transitions play a major role for the hunting of new physics (NP). Though the SM gauge interactions are lepton flavor universal, the infringement of lepton universality  has been noticed in numerous  semileptonic $B$ decays. Integrating the data of all the experiments, the world average values  of the lepton non-universality (LNU)  parameters: $R_{D^{(*)}}={\rm Br}(\bar B \to \bar D^{(*)} \tau \bar \nu_\tau)/{\rm Br}(\bar B \to \bar D^{(*)} l \bar \nu_l)$ \cite{Lees:2012xj,Lees:2013uzd,Huschle:2015rga,Hirose:2016wfn,Aaij:2015yra,Aaij:2017uff,Aaij:2017deq,Belle:2019rba,LHCb:2023zxo} and $R_{J/\psi}={\rm Br}(B_c \to J/\psi  \tau \bar \nu_\tau)/{\rm Br}(B_c \to J/\psi l \bar  \nu_l) $ \cite{lhcb-new,Dutta:2017xmj} depict a discrepancy of $1.98\sigma~(2.15\sigma)$ and $2\sigma$ respectively from their corresponding SM values.
The current data for experimental values as well as the  corresponding SM predictions of various parameters along with their  deviations are presented in Table \ref{LNU-data}\,. 
\begin{table}[htb]
\begin{center}
\caption{Current measured experimental and SM values of $R_D$, $R_{D^\star}$, $R_{J/\psi}$, $P_\tau^{D^\star}$ and $\mathcal{B}{(B_c \to \tau\bar{\nu})}$.} \label{LNU-data}
\begin{tabular}{|c|c|c|c|}
\hline

LNU parameters~& ~Experimental value~&~ SM prediction ~&~ Deviation~\\
 \hline
$R_D $ ~& ~$0.356\pm 0.029$~\cite{Belle:2019rba,Huschle:2015rga,LHCb:2023zxo}& ~$0.298 \pm 0.004$ ~\cite{HFLAV,MILC:2015uhg, Na:2015kha,Aoki:2016frl,Bigi:2016mdz} & ~$1.98\sigma$~\\
\hline
 $R_{D^*}$ ~& ~ $0.284 \pm 0.013$~\cite{Lees:2012xj,Belle:2019rba,Huschle:2015rga,LHCb:2023zxo} ~ &~$ 0.254 \pm 0.005$~ \cite{Hirose:2016wfn,Aaij:2015yra,Aaij:2017uff,Aaij:2017deq}  &~$ 2.15 \sigma$\\
 \hline
 $ R_{J/\psi} $ ~&~$ 0.71\pm 0.17 \pm 0.18$~\cite{lhcb-new} &~ $0.258\pm 0.004$ ~\cite{Harrison:2020nrv,Cohen:2018dgz}&~ $2\sigma$\\
 \hline
$P_{\tau}^{D^*} $~&~$ -0.38 \pm 0.51 \pm 0.21$ ~\cite{Hirose:2016wfn} &~ $-0.497 \pm 0.013$ \cite{Tanaka:2012nw,Blanke:2018yud}&~ $\approx (1.5\sigma-1.6 \sigma)$\\

\hline
${\rm Br} (B_c^+ \to \tau^+ \nu_\tau)$~&~ &~ $-0.497 \pm 0.013$ \cite{Aoki:2016frl}&~ \\
 \hline
\end{tabular}
\end{center}
\end{table}
 It is generally argued that, the third generation leptons involving processes are more sensitive to NP as compared to the first two generations, because of it being the heaviest lepton. Due to its resonably large mass, it has the largest coupling in the SM. $B$ meson decays to the $\tau$ is mediated by a $W$ boson in SM and has benefactions from new scalar particles, leptoquarks or some additional states like new vector bosons in other models of NP. These new  states influence the semileptonic $b \to u$ and $b \to c$ decay channels providing hints towards NP beyond the SM. As the LNU parameters  are the ratio of branching fractions, the uncertainties arising due to the CKM matrix elements and  hadronic form factors are expected to be  reduced, as they  cancelled out to a great extent in the ratio. Hence, these deviations of various LNU parameters hint towards the possible interplay of new physics  in an unambiguous manner. Using the full data sample of $772 \times 10^6$ $B\overline{B}$ pairs, the Belle collaboration has recently reported the measurement of $\tau$  polarization in the $B \to D^* \tau \bar{\nu}$ decay. The measured value, $P_{\tau}^{D^*} = -0.38 \pm 0.51 \pm 0.21$ is consistent with its SM prediction of $-0.497 \pm 0.013$ \cite{Tanaka:2012nw,Blanke:2018yud}. The requirement that the branching fraction $\mathcal{B}(B_c\to \tau\nu)$ remains less than or equal to $5\%$ can be relaxed up to $30\%$ when considering the upper bound of $\tau_{B_c}$~\cite{Alonso:2016oyd}. However, more recent data from the LEP experiment at the Z peak has imposed a stricter constraint, demanding that $\mathcal{B}(B_c\to \tau\nu)$ be less than or equal to $10\%$~\cite{Akeroyd:2017mhr}. This constraint is considerably stronger than the one obtained from the $B_c$ meson lifetime, which allowed for a $\mathcal{B}(B_c\to \tau\nu)$ of up to $30\%$. The predicted   branching fraction  in the SM   is given as ${\rm Br} (B_c^+ \to \tau^+ \nu_\tau)|^{\rm SM}= (3.6 \pm 0.14) \times 10^{-2}\,$\cite{Aoki:2016frl}
and its current  experimental upper limit is \cite{Aoki:2016frl}
$ {\rm Br} (B_c^+ \to \tau^+ \nu_\tau)|^{\rm Expt} < 30\%\;.$

Semileptonic decays like $\Lambda_b \to \Lambda_c \tau \bar \nu_\tau$, $ B_c^+ \to \eta_c \tau^+   \nu_\tau$,  and $ B \to D^{**} \tau \bar \nu_\tau$, mediated through the quark level transition $b \to c \tau \bar \nu_\tau$  are quite interesting, as they provide alternative avenues to probe the presence of NP. Since, these are tree level processes,  the NP contributions have to be significantly large enough to provide visible impact. For this, it would require new particles to be either light or strongly coupled to the SM particles. All the above said channels are mediated by $ b\rightarrow c$ transition involving the same CKM matrix element $ V_{cb}$. So a detailed comprehensive analysis of these channels will help in determining the value of  $|V_{cb}|$ in SM as well as in the presence of NP  \cite{Bhattacharya:2016zcw,Alonso:2016oyd,Alonso:2017ktd,Jung:2018lfu}. Thus, as in $B$ decays one can also scrutinize the presence of  lepton universality violation in the corresponding semileptonic baryon decays $\Lambda_b \to \Lambda_c  l \bar \nu_l$ to corroborate the results from $B$ sector and thus, to  probe the structure of NP \cite{Aaij:2011jp,Aaij:2014jyk,Atasi:2018}. The initial detection of the $\Lambda_b \to \Lambda_c  \bar{\tau} \nu_\tau $ decay, a semileptonic decay of the b-baryon, is reported with a significance of $6.1 \sigma $. This observation is based on a data sample corresponding to $3 fb^{-1}$ of integrated luminosity collected by the LHCb experiment at the LHC. The ratio of semileptonic branching fractions $R({\Lambda^+} _c ) = Br({\Lambda^0}_b \to {\Lambda^+}_ c \tau \bar{\nu_\tau} )/Br({\Lambda^0}_b \to {\Lambda^+}_ c \mu \bar{\nu_\mu} )$ is calculated to be $ 0.242 \pm 0.026 \pm 0.040 \pm 0.059$, with the last term representing the contribution of the external branching fraction uncertainty from the channel ${\Lambda^0}_b \to {\Lambda^+}_ c \mu \bar{\nu_\mu}$ ~\cite{LHCb:2022piu}. This result is consistent with the prediction of the standard model.

  Recently, $BABAR$, LHCb, and Belle showed deviations from the SM predictions in semi-tauonic decays differentiated to the $l=e,\, \mu$ light lepton in the final states~\cite{Lees:2012xj,Huschle:2015rga, Aaij:2015yra,Lees:2013uzd}. Excited charmed mesons produced from semi-leptonic decays into light leptons in the final states constitute significant background to the measurement of $R_{D^{(*)}}$, and hence their finer understanding is required for the precise determination of these ratios. The study of $B$ meson decays into four excited charmed mesons $(D^{**})$ i.e., $ B \to D^{**} l\bar \nu_l~ (D^{**}\in {D_0^*, D_1^*, D_1, D_2})$ are essential to boost the precision measurement of lepton non universality parameters in the SM as well as in  model independent framework for the NP contributions \cite{Bernlochner:2016bci,HFT:2014bci,Bernlochner:2017bci}. Heavy quark symmetry~\cite{Isgur:1989vq} gives some model independent forecasts for $\lqcd/m_{c,b}$ corrections including exclusive semileptonic $B$ decays to excited charmed mesons \cite{Leibovich:1997tu}. The essential outcome was that the $\lqcd/m_{c,b}$ corrections to semileptonic form factors at zero recoil are set on by the masses of orbitally excited charmed mesons~\cite{ Leibovich:1997em,Leibovich:1997tu}. By extending the operator structure of Lagrangian beyond standard model, we would like to investigate  all the above discussed  decay processes involving  $b \to c \tau \bar \nu_\tau$ quark level transition in a  model independent way, in this work . In this approach, we  find additional  Wilson coefficients contribution to the SM coefficients. From the $\chi^2$ fit of $R_{D}$, $R_{D^{*}}$, $P_{\tau}^{D^*}$, $R_{J/\Psi}$ and the upper limit on Br($B_c^+ \to \tau^+ \bar \nu_\tau$)  we constrain the new parameters. Using the constrained new couplings, the branching fractions, forward-backward asymmetries, LNU ratios and polarization asymmetries of $b \to c \tau \bar \nu_\tau$ decay modes  are investigated in this analysis. 

 The outline of our paper is as follows. We consider the most general effective Hamiltonian and the theoretical framework for the analysis  of $b \to c \tau  \bar \nu_\tau$ transition, in section II. In section III, the methodology used to constrain the new coefficients is discussed. Section IV shows the numerical analysis of branching ratios and various angular observable of $\Lambda_b \to \Lambda_c \tau \bar \nu_\tau$ decay processes. The  $B_c^+ \to \eta_c \tau^+ \nu_\tau$  and $ B \to D^{**} \tau \bar \nu_\tau$ processes are analyzed in sections V and VI respectively. We summarize our results in Section VII.

\section{Theoretical framework}
 Assuming the neutrino to be left chiral,  the effective Hamiltonian for $b\rightarrow c\tau\bar{\nu}$ transition, containing all possible Lorentz structures, is given as \cite{Freytsis:2015qca}
\begin{equation}
H_{eff}= 2\sqrt{2} G_F V_{cb}\left[O_{V_L} + \frac{\sqrt{2}}{4 G_F V_{cb}} \frac{1}{\Lambda^2} \left\lbrace \sum_i \left(C_i O_i +
 C^{'}_i O^{'}_i + C^{''}_i O^{''}_i \right) \right\rbrace \right]\;,
\label{effH}
\end{equation}
where $V_{cb}$ is the CKM 
matrix element, $G_F$ is the Fermi coupling constant and $O_{V_L}$ is the SM operator which has the usual $(V-A) \times (V -A)$ structure. The explicit forms of the four-fermion operators $O_i$, $O^{'}_i$ and $O^{''}_i$ are given in the Table \ref{tab1}. The couplings $C_i$, $C^{'}_i$ and $C^{''}_i$ represent the respective Wilson coefficients of the NP operators in which NP effects are encoded. The Wilson coefficients serve as a way to quantify the influence of new physics interactions, and their values can be restricted based on experimental observations of flavor-related phenomena in transitions involving particles like $b\to c\ell\nu$. Analyzing these Wilson coefficients and the limitations imposed on them enables researchers to investigate various new physics Lorentz structures and determine whether they align with the observed deviations from the predictions of the SM. The table also gives the Fierz transformed forms of primed and double primed operators in terms of the unprimed operators. For later convenience, we define $(2\sqrt{2} G_F V_{cb}\Lambda^2)^{-1} \equiv \alpha$. We set the new physics scale $\Lambda$ to be 1 TeV, which leads to $\alpha = 0.749$.
\begin{table}[h!]
\centering
\begin{tabular}{|l|ccc|}
  \hline
	& Operator & & Fierz identity\\
  \hline
$O_{V_L}$   & $(\bar{c} \gamma_\mu P_L b)\,(\bar{\tau} \gamma^\mu P_L \nu)$ & & \\
$O_{V_R}$   & $(\bar{c} \gamma_\mu P_R b)\,(\bar{\tau} \gamma^\mu P_L \nu)$ & & \\
  $O_{S_R}$   & $(\bar{c} P_R b)\,(\bar{\tau} P_L \nu)$ & & \\
  $O_{S_L}$   & $(\bar{c} P_L b)\,(\bar{\tau} P_L \nu)$ & &\\
  $O_T$       & $(\bar{c}\sigma^{\mu\nu}P_L b)\,(\bar{\tau}\sigma_{\mu\nu}P_L \nu)$ & &\\[2pt]
  \hline
 $O^{'}_{V_L}$ &$(\bar{\tau} \gamma_\mu P_L b)\,(\bar{c} \gamma^\mu P_L \nu)$ &
   $\longleftrightarrow$ &$(\bar{c} \gamma_\mu P_L b)\,(\bar{\tau} \gamma^\mu P_L \nu) = O_{V_L}$\\
$O^{'}_{V_R}$  & $(\bar{\tau} \gamma_\mu P_R b)\,(\bar{c} \gamma^\mu P_L \nu)$ &
    $\longleftrightarrow$ & $-2 (\bar{c} P_R b)\,(\bar{\tau} P_L \nu)= -2 O_{S_R} $ \\
 $O^{'}_{S_R}$  & $(\bar{\tau} P_R b)\,(\bar{c} P_L \nu)$ &
    $\longleftrightarrow$ & $-\frac{1}{2}(\bar{c} \gamma_\mu P_R b)\,(\bar{\tau} \gamma^\mu P_L \nu) = -\frac{1}{2}O_{V_R}$ \\
  $O^{'}_{S_L}$  & $(\bar{\tau} P_L b)\,(\bar{c} P_L \nu)$ &
    $\longleftrightarrow$ & $-\frac{1}{2}(\bar{c} P_L b)\,(\bar{\tau} P_L \nu) - \frac{1}{8}(\bar{c}\sigma^{\mu\nu}P_L b)\,(\bar{\tau}\sigma_{\mu\nu}P_L \nu) = -\frac{1}{2}O_{S_L}-\frac{1}{8}O_{T}$\\
$O^{'}_T$      & $(\bar{\tau}\sigma^{\mu\nu}P_L b)\,(\bar{c}\sigma_{\mu\nu}P_L \nu)$ &
    $\longleftrightarrow$ & $-6(\bar{c} P_L b)\,(\bar{\tau} P_L \nu) + \frac{1}{2}(\bar{c}\sigma^{\mu\nu}P_L b)\,(\bar{\tau}\sigma_{\mu\nu}P_L \nu) = -6O_{S_L}+\frac{1}{2}O_{T}$ \\[2pt]
  \hline
 $O^{''}_{V_L}$ & $(\bar{\tau} \gamma_\mu P_L c^c)\,(\bar{b}^c \gamma^\mu P_L \nu)$ &
    $\longleftrightarrow$ & $-(\bar{c} \gamma_\mu P_R b)\,(\bar{\tau} \gamma^\mu P_L \nu)= -O_{V_R}$  \\
  $O^{''}_{V_R}$ & $(\bar{\tau} \gamma_\mu P_R c^c)\,(\bar{b}^c \gamma^\mu P_L \nu)$ &
    $\longleftrightarrow$ & $-2(\bar{c} P_R b)\,(\bar{\tau} P_L \nu) = -2 O_{S_R}$ \\
$O^{''}_{S_R}$ &$(\bar{\tau} P_R c^c)\,(\bar{b}^c P_L \nu)$ &
$\longleftrightarrow$ & $\frac{1}{2}(\bar{c} \gamma_\mu P_L b)\,(\bar{\tau} \gamma^\mu P_L \nu) = \frac{1}{2}O_{V_L}$ \\
$O^{''}_{S_L}$ & $(\bar{\tau} P_L c^c)\,(\bar{b}^c P_L \nu)$ &
    $\longleftrightarrow$ & $-\frac{1}{2}(\bar{c} P_L b)\,(\bar{\tau} P_L \nu) + \frac{1}{8}(\bar{c}\sigma^{\mu\nu}P_L b)\,(\bar{\tau}\sigma_{\mu\nu}P_L \nu) = -\frac{1}{2}O_{S_L}+\frac{1}{8}O_{T}$ \\
 $O^{''}_T$     & $(\bar{\tau}\sigma^{\mu\nu}P_L c^c)\,(\bar{b}^c\sigma_{\mu\nu}P_L \nu)$ &
    $\longleftrightarrow$ & $-6(\bar{c} P_L b)\,(\bar{\tau} P_L \nu) - \frac{1}{2}(\bar{c}\sigma^{\mu\nu}P_L b)\,(\bar{\tau}\sigma_{\mu\nu}P_L \nu) = -6O_{S_L}-\frac{1}{2}O_{T}$  \\[2pt]
  \hline
\end{tabular}
\caption{All possible four-fermion operators that can contribute to $ b \to
c \tau\bar{\nu}$ transition.}
\label{tab1}
\end{table}
 The primed and double primed operators are products of quark-lepton bilinears. They arise naturally in models containing leptoquarks~\cite{Davidson:1993qk,Sakaki:2013bfa,Fajfer:2015ycq,Bauer:2015knc,Barbieri:2015yvd,Dorsner:2016wpm,Li:2016vvp,Sahoo:2016pet,Bhattacharya:2016mcc,Barbieri:2016las,
Chen:2017hir,Crivellin:2017zlb,Alok:2017jaf,Calibbi:2017qbu}. Models with leptoquarks of charge $2/3$, for example the model in ref.~\cite{Fajfer:2015ycq}, can give rise to primed operators. The double primed operators occur due to the exchange of charge $1/3$ leptoquarks, such as those in the models of ref.~\cite{Davidson:2010uu}. For this reason we have explicitly included these operators in our analysis even though they are linear combinations of unprimed operators. 
 Though, here we are not considering any specific new physics model, rather focusing on a generalized effective field theory (EFT) formalism, with all possible Lorentz structures in the effective Hamiltonian, we present here some well-motivated new physics scenarios, which could contribute to some of these combinations of NP coefficients.
\begin{itemize}
\item Models with an additional vector boson $W_L'$ that couples to left-handed fermions will contribute to $C_{V_L}$.

\item The scalar leptoquark $S_3(\bar 3, 3,1/3)$ and the vector leptoquark $U_3(3, 3,2/3)$ can also contribute to $C_{V_L}$.
\item The scalar leptoquark $R_2(3,2,7/6)$ can provide additional contributions to $C_{S_L}$ and $C_T$, which are related as $C_{S_L}=4C_T$.

\item The $S_1(\bar 3, 1,1/3)$ scalar leptoquark can contribute to $C_{V_L}$ and $C_{S_L}=-4C_T$ coefficients.

\item The $U_1( 3, 1,1/3)$ vector leptoquark provides additional contributions to $C_{V_L}$ and $C_{S_R}$ coefficients.

\item The coefficients $C_{S_L}$ and $C_{S_R}$ can be generated by models with extra charged scalars, e.g., two Higgs doublet models.

\item Models with extra $Z'$ contribute to $C_{V_L}$ and $C_{V_{R}}$ coefficients.
\end{itemize} 
\section{Constraints on new couplings} 
\subsection{Input Parameters:}
 To ensure transparency, we offer a detailed overview of the input parameters used in our calculations, which can be found in Table \ref{inputs_parameter}. This table includes information on the masses of various hadrons and leptons, as well as the relevant quark masses, all evaluated at the renormalization scale $\mu = m_b$. It's worth mentioning that in this analysis, we do not consider the uncertainties related to the mass parameters and decay lifetimes of the hadrons. Nevertheless, we meticulously address the uncertainties stemming from input parameters related to form factors and the CKM matrix element  $|V_{cb}|$. In our analysis, we have computed the $1\sigma$ uncertainty values for all the measurements related to different decay channels. However, in the figures presented, we have chosen to display the $1\sigma$ error range only for the SM results. This decision was made because our study involves 18 new physics parameters, and including all of them in the figures would lead to a cluttered and potentially confusing representation. These uncertainties significantly impact our computations and are thoughtfully incorporated to ensure a comprehensive and rigorous analysis.
  \begin{table}[htb!]
\centering 
\begin{tabular}{|l|l||l|l|}
\hline
Parameters  & Values & Parameters  & Values\\
\hline \hline
$m_{B^-} $ & 5.27931 & $m_{B_c}$ & 6.2751\\
$m_{\Lambda_b}$ & 5.61951  & $m_{\Lambda_c} $ & 2.28646 \\
$ m_{J/\Psi}$ & $ 3.0969$ &  $m_{{D^{\ast}}^0}$ & 2.00685\\ 

$\tau_{B_c}$  & $0.507 \times 10^{-12}$ & $f_{B_c}$ & $0.434(0.015)$ \\
$\tau_{B^-}$ & $1.638 \times 10^{-12}$ &$V_{cb}$     & 0.0409(11)  \\
$\tau_{\Lambda_b}$  & $(1.466 \pm 0.010)\times 10^{-12}$ & $G_F$ & $1.1663787\times10^{-5}$ \\
$m_e$        & $0.5109989461\times10^{-3}$ & $m_\tau$ & 1.77682\\
$m_b$        & 4.18          &  $m_c$ & 0.91\\

\hline \hline
\end{tabular}
\caption{The analysis relies on input parameters as specified in~\cite{ParticleDataGroup:2022pth}, where masses are expressed in GeV units, and lifetimes are presented in seconds.}
\label{inputs_parameter}
\end{table}

  \subsection{Chi Square Fitting:}
 Using the effective Hamiltonian given in Eq.~(\ref{effH}), we compute the observables $R_D$, $R_{D^*}$, $R_{J/\psi}$,  $P_{\tau}$ and the upper limit on Br($B_c^+ \to \tau^+ \nu_\tau$) as functions of the various Wilson coefficients. We offer an exhaustive table, labeled as  Table \ref{LNU-data}, which furnishes the numerical data for these measurements. During the fitting procedure, we are not taking into account the correlation between the observables $R_D$ and $R_{D^{\star}}$. By fitting these expressions to the measured values of the observables, we obtain the values of WCs which are consistent with the data. Here we consider either one NP operator or a combination of two similar operators (for example, [$O_{V_L}$, $O_{V_R}$], [$O_{S_L}$, $O_{S_R}$], [$O^{'}_{V_L}$, $O^{'}_{V_R}$] and [$O^{''}_{S_L}$, $O^{''}_{S_R}$]) at a time while making the fit to the experimental observables. 
Here $ \chi^2$ includes experimental and SM theory uncertainties. For a given NP operator, the most likely value of its WC is obtained by minimizing the $\chi^2$. The corresponding $\chi^2$ is defined as
\bea
\chi^2(C_{\rm{eff}}^i)=\sum \frac{(\mathcal{O}^{\rm th}(C_{\rm{eff}}^i)-\mathcal{O}^{\rm Expt})^2}{(\Delta \mathcal{O}^{\rm Expt})^2+(\Delta \mathcal{O}^{\rm SM})^2}\,.
\label{chi2}
\eea
Here ${\cal O}^{\rm th}(C_{\rm{eff}}^i)$ are the total theoretical predictions for the observables which depend upon the effective NP WCs $C_{\rm{eff}}^i$ and ${\cal O}^{\rm Expt}$ represent the corresponding measured central values. $\Delta \mathcal{O}^{\rm Expt}$ and $\Delta \mathcal{O}^{\rm SM}$ are the experimental and SM uncertainties of the observables respectively. The expressions for various $C_{\rm{eff}}^i$, as linear combinations of $C_i$, $C^{'}_i$ and $C^{''}_i$, are defined below in eq.~(\ref{Cieff}).
\bea
C^{\rm{eff}}_{S_L} &=& -\alpha \left(6C^{'}_{T}+6C^{''}_{T}-C_{S_L} +0.5C^{'}_{S_L}+0.5C^{''}_{S_L}\right), \nonumber \\
C^{\rm{eff}}_{S_R} &=& -\alpha \left(2C^{'}_{V_R}+2C^{''}_{V_R}-C_{S_R}\right), \nonumber \\
C^{\rm{eff}}_{V_L} &=&  \alpha \left(C_{V_L}+ 0.5C^{''}_{S_R} + C^{'}_{V_L} \right), \nonumber \\
C^{\rm{eff}}_{V_R} &=& -\alpha \left(C^{''}_{V_L}-C_{V_R} + 0.5C^{'}_{S_R}  \right), \nonumber \\
C^{\rm{eff}}_T &=& \alpha \left(C_T+0.5 C^{'}_T -0.5C^{''}_T- 0.125C^{'}_{S_L}+0.125C^{''}_{S_L}\right).
\label{Cieff}
\eea
The expressions for $R^{\rm th}_D$, $R^{\rm th}_{D^*}$, $R^{\rm th}_{J/\psi}$, $P^{\rm th}_{\tau}$ and ${\rm Br}(B_c^+ \to \tau^+ \nu_\tau)$ in terms of $C^{\rm{eff}}_i$ are \cite{Alok:2017qsi}
\bea
R^{\rm th}_D &=& 0.398\mid C^{\rm{eff}}_{S_L}+C^{\rm{eff}}_{S_R}\mid^2 + 0.297\mid 1+C^{\rm{eff}}_{V_L}+C^{\rm{eff}}_{V_R}\mid^2  \nonumber \\
& & + 0.509~{\rm  Re}\left[\left(1+C^{\rm{eff}}_{V_L}+C^{\rm{eff}}_{V_R}\right)\left(C^{\rm{eff}*}_{{S_L}}+C^{\rm{eff}*}_{{S_R}}\right)\right] + 0.140\mid C^{\rm{eff}}_T\mid^2  \nonumber \\
& & + 0.244~ {\rm Re}\left[\left(1+C^{\rm{eff}}_{V_L}+C^{\rm{eff}}_{V_R}\right)C^{\rm{eff}*}_{T}\right],
\eea
\bea
R^{\rm th}_{D^*} & =&  0.011\mid C^{\rm{eff}}_{S_R}-C^{\rm{eff}}_{S_L}\mid^2 - 0.449 Re\left[\left(1+C^{\rm{eff}}_{V_L}\right)C^{\rm{eff}*}_{{V_R}}\right]\nonumber\\
& &+  0.253\left(\mid 1+C^{\rm{eff}}_{V_L}\mid^2 + \mid C^{\rm{eff}}_{V_R}\mid^2\right) + 3.077\mid C^{\rm{eff}}_{T}\mid^2 \nonumber \\
& &  - 1.055~ {\rm Re}\left[\left(1+C^{\rm{eff}}_{V_L}\right)C^{\rm{eff}*}_{T}\right] + 1.450 Re\left[C_{\rm{eff}}^{V_R}C_{\rm{eff}}^{T*}\right] \nonumber\\
& & + 0.030~{\rm  Re}\left[\left(C^{\rm{eff}*}_{{S_R}}-C^{\rm{eff}*}_{{S_L}}\right)\left(1+C^{\rm{eff}}_{V_L}-C^{\rm{eff}}_{V_R}\right)\right],
\eea
\bea
P^{\rm th}_{\tau} &=& \left\lbrace  4.089\mid C^{\rm{eff}}_T\mid^2 - 
   2.985 \left(\mid 1 + C^{\rm{eff}}_{V_L}\mid^2 + \mid C^{\rm{eff}}_{V_R}\mid^2\right)\right. + 0.252\mid C^{\rm{eff}}_{S_L} - C^{\rm{eff}}_{S_R}\mid^2 \nonumber\\
& &   0.716 ~{\rm Re}\left[\left(C^{\rm{eff}*}_{{S_R}}-C^{\rm{eff}*}_{{S_L}}\right) \left(1 + C^{\rm{eff}}_{V_L} - C^{\rm{eff}}_{V_R}\right)\right]  +8.298 ~{\rm Re}\left[C^{\rm{eff}*}_{T} \left(1 + C^{\rm{eff}}_{V_L}\right)\right] \nonumber\\
& & \left. + 5.136~{\rm  Re}\left[\left(1 + C^{\rm{eff}}_{V_L}\right) C^{\rm{eff}*}_{{V_R}}\right] -11.410~{\rm  Re}\left[C^{\rm{eff}*}_{T} C^{\rm{eff}}_{V_R}\right] \right\rbrace /
   \left\lbrace 0.252 \mid C^{\rm{eff}}_{S_L} - C^{\rm{eff}}_{S_R}\mid^2 \right.\nonumber\\
   &&+ 5.983 \left(\mid 1 + C^{\rm{eff}}_{V_L}\mid^2 + \mid C^{\rm{eff}}_{V_R}\mid^2 \right)  +  72.609 \mid C_{\rm{eff}}^T\mid^2  - 24.894 ~{\rm Re}\left[C^{\rm{eff}*}_{T} \left(1 + C^{\rm{eff}}_{V_L}\right)\right]\nonumber\\  
   &&   + 
   34.232~{\rm Re}\left[C^{\rm{eff}*}_{T} C^{\rm{eff}}_{V_R}\right] + 0.716~{\rm Re}\left[\left(C^{\rm{eff}*}_{{S_R}} - C^{\rm{eff}*}_{{S_L}}\right) \left(1 + C^{\rm{eff}}_{V_L} - C^{\rm{eff}}_{V_R}\right)\right] \nonumber\\
   & & \left. - 10.625~{\rm Re}\left[ C^{\rm{eff}*}_{{V_R}}\left(1 + C^{\rm{eff}}_{V_L}\right) \right]\right\rbrace,
   \eea
\bea
R^{\rm th}_{J/\psi} & =&  0.014\mid C^{\rm{eff}}_{S_R}-C^{\rm{eff}}_{S_L}\mid^2  - 0.559 ~{\rm Re}\left[\left(1+C^{\rm{eff}}_{V_L}\right)C^{\rm{eff}*}_{{V_R}}\right]
+  0.289\left(\mid 1+C^{\rm{eff}}_{V_L}\mid^2 + \mid C^{\rm{eff}}_{V_R}\mid^2\right) \nonumber\\
&&+ 3.095\mid C^{\rm{eff}}_{T}\mid^2 
- 1.421~{\rm Re}\left[C^{\rm{eff}*}_{T}\left(1+C^{\rm{eff}}_{V_L}\right)\right] + 1.562~{\rm Re}\left[C^{\rm{eff}*}_{T} C^{\rm{eff}}_{V_R}\right] \nonumber\\
& & + 0.041~{\rm Re}\left[\left(1+C^{\rm{eff}}_{V_L}-C^{\rm{eff}}_{V_R}\right)\left(C^{\rm{eff}*}_{{S_R}}-C^{\rm{eff}*}_{{S_L}}\right)\right],
\eea
 and
 \bea
{\rm Br}(B_c^+ \to \tau^+ \nu_\tau) &=& \frac{G^2_F \vert V_{cb}\vert^2 m_{B_c} f^2_{B_c} m^2_{\tau}\tau^{\rm exp}_{B_c}}{8\pi}\left(1-\frac{m^2_{\tau}}{m^2_{B_c}}\right)^2 \nonumber\\
& &\times  \left| 1+\frac{m^2_{B_c}}{m_{\tau}(m_b+m_c)}(C^{\rm{eff}}_{S_R}-C^{\rm{eff}}_{S_L})+ C^{\rm{eff}}_{V_L}-C^{\rm{eff}}_{V_R}\right|^2,
\eea

 \begin{table}[htbp]
\centering
\tabcolsep 6pt

\begin{tabular}{|c|c|c|c|c|}
\hline
Cases & NP Coefficient(s)  &  Best fit value(s)  & $\chi^2_{\rm min}/{\rm d.o.f}$ & \emph{pull} \\
\hline
 Case  I & $C_{V_L}$  &  $ -2.839 $ & 0.56 & 3.54 \\

 ~&$C_{T}$  &  $-0.069 $ & 1.50 & 2.96\\

 &$C'_{V_L}$ & $-2.839$& 0.56 & 3.54\\

&$C''_{S_L}$ & $-0.584$ & 0.57 & 3.53\\

 &$C''_{S_R}$ & $-5.678$ & 0.56 & 3.54\\
\hline
Case  II &$(C_{V_L},\, C_{S_L})$ & $(1.642,-0.219)$ &1.48 & 3.216\\
A&$(C_{V_R},\, C_{S_R})$ & $(-1.142,-2.537)$ & 1.09 & 3.39\\
&$(C_{S_L},\, C_{V_R})$ & $(-2.644,-0.962)$ & 1.10 & 3.40\\
\hline
&$(C'_{V_L},\, C'_{S_L})$  &  $(-1.969, -5.626)$ & 0.71 &3.56 \\
&$(C'_{V_L},\, C'_{S_R})$  &  $(-4.223, -0.207)$  & 0.90 & 3.47 \\
&$(C'_{V_L},\, C'_{T})$ & $(-4.342, -0.044)$ &0.96 & 3.45\\
B&$(C'_{V_L},\, C'_{V_R})$ & $(-4.512, -1.794)$ & 1.06 &3.41\\
&$(C'_{S_L},\, C'_{V_R})$ & $(3.580,-1.642)$ & 0.77& 3.53 \\
&$(C'_{S_L},\, C'_{T})$ & $(2.726, -0.479)$ & 0.74 & 3.54\\
&$(C'_{V_R},\, C'_{S_R})$ & $(1.270, 2.272)$ & 0.51 & 3.64\\
\hline
&$(C''_{V_L},\, C''_{V_R})$ & $(1.142, 1.268)$ &1.09& 3.40\\ 
C&$(C''_{V_L},\, C''_{S_L})$  & $(0.397, -3.584)$  & 1.21 & 3.34\\
&$(C''_{V_R},\, C''_{S_R})$  &  $(0.129, 3.35)$  & 1.12 & 3.20 \\
\hline
\end{tabular}
\caption{Best fit values of new WCs for all possible cases. We also show the  pull values using \emph{pull} = $\sqrt{\chi^2_{SM}-\chi^2_{min}}$ . Here we list solutions with $\chi^2_{\rm min}/{\rm d.o.f}\lesssim 1.5$. For the SM, we have $\chi^2_{\rm SM} = 14.806$.}
\label{Tab:Best-fit}
\end{table}
 In Table  \ref{Tab:Best-fit}\,, Case I contains $5$ selected individual new coefficients  whose $\chi^2_{\rm min}/{\rm d.o.f}\lesssim 1.5$. Here we have also presented the pull values,  defined as: pull=$\sqrt{\chi^2_{\rm SM}-\chi^2_{\rm min}}$\,. Since we have taken five observables in the presence of one new parameter, the degrees of freedom in this case is $4$. We find $\chi^2_{\rm SM}=14.806$ for the SM. We know $O_{V_L}$ is the Fierz transform of both  $O''_{S_R}$  and $O'_{V_L}$. So in Case  I, we can see that the solutions with  $C''_{S_R}$  and $C'_{V_L}$ are degenerate with the $C_{V_L}$ solutions.
 
\begin{figure}[htbp]
\includegraphics[width=0.25\textwidth]{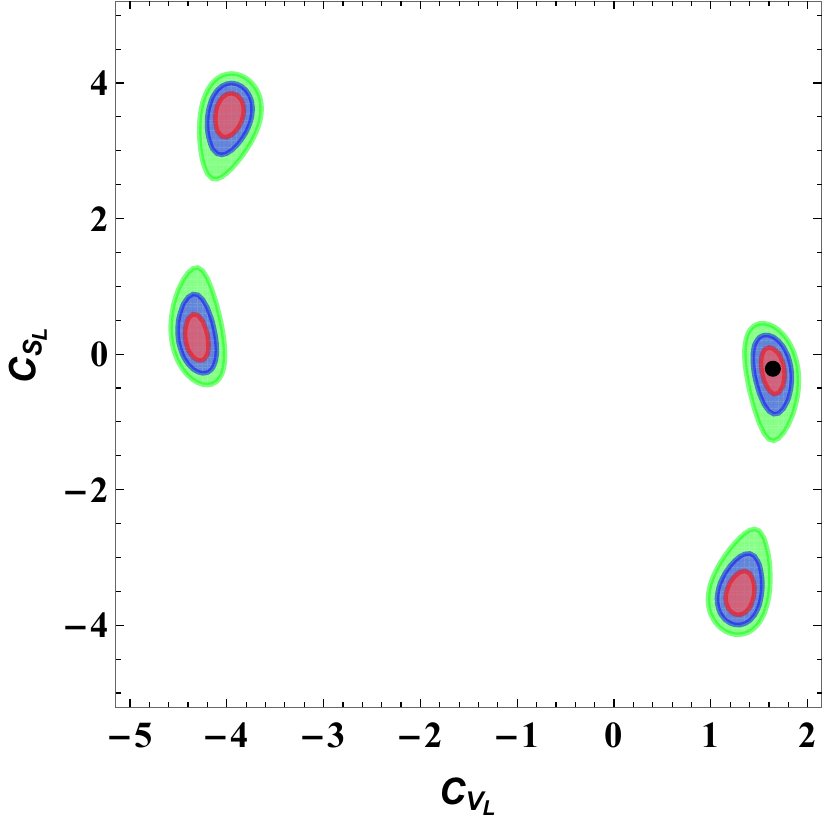}
\quad
\includegraphics[width=0.25\textwidth]{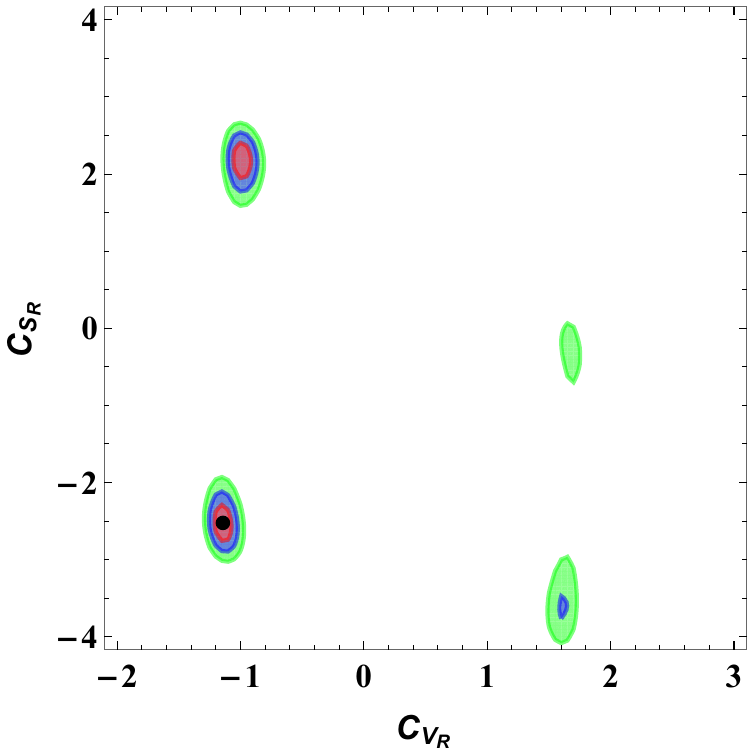}
\quad
\includegraphics[width=0.25\textwidth]{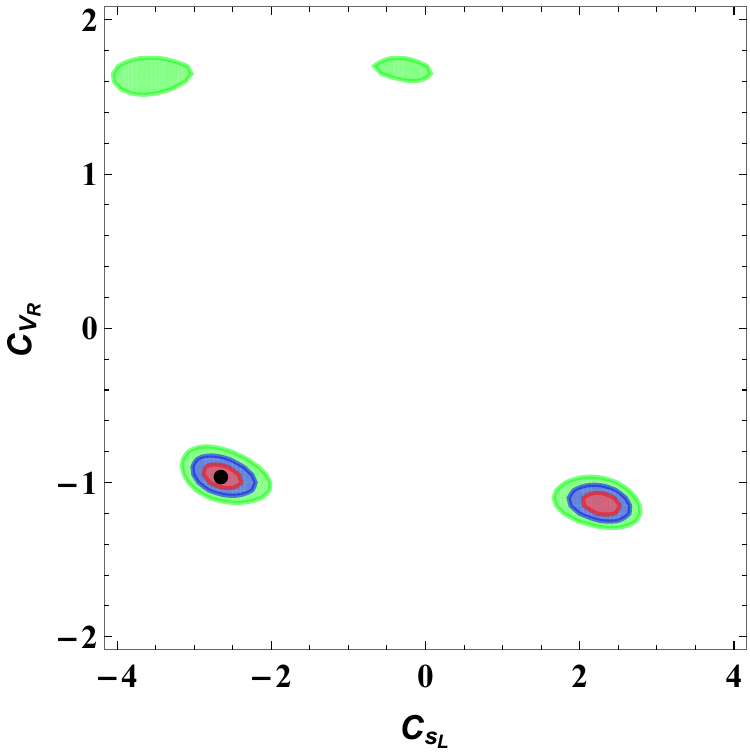}
\quad
\includegraphics[width=0.25\textwidth]{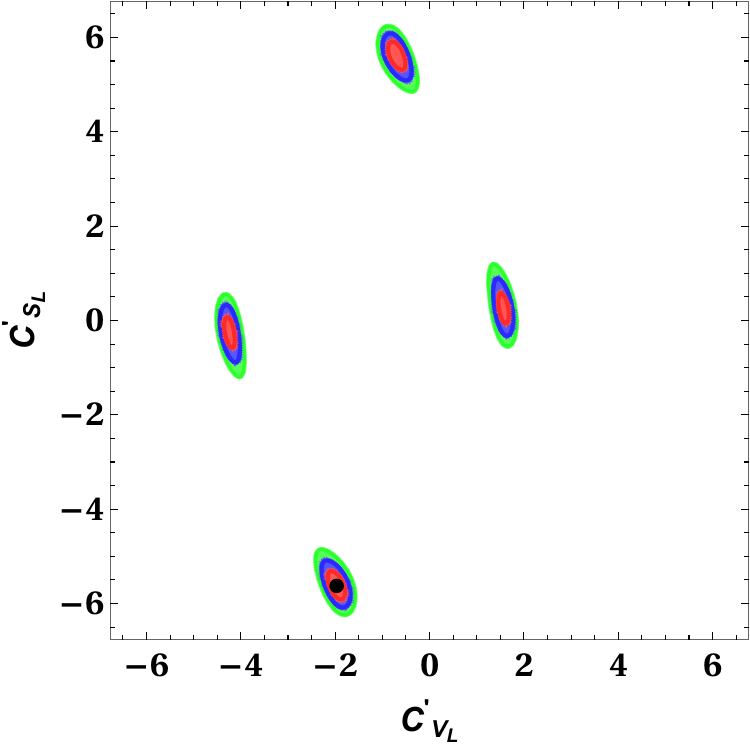}
\quad
\includegraphics[width=0.25\textwidth]{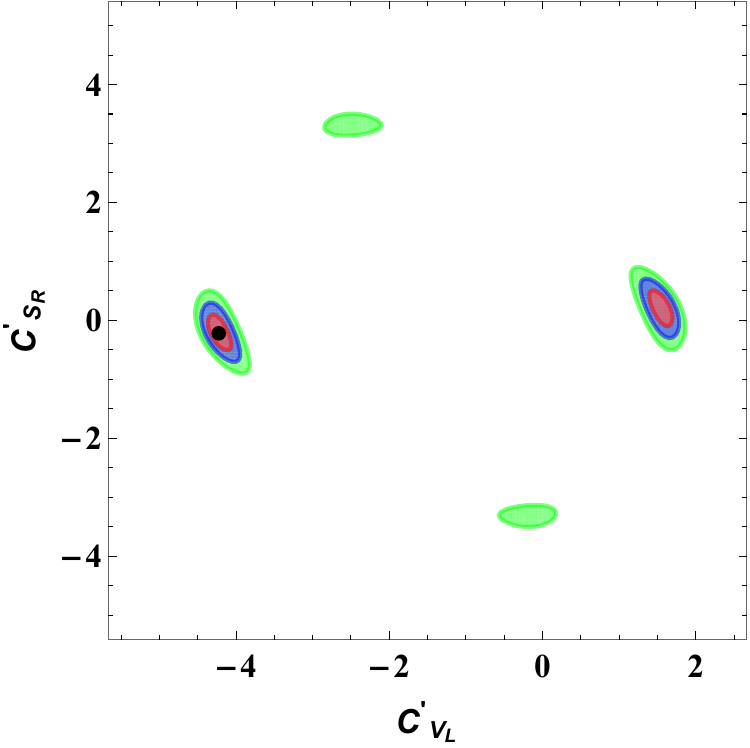}
\quad
\includegraphics[width=0.25\textwidth]{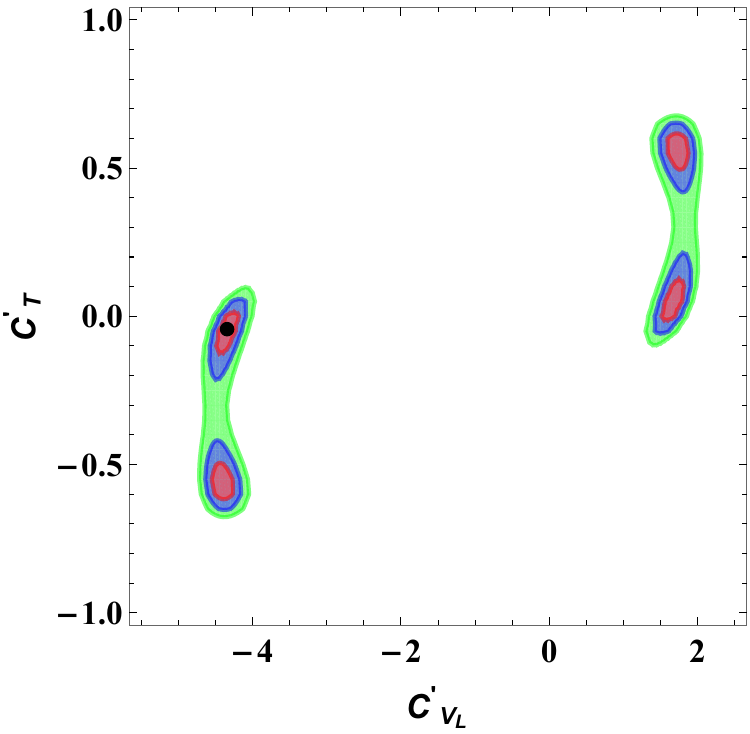}
\quad
\includegraphics[width=0.25\textwidth]{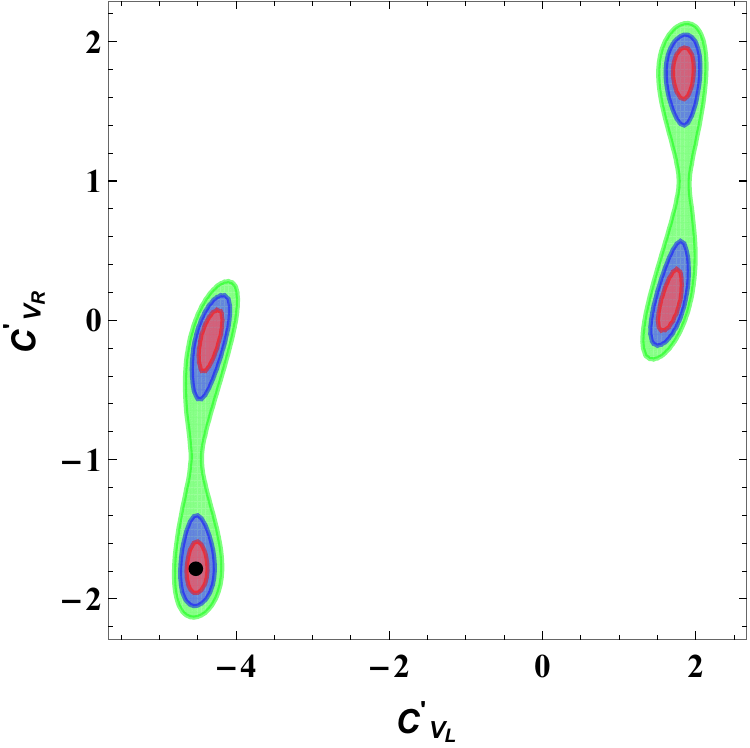}
\quad
\includegraphics[width=0.25\textwidth]{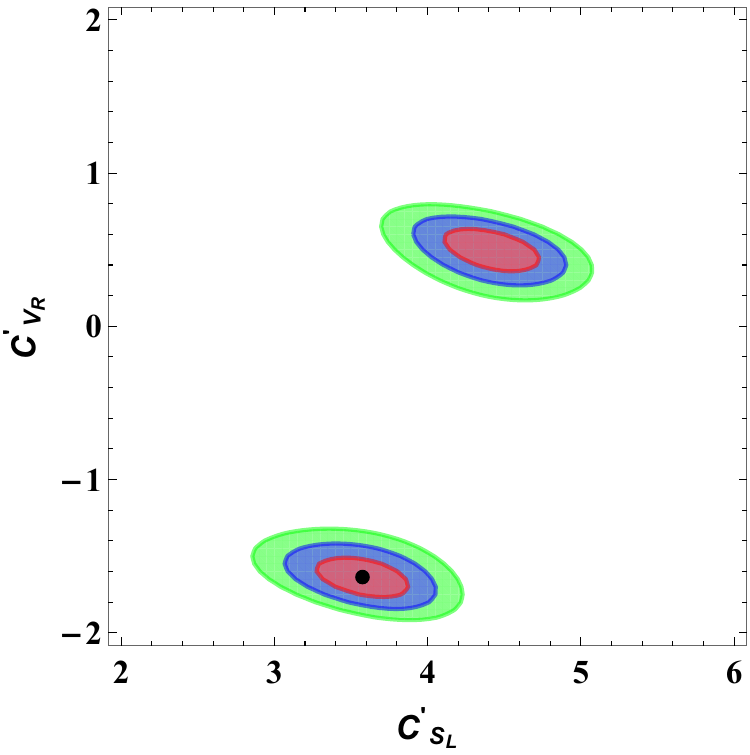}
\quad
\includegraphics[width=0.25\textwidth]{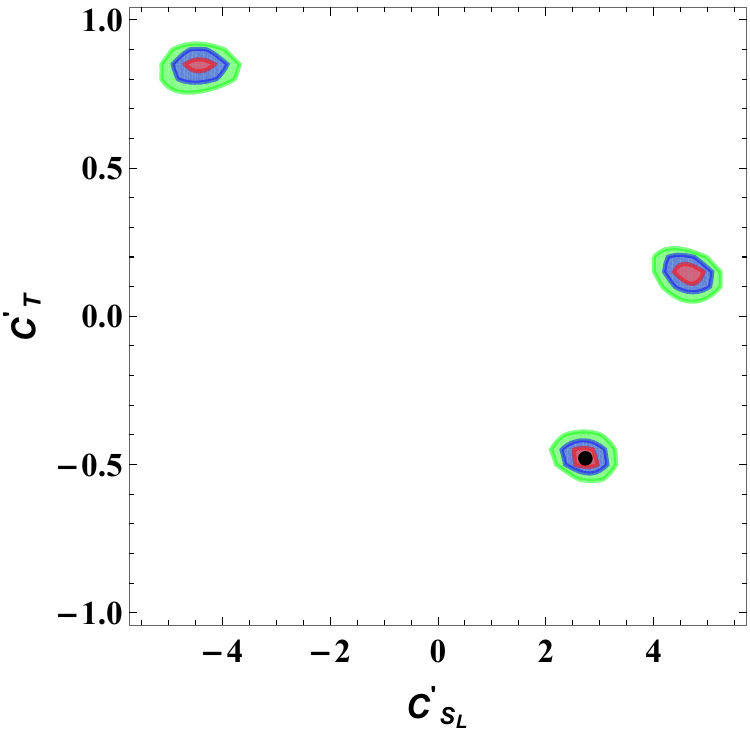}
\quad
\includegraphics[width=0.25\textwidth]{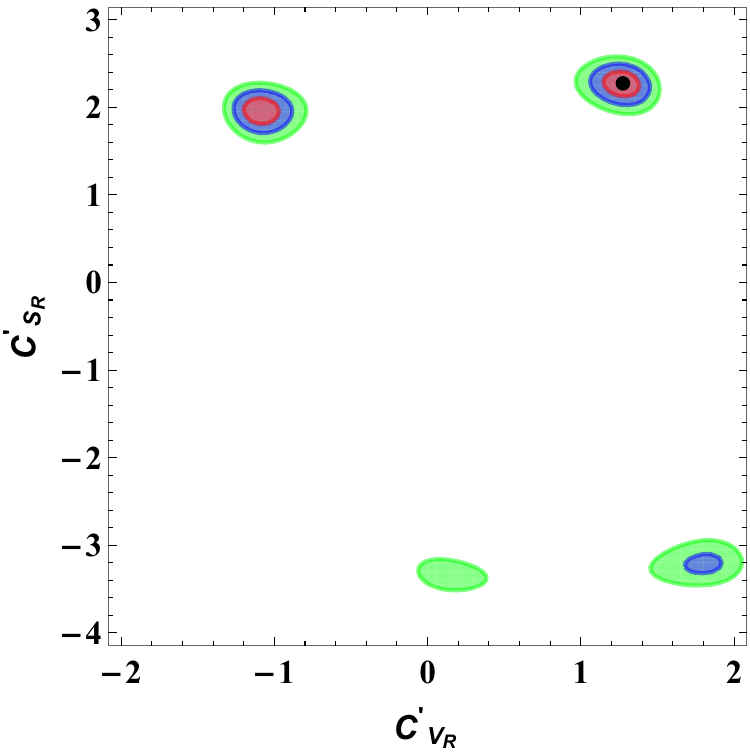}
\quad
\includegraphics[width=0.25\textwidth]{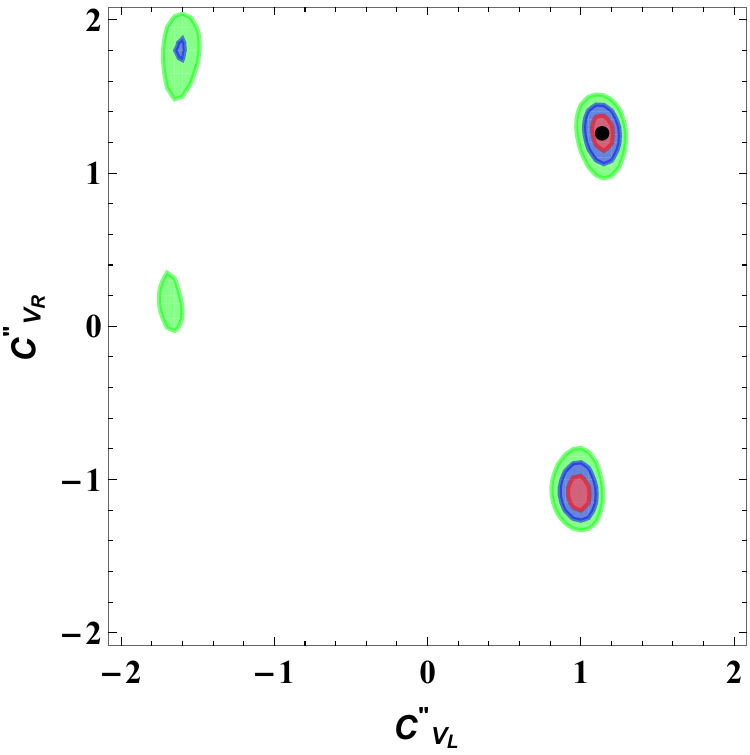}
\quad
\includegraphics[width=0.25\textwidth]{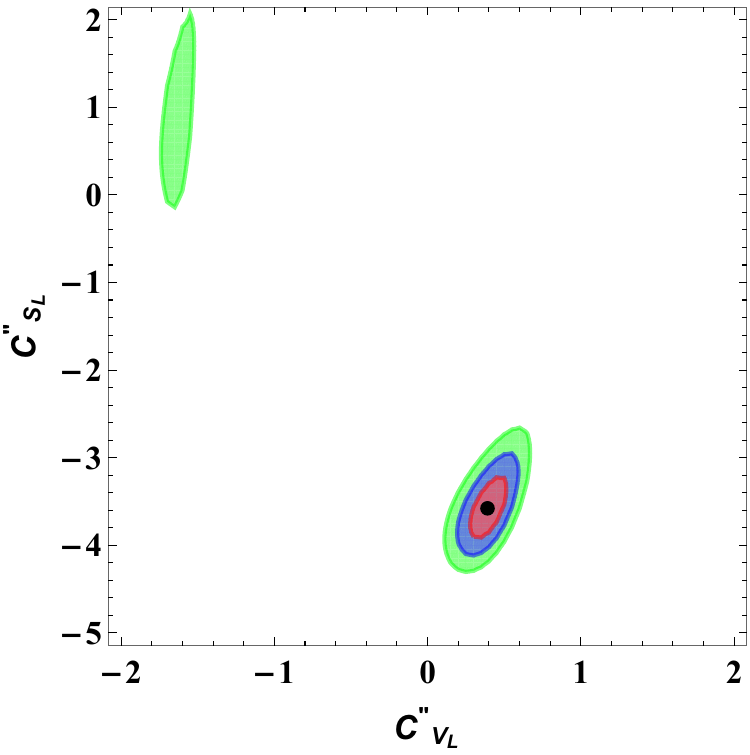}
\quad
\includegraphics[width=0.25\textwidth]{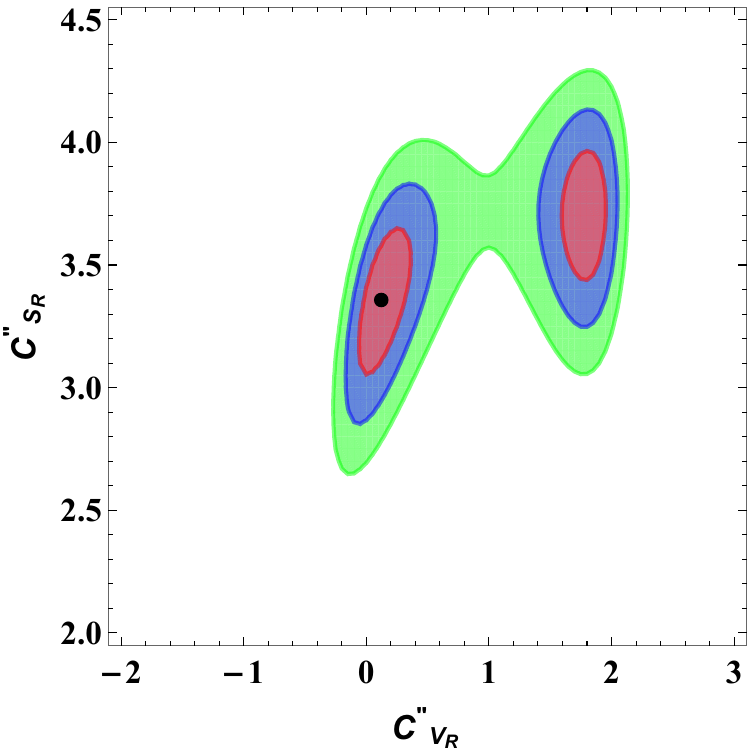}
\caption{For case II, constraints on various combination of new coefficients obtained from $\chi^2$ fitting. Here red, blue and green colours represent the $1\sigma$, $2\sigma$ and $3\sigma$ contours and the black dots represent the best-fit values.}\label{Fig:Case-C}
\end{figure}

  Next in case II, It contains different possible combination of two Wilson coefficients such as ($C_{i} ~\&~ C_{j}$),  ($C'_{i}~ \&~ C'_{j}$) and  ($C''_{i}~\&~C''_{j}$). Thirteen possible sets of new coefficients of the constrained plots are displayed in Fig.\ref{Fig:Case-C}\,. Out of all the considered combinations, here we display only those combinations whose $\chi^2_{\rm min}/{\rm d.o.f}\lesssim 1.5$, which give a good fit to us. The NP bounds arising due to ($C'_{V_L}~\&~C'_{S_L}$), ($C'_{V_L}~\& ~C'_{S_R}$), ($C'_{V_L}~\&~C'_{T}$), ($C'_{S_L}~\&~C'_{V_R}$), ($C'_{S_L}~\&~C'_{T}$) and ($C'_{V_R}~\&~C'_{S_R}$) are most acceptable as compared to others as they have  $\chi^2_{\rm min}/{\rm d.o.f}\lesssim 1$.  Here red, blue and green colours represent the $1\sigma$, $2\sigma$ and $3\sigma$ contours and the black dots represent the best-fit values.
 
 \section{Numerical Analysis For  $\Lambda_b \to \Lambda_c \tau \bar \nu_\tau$ decay processes. }
In this section, we discus the decay rate and angular observables of $\Lambda_b \to \Lambda_c \tau \bar \nu_\tau$ process mediated by $b \to c \tau \bar \nu_\tau$ transition.   The double differential decay rate, in the presence of NP for $\Lambda_b \to \Lambda_c \tau \bar \nu_\tau$ decay process with respect to $q^2$ and $\cos \theta_l$ ($\theta_l$ is the angle between the directions of parent  $\Lambda_b$ baryon and the $\tau$ in the dilepton rest frame) is  \cite{Shivashankara:2015cta,Li:2016pdv}
 \bea \label{decayrate}
\frac{d\Gamma}{dq^2 d\cos \theta_l}&=&\frac{G_F^2 |V_{c b}|^2 q^2  \sqrt{\lambda(q^2)}}{2^{10} \pi^3 M_{\Lambda_b}^3}\, \left(1-\frac{m_\tau^2}{q^2}\right)^2\left[A_1+\frac{m_\tau^2}{q^2}A_{2}+2A_3+\frac{1}{4} A_4 +\frac{4m_\tau}{\sqrt{q^2}}\left( A_5+A_6 \right)+A_7\right],\hspace*{0.7 true cm}
\eea
 where $\lambda_{\Lambda_b} =\lambda (M_{\Lambda_b}^2, M_{\Lambda_c}^2, q^2)$ and $M_{\Lambda_b}~(M_{\Lambda_c})$ is the mass of the $\Lambda_b~(\Lambda_c)$ baryon. The mathematical expressions for $   {A_i} $'s \cite{Shivashankara:2015cta} depend upon the helicity amplitudes in terms of different form factors and the new physics couplings, which are given below.  
\bea
A_1&=&(1-\cos\theta_l)^2
H_{\frac{1}{2},+}^2+2\sin^2\theta_l\big(H_{\frac{1}{2},0}^2+H_{-\frac{1}{2},0}^2\big)+(1+\cos\theta_l)^2H_{-\frac{1}{2},-}^2\,,\nn\\
A_2&=&\sin^2\theta_l
\big(H_{\frac{1}{2},+}^2+H_{-\frac{1}{2},-}^2\big)+2\cos^2\theta_l\big(H_{\frac{1}{2},0}^2+H_{-\frac{1}{2},0}^2\big)-4\cos\theta_l \big(H_{\frac{1}{2},0}H_{\frac{1}{2},t}+H_{-\frac{1}{2},0} H_{-\frac{1}{2},t}\big)\nonumber\\&&+2\big(H_{\frac{1}{2},t}^2+H_{-\frac{1}{2},t}^2\big)\ \,,\nn\\
A_3&=&H^{SP^2}_{-\frac{1}{2},0}+H^{SP^2}_{\frac{1}{2},0}\,,\nn \\
A_4&=&\sin^2\theta_l\big(
H_{\frac{1}{2},+,0}^{T2}\!+\!H_{\frac{1}{2},+,t}^{T^2}+H_{-\frac{1}{2},-,t}^{T^2}+H_{-\frac{1}{2},0,-}^{T2}+2H_{-\frac{1}{2},0,-}^TH_{-\frac{1}{2},-,t}^T
+2H_{\frac{1}{2},+,0}^TH_{\frac{1}{2},+,t}^T\big)\nn\\ &&
+\frac{m_\tau^2}{q^2}\Big[2\sin^2\theta_l\big(H_{\frac{1}{2},+,-}^{T^2}+H_{\frac{1}{2},0,t}^{T^2}+
H_{-\frac{1}{2},+,-}^{T^2}+H_{-\frac{1}{2},0,t}^{T^2}+2H_{\frac{1}{2},+,-}^TH_{\frac{1}{2},0,t}^T+
2H_{-\frac{1}{2},+,-}^TH_{-\frac{1}{2},0,t}^T\big)\nn\\&&+(1-\cos\theta_l)^2\big(H_{\frac{1}{2},+,0}^{T^2}+H_{\frac{1}{2},+,t}^{T^2}+
2H_{\frac{1}{2},+,0}^TH_{\frac{1}{2},+,t}^T\big)\nn\\&&+(1+\cos\theta_l)^2\big(H_{-\frac{1}{2},0,-}^{T^2}+H_{-\frac{1}{2},-,t}^{T^2}+2H_{-\frac{1}{2},0,-}^T H_{-\frac{1}{2},-,t}^T\big)\Big]\nn \\ &&+2\cos^2\theta_l\big(H_{\frac{1}{2},+,-}^{T^2}\!+\!H_{\frac{1}{2},0,t}^{T^2}
+H_{-\frac{1}{2},+,-}^{T^2}\!+H_{-\frac{1}{2},0,t}^{T^2}+
2H_{\frac{1}{2},+,-}^TH_{\frac{1}{2},0,t}^T+2H_{-\frac{1}{2},+,-}^TH_{-\frac{1}{2},0,t}^T\big)\,,\nn\\
A_5&=&\big(H_{\frac{1}{2},t}H^{SP}_{\frac{1}{2},0}+H_{-\frac{1}{2},t}H^{SP}_{-\frac{1}{2},0}\big)-\cos\theta_\ell\big(H_{\frac{1}{2},0}H^{SP}_{\frac{1}{2},0}+H_{-\frac{1}{2},0}H^{SP}_{-\frac{1}{2},0}\big)
\,,\nn \\
A_6&=&-\frac{\cos\theta_l}{2}\big(H_{\frac{1}{2},t}H^T_{\frac{1}{2},+,-}+
H_{\frac{1}{2},t}H^T_{\frac{1}{2},0,t}+H_{-\frac{1}{2},t}H^T_{-\frac{1}{2},+,-}+
H_{-\frac{1}{2},t}H^T_{-\frac{1}{2},0,t}\big)\nn\\&&+\frac{\cos^2\theta_l}{2}\big(H_{\frac{1}{2},0}H^T_{\frac{1}{2},+,-}+
H_{\frac{1}{2},0}H^T_{\frac{1}{2},0,t}+H_{-\frac{1}{2},0}H^T_{-\frac{1}{2},+,-}+
H_{-\frac{1}{2},0}H^T_{-\frac{1}{2},0,t}\big)\nn\\&&+\frac{(1+\cos\theta_l)^2}{4}\big(H_{-\frac{1}{2},-}H^T_{-\frac{1}{2},0,-}
+H_{-\frac{1}{2},-}H^T_{-\frac{1}{2},-,t}\big)\nn\\&&+\frac{(1-\cos\theta_l)^2}{4}\big(H_{\frac{1}{2},+}H^T_{\frac{1}{2},+,0}
+H_{\frac{1}{2},+}H^T_{\frac{1}{2},+,t}\big)\nn\\&&+\frac{\sin^2\theta_l}{4}\big(2H_{\frac{1}{2},0}H^T_{\frac{1}{2},+,-}+2H_{\frac{1}{2},0}H^T_{\frac{1}{2},0,t}
+2H_{-\frac{1}{2},0}H^T_{-\frac{1}{2},+,-}+2H_{-\frac{1}{2},0}H^T_{-\frac{1}{2},0,t}\nn\\
&&+ H_{\frac{1}{2},+}H^T_{\frac{1}{2},+,0}+
H_{\frac{1}{2},+}H^T_{\frac{1}{2},+,t}+H_{-\frac{1}{2},-}H^T_{-\frac{1}{2},0,-}+
H_{-\frac{1}{2},-}H^T_{-\frac{1}{2},-,t}\big)\,,\nn \\
A_7&=&-2\cos\theta_l\Big(H^{SP}_{-\frac{1}{2},0}H^T_{-\frac{1}{2},+,-}
+H^{SP}_{-\frac{1}{2},0}H^T_{-\frac{1}{2},0,t}+H^{SP}_{\frac{1}{2},0}H^T_{\frac{1}{2},+,-}+H^{SP}_{\frac{1}{2},0}
H^T_{\frac{1}{2},0,t}\Big)\,,
\eea
with 
\bea
&& H^{S}_{\lambda_{\Lambda_{c}},\lambda_{NP}} =  H^{S}_{-\lambda_{\Lambda_{c}},-\lambda_{NP}},~~~H^{SP}_{\lambda_{\Lambda_c},\lambda=0}=H^S_{\lambda_{\Lambda_c},\lambda=0}+H^P_{\lambda_{\Lambda_c},\lambda=0}, ~~
H^{P}_{\lambda_{\Lambda_{c}},\lambda_{NP}} =  -H^{P}_{-\lambda_{\Lambda_{c}},-\lambda_{NP}},\nn \\
&&H_{\lambda_{\Lambda_{c}},\lambda,\lambda}^T=0~~~~~H^{T}_{\lambda_{\Lambda_{c}},\lambda,\lambda^\prime}=-H^{T}_{\lambda_{\Lambda_{c}},\lambda^\prime,\lambda}\nn \\
&& H_{\lambda_{\Lambda_{c}},\lambda_{W}}^V=H_{-\lambda_{\Lambda_{c}},-\lambda_{W}}^V,~~H^{VA}_{\lambda_{\Lambda_c},\lambda}=H^V_{\lambda_{\Lambda_c},\lambda}-H^A_{\lambda_{\Lambda_c},\lambda},~~
H_{\lambda_{\Lambda_{c}},\lambda_{W}}^A=-H_{-\lambda_{\Lambda_{c}},-\lambda_{W}}^A. \nn\\
\eea
The detailed expressions for helicity amplitudes are given in Appendix A.
 After integrating out $\cos \theta_l$ in Eqn. (\ref{decayrate}), we obtain the differential decay rate with respect to $q^2$. 
Besides the branching ratio, other interesting observables in this decay process are
\begin{itemize}
\item Forward-backward asymmetry parameter:
\bea
A_{FB}(q^2)=\left ( \int_{-1}^0 d \cos \theta_l \frac{d^2 \Gamma}{d q^2 d \cos \theta_l}- \int_0^1 d \cos \theta_l \frac{d^2 \Gamma}{d q^2 d \cos \theta_l}\right )\Big {/}\frac{d \Gamma}{d q^2}\,.
\eea

\item Longitudinal lepton and hadron polarization asymmetry parameters:
\bea
P_L^{\tau}(q^2)=\frac{{\rm d}\Gamma^{\lambda_{\tau}=1/2}/{\rm d}q^2-
    {\rm d}\Gamma^{\lambda_{\tau}=-1/2}/{\rm d}q^2}{{\rm d}\Gamma/{\rm d}q^2}\,,\nonumber\\    
P_L^{\Lambda_c}(q^2)=\frac{{\rm d}\Gamma^{\lambda_2=1/2}/{\rm d}q^2-
    {\rm d}\Gamma^{\lambda_2=-1/2}/{\rm d}q^2}{{\rm d}\Gamma/{\rm d}q^2}\,,
\label{Pol}
\eea
where ${\rm d}\Gamma^{\lambda_\tau=\pm 1/2}$ and ${\rm d}\Gamma^{\lambda_2=\pm 1/2}$  are the individual helicity-dependent differential decay rates, whose detailed expressions are given in reference \cite {Li:2016pdv}.
\item Lepton non-universality parameter:
\bea
&& R_{\Lambda_c}=\frac{{\rm Br}(\Lambda_b \to\Lambda_c \tau^- \bar{\nu}_\tau)}{{\rm Br}( \Lambda_b \to\Lambda_c\, l^- \bar{\nu}_l)} \,, ~~~l=e,\mu.
\eea
\end{itemize}

 Using the best fit values of individual new Wilson coefficients from Table \ref{Tab:Best-fit}\,, we show the graphical representations of various observables of $\Lambda_b \to\Lambda_c \tau \bar{\nu}_\tau$ process below.

\subsection{Case I}
\begin{figure}[htb]
\includegraphics[scale=0.3]{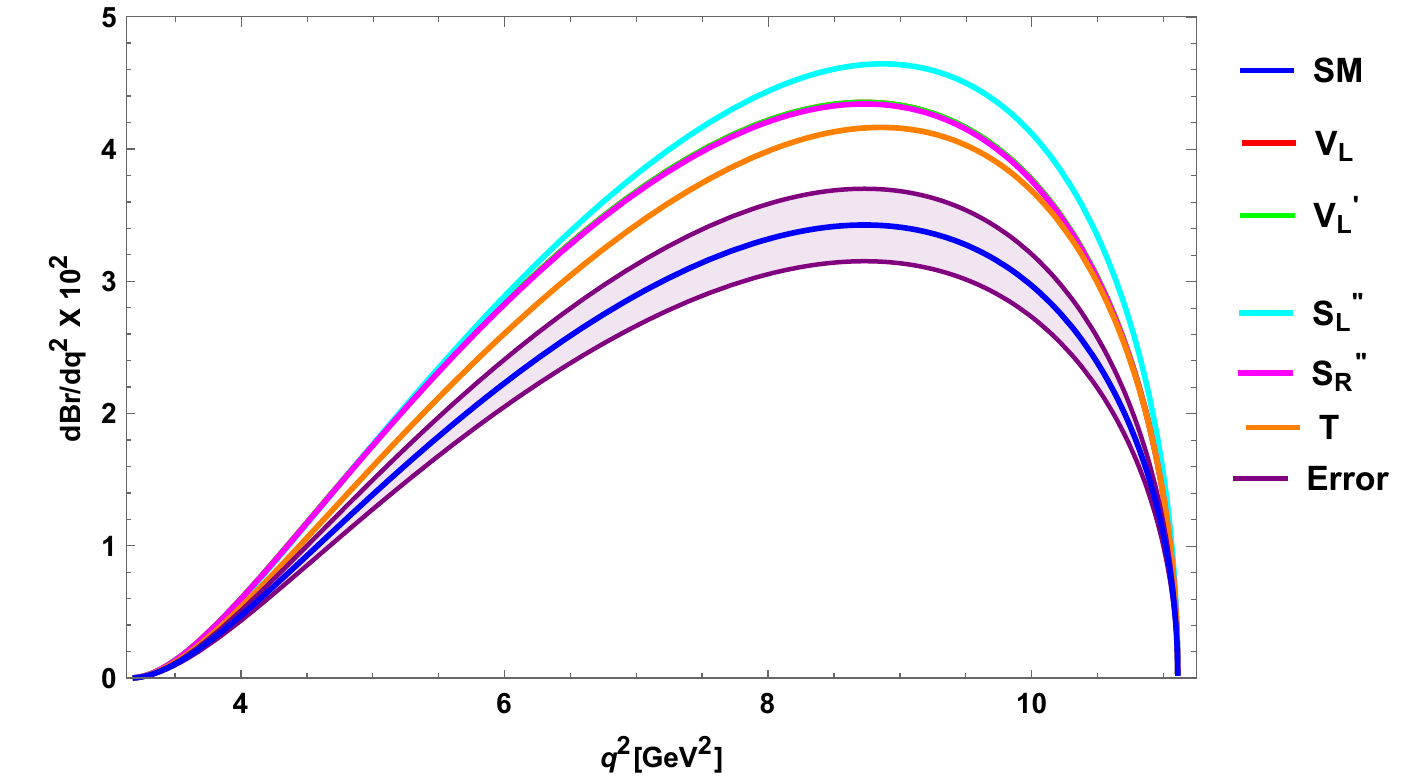}
\quad
\includegraphics[scale=0.3]{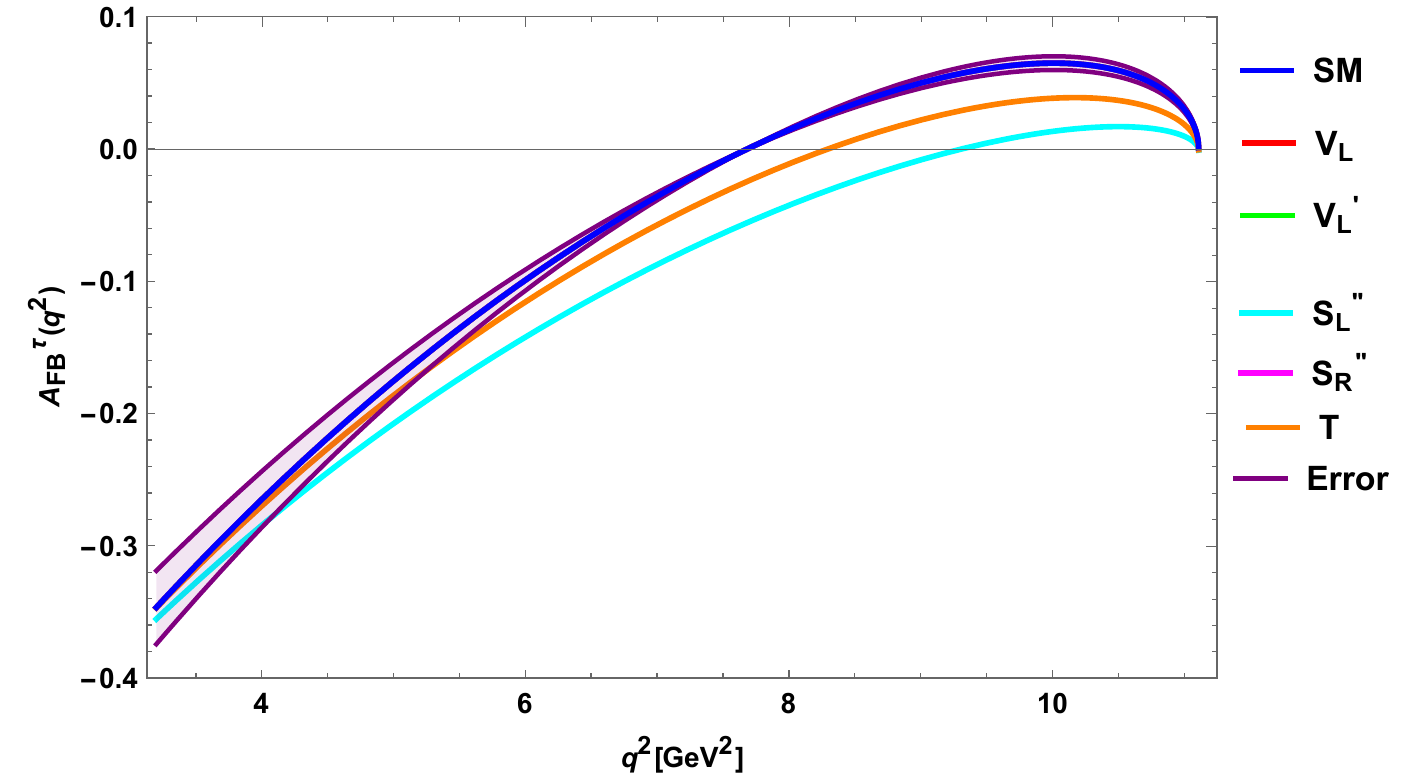}
\quad
\includegraphics[scale=0.3]{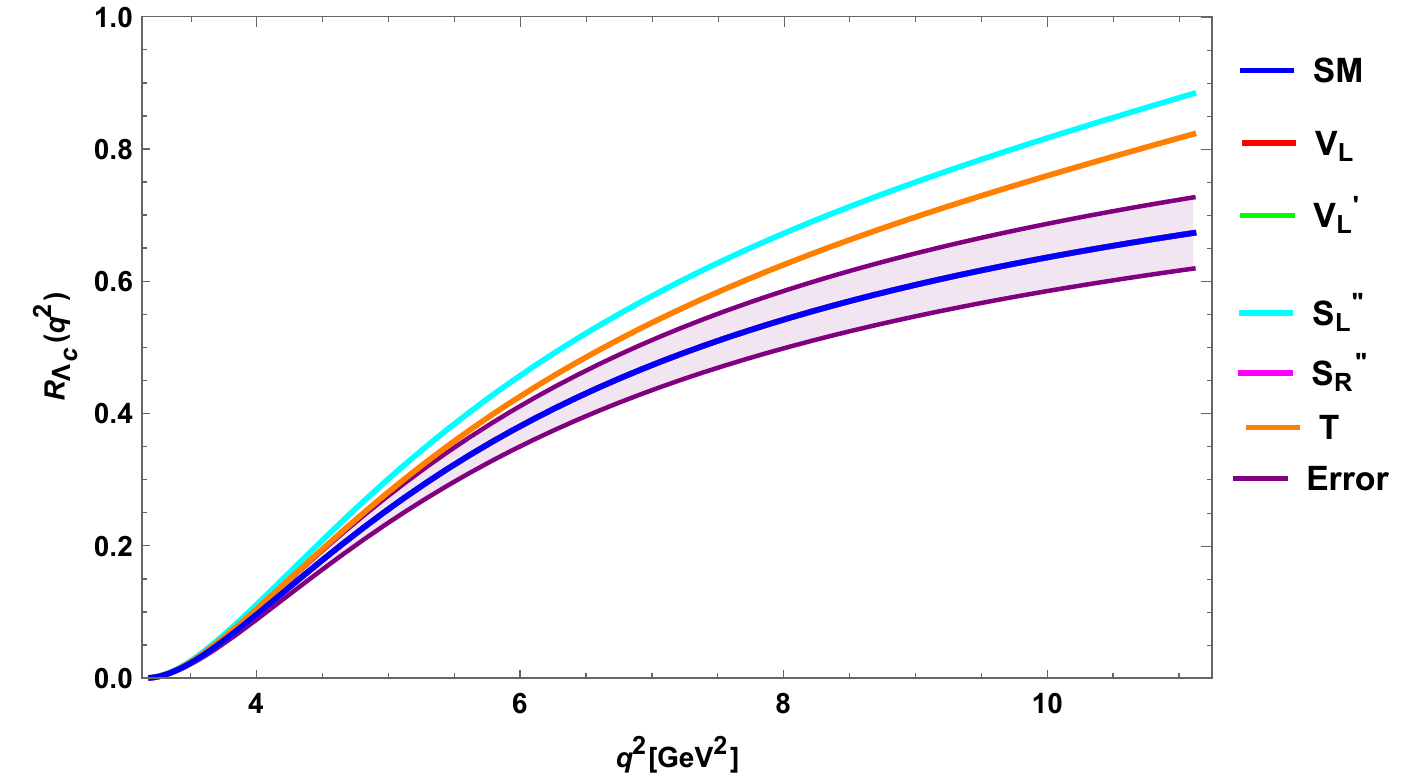}
\quad
\includegraphics[scale=0.3]{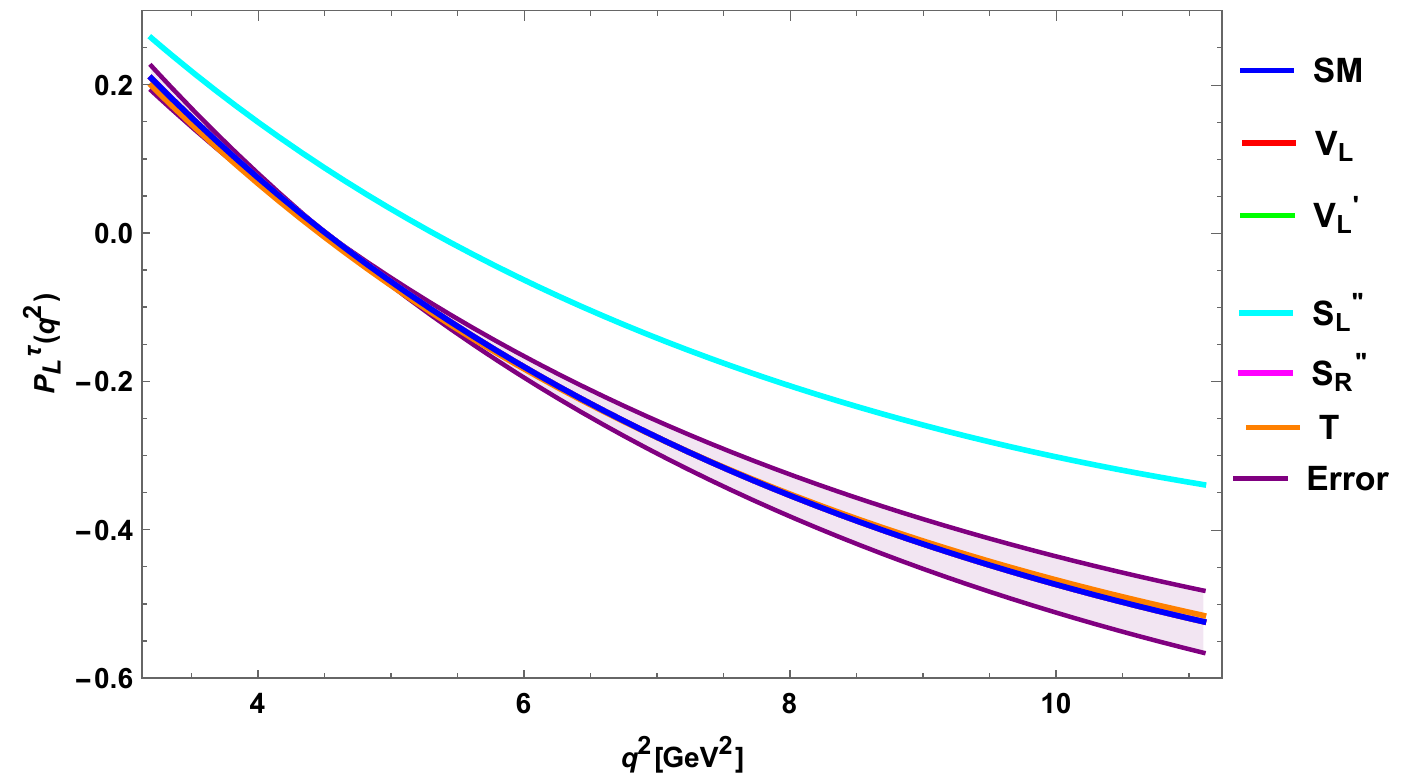}
\quad
\includegraphics[scale=0.3]{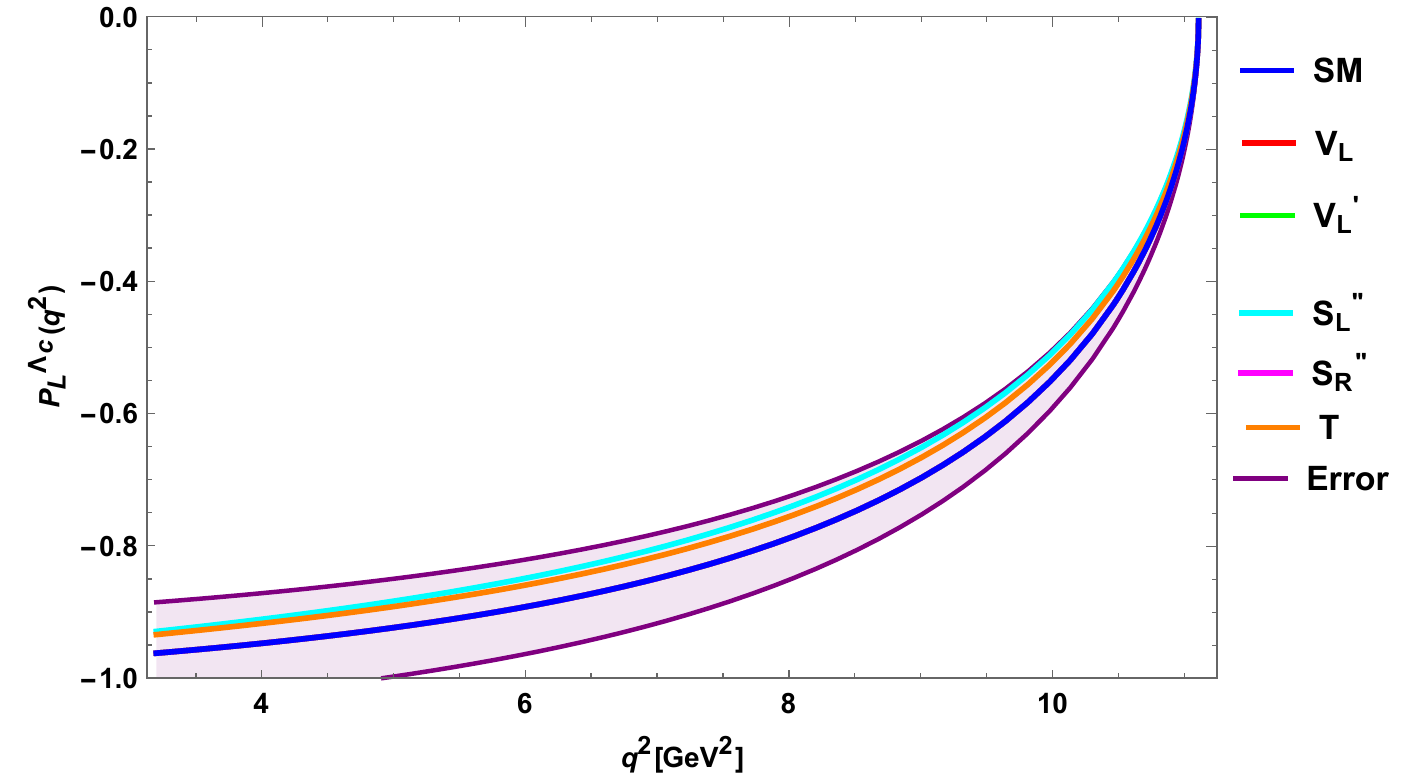}
\caption{ The branching ratio (top-left), forward-backward asymmetry (top-right), $R_{\Lambda_c}$ (middle-left), tau (middle-right) and $\Lambda_c$  (bottom) longitudinal polarization asymmetry of $\Lambda_b \to \Lambda_c  \tau \bar \nu_\tau$   for case I, where the legends $V_L, V_L', \cdots$ represent the contributions arising from the new coefficients $C_{V_L}, C_{V_L}', \cdots$ respectively.}\label{Fig:Case-I Lambdab}
\end{figure}

Taking into consideration the presence of a single  Wilson coefficient at one time, whose prophesied best-fit values are provided in Table  \ref{Tab:Best-fit}\,, we display the branching ratio (top-left panel), forward-backward asymmetry (top-right panel), $R_{\Lambda_c}$ (middle-left  panel), longitudinal polarization asymmetry of tau (middle-right panel) and $\Lambda_c$  (bottom panel)  of $\Lambda_b \to\Lambda_c \tau \bar{\nu}_\tau$ decay process for case I, in Figure[\ref{Fig:Case-I Lambdab}]. In these plots blue lines represent the SM predictions, Purple band represents the $1\sigma$ uncertainties for SM, and red, green, cyan blue, pink and orange lines are obtained by using the  $ C_{V_L}, C_{V_L}', C_{S_L}'',C_{S_R}''$ and $C_T$ coefficients sequentially. The graphical representation for branching ratio, reveals that all the coefficients convey significant deviations from SM prediction. Here we can spot that due to the degeneracy of $ C_{V_L}, C_{V_L}' $ and $C_{S_R}''$ the  corresponding results  coexist with one line and give reasonable deviations from the SM prediction. From the forward-backward asymmetry plot, we conclude that $S_L''$ and $C_T$ indicate sizeable deviation from SM result with the  zero crossing limits shifted towards high-$q^2$ regime. However, the effect of $ C_{V_L}, C_{V_L}'$ and $S_R''$ are  consistent with the SM results. In the lepton non universality parameter $ R_{\Lambda_c}$, $S_L''$ and $C_T$ give an acceptable deviation from SM prediction and discrepancies of  $ C_{V_L}, C_{V_L}'$ and $C_{S_R}''$ are negligible. In case of $ P_L^{\tau}$, $S_L''$ gives profound deviation whereas other coefficients show no deviations.  In $ P_L^{\Lambda_c} $ asymmetry $C_{S_L}''$ and $C_T$ give marginal deviation compared to other observable plots, whereas other coefficients show no deviations. Taking the $1\sigma$ uncertainties of the parameters using in the calculations, in each table we show the central values and $1\sigma$ uncertainties for the  observable. Here the calculated values of branching ratio and angular observable of  $\Lambda_b \to \Lambda_c  \tau \bar \nu_\tau$ process in the SM and in the presence of new complex Wilson coefficients for case I is presented in the Table \ref{Tab:CI-Lambdab}\,.
\begin{table}[htb]
\scriptsize
\caption{Calculated values of branching ratio and angular observables of  $\Lambda_b \to \Lambda_c \tau \bar \nu_\tau$ process in the SM and in the presence of new complex Wilson coefficients (case I ).}\label{Tab:CI-Lambdab}
\begin{center}
\begin{tabular}{|c|c|c|c|c|c|c|}
\hline
 ~Observables~&~Values for SM~&~values for $C_{V_L}$~&~Values for $C_{V_L}'$~&~values for $C_{S_L}^{''}$~&~Values for $C_{S_R}^{''}$~&~Values for $C_T$~\\
\hline
~ Br$(\Lambda_b \to \Lambda_c  \tau \bar \nu_\tau)$~&~$0.017 \pm 0.004$~&~$0.022 \pm 0.002$~&~$0.022 \pm 0.002$~&~$0.023 \pm 0.005$~&~$0.022 \pm 0.002$~&~$0.021 \pm 0.006$~\\
\hline
~ Br$(\Lambda_b \to \Lambda_c  \mu \bar \nu_\tau)$~&~$0.049 \pm 0.006$~&~$0.063 \pm 0.002$~&~$0.063 \pm 0.002$~&~$0.053 \pm 0.003$~&~$0.063 \pm 0.002$~&~$0.052 \pm 0.002$~\\
\hline

~$ A_{FB}^{\Lambda_c}$~&~$0.09 \pm 0.006$~&~$0.09 \pm 0.003$~&~$0.09\pm 0.003$~&~$0.12\pm 0.027$~&~$0.09\pm 0.003$~&~$0.102\pm 0.032$~\\
\hline
~$ R_{\Lambda_c}$~&~$0.352 \pm 0.031$~&~$0.352\pm 0.042$~&~$0.352\pm 0.042$~&~$0.436\pm 0.029$~&~$0.352\pm 0.042$~&~$0.405\pm 0.027$~\\
\hline

~$ P_L^{\Lambda_c} $~&~$-0.796 \pm 0.820$~&~$-0.796 \pm 0.820$~&~$-0.796 \pm 0.084$~&~$-0.752 \pm 0.691$~&~$-0.796 \pm 0.820$~&~$-0.765 \pm 0.581$~\\
\hline
~$ P_L^{\tau} $~&~$-0.207 \pm 0.140$~&~$-0.207 \pm 0.141$~&~$-0.207 \pm 0.141$~&~$-0.086 \pm 0.049$~&~$-0.207 \pm 0.320$~&~$-0.209 \pm 0.512$~\\
\hline
\end{tabular}
\end{center}
\end{table}
\subsection{Case II}
\subsubsection{A: Presence of $C_i ~\& ~C_j$ coefficients}
\begin{figure}[htb]
\includegraphics[scale=0.3]{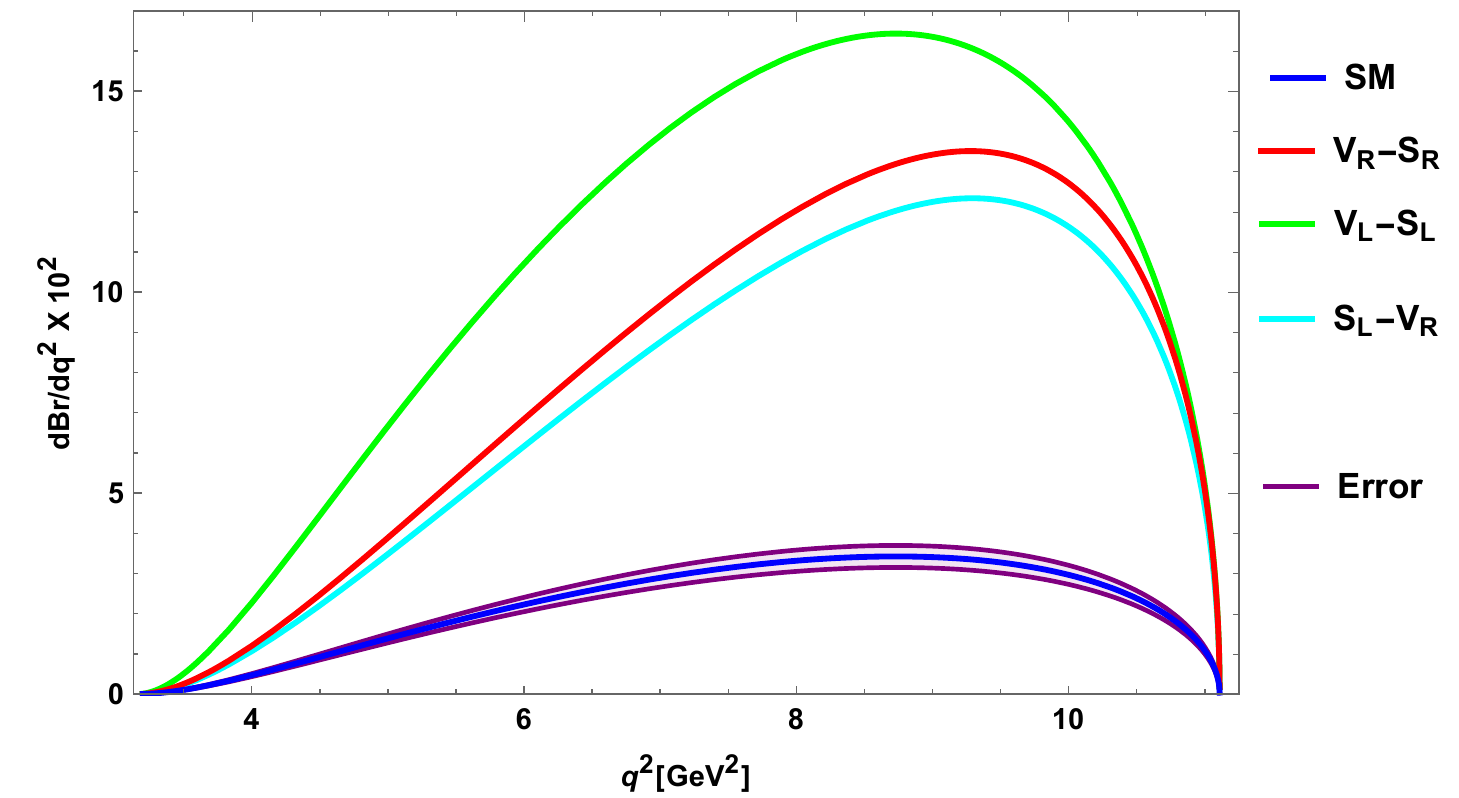}
\quad
\includegraphics[scale=0.3]{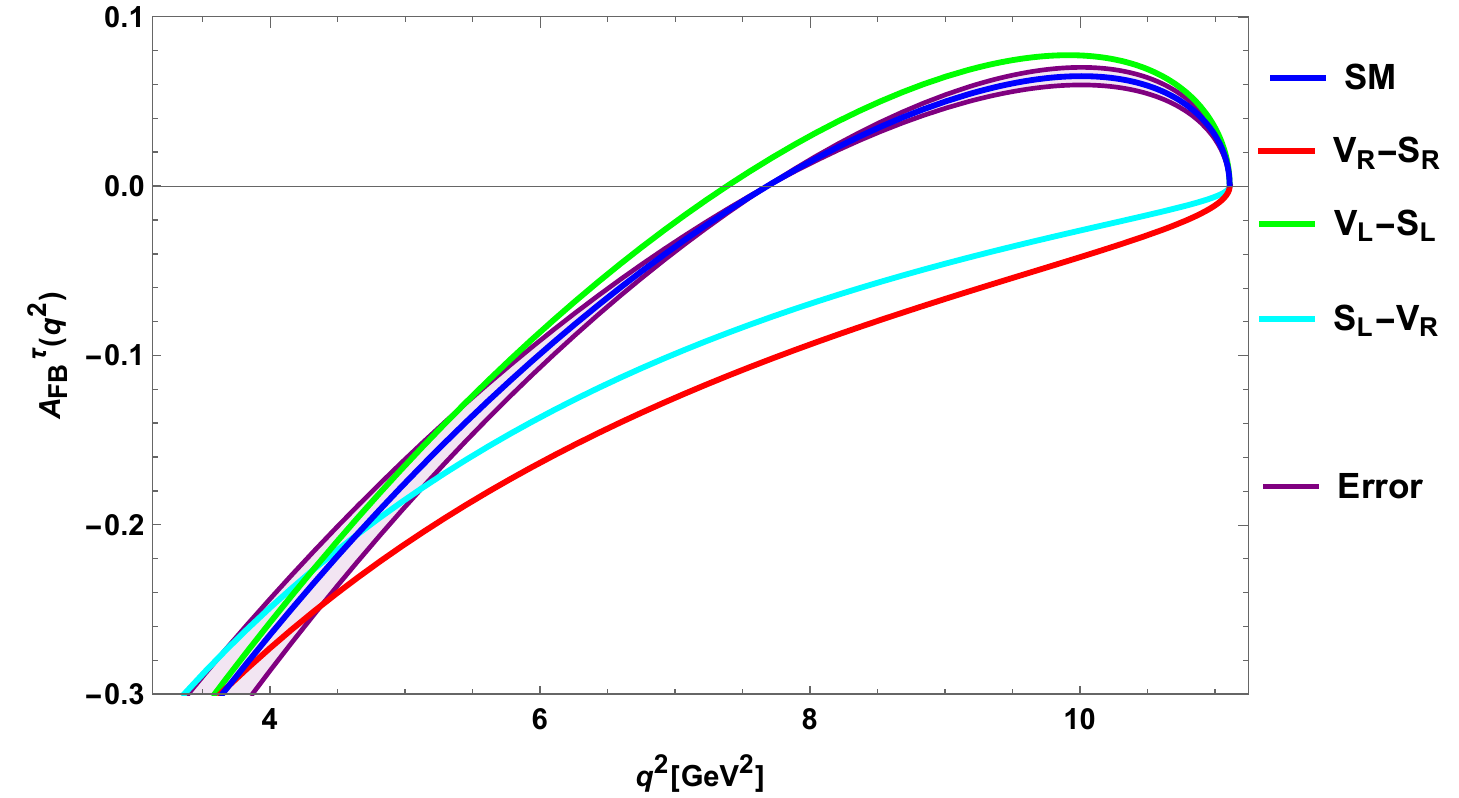}
\quad
\includegraphics[scale=0.3]{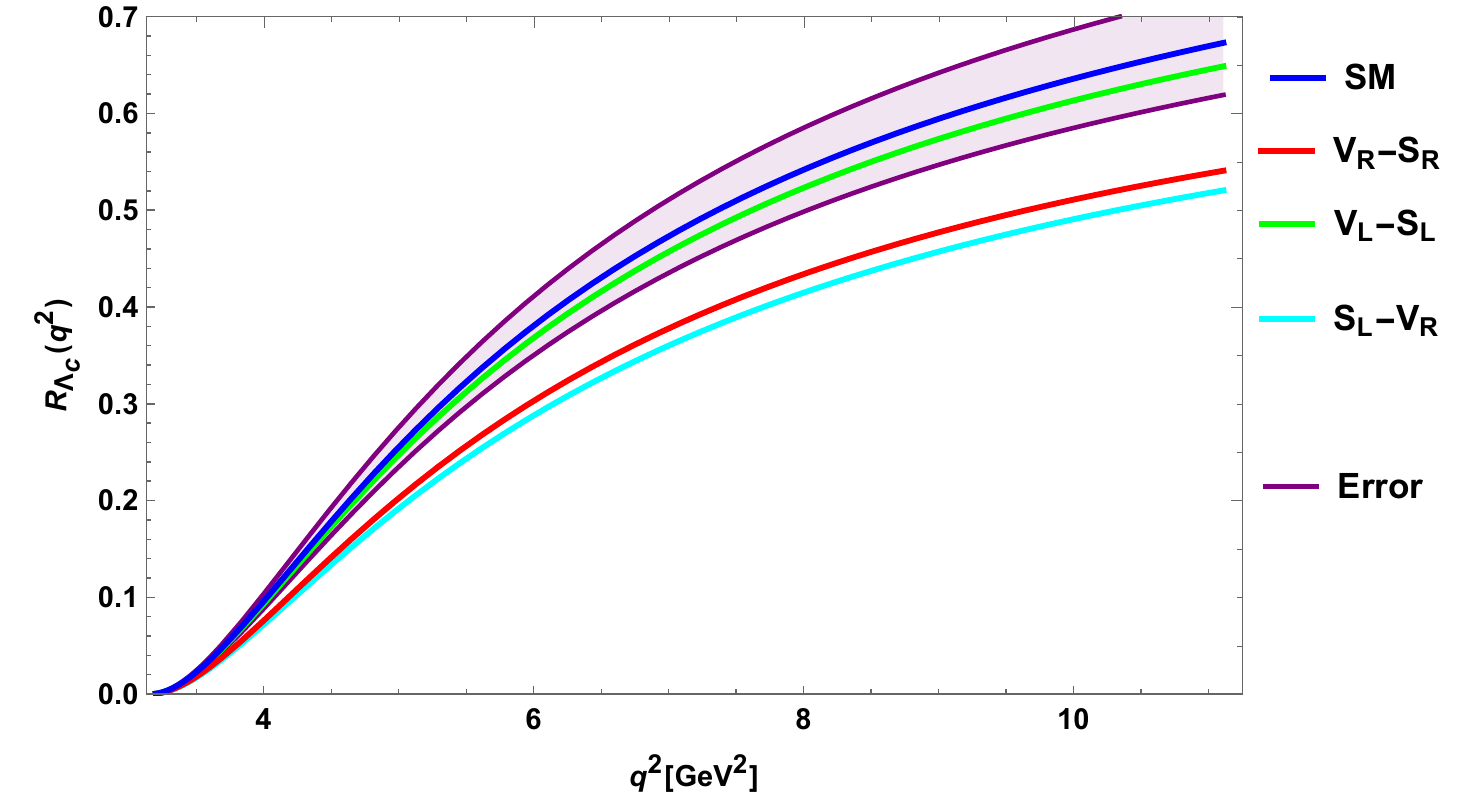}
\quad
\includegraphics[scale=0.3]{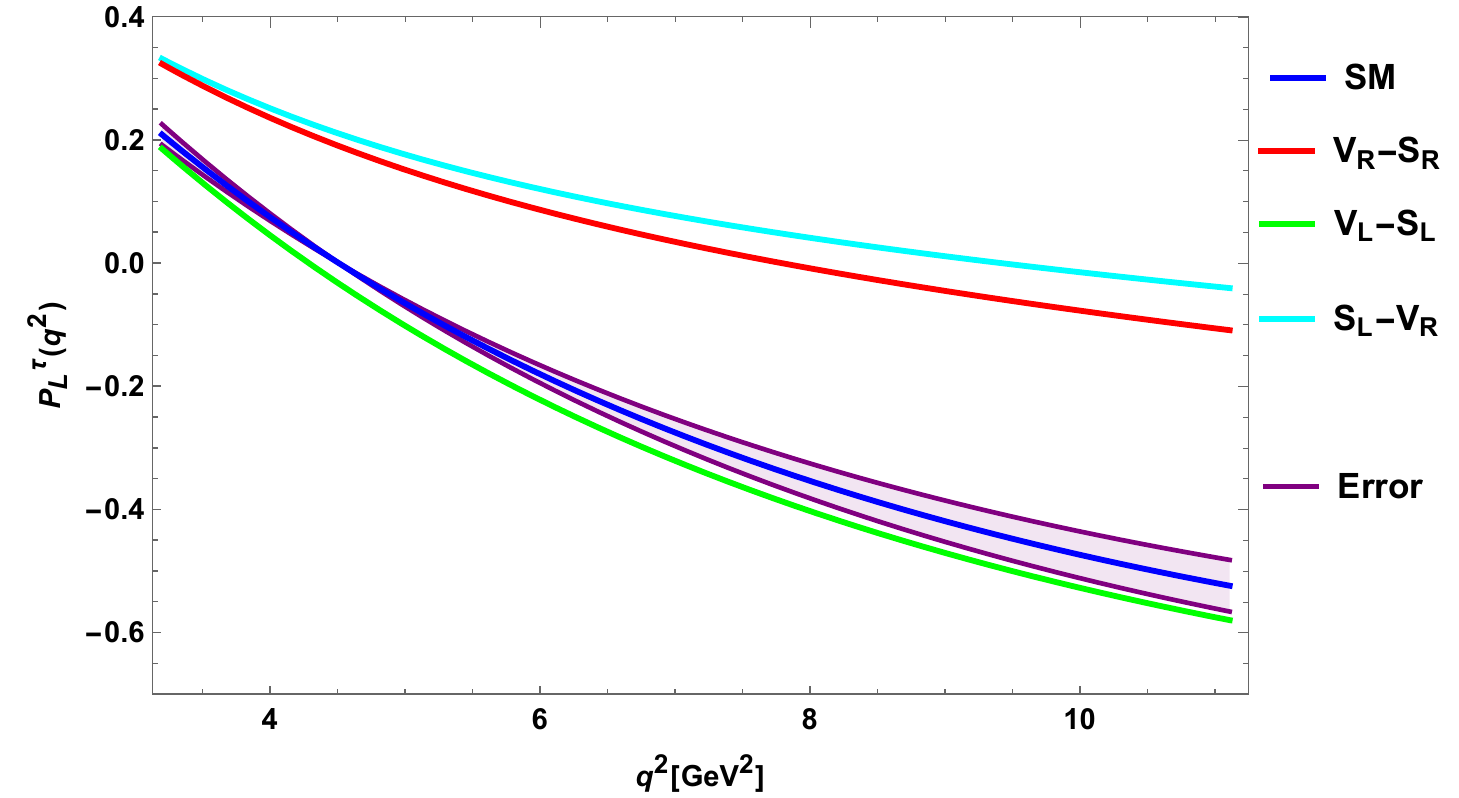}
\quad
\includegraphics[scale=0.3]{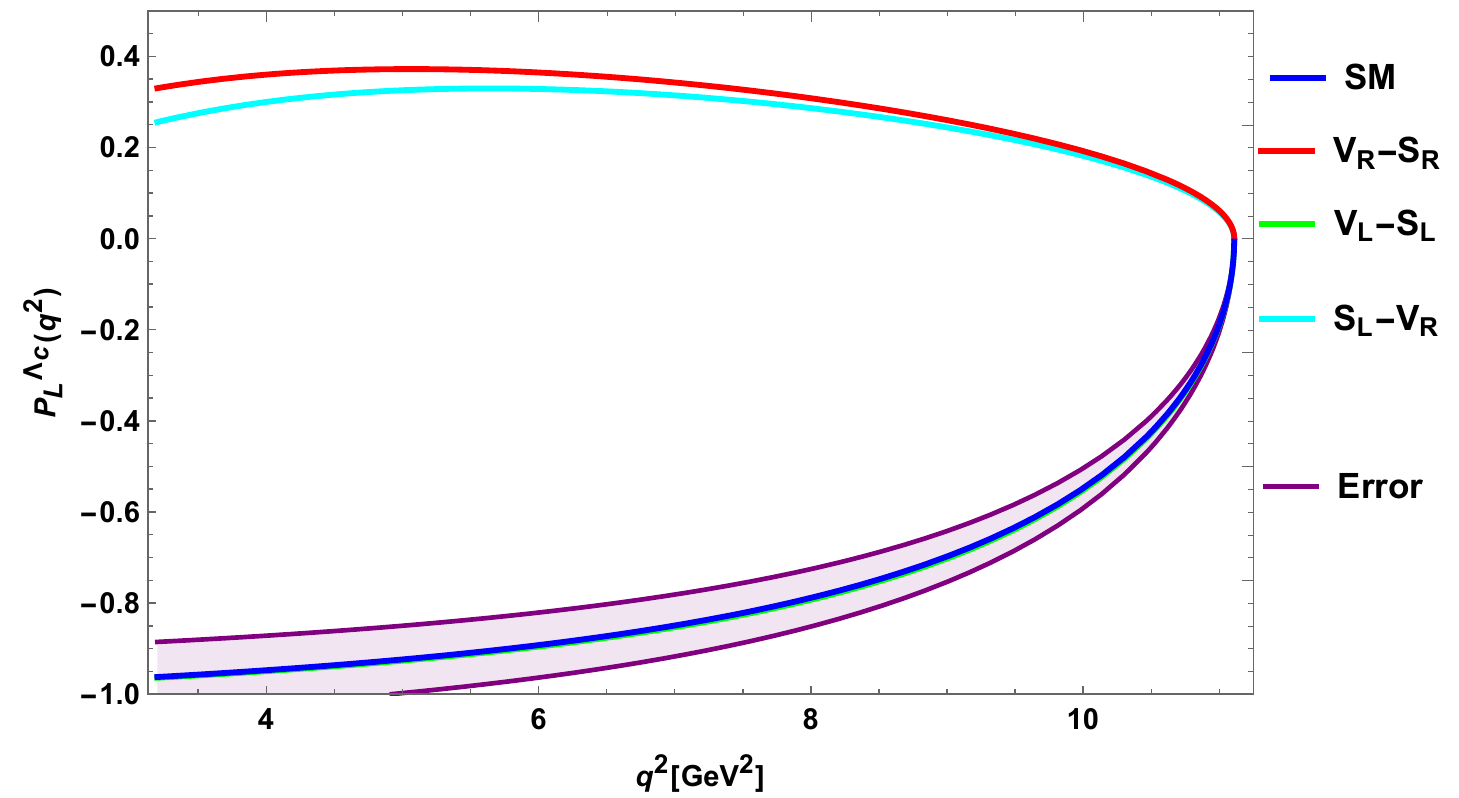}
\caption{ The branching ratio (top-left), forward-backward asymmetry (top-right), $R_{\Lambda_c}$ (middle-left), tau (middle-right) and $\Lambda_c$  (bottom) longitudinal polarization asymmetry of $\Lambda_b \to \Lambda_c  \tau \bar \nu_\tau$   for case  IIA, where the legends $V_R-S_R$, $V_L-S_L$, $S_L-V_R$ represent the combination of new coefficients $(C_{V_R}, C_{S_R})$, $(C_{V_L},C_{S_L})$, $(C_{S_L},C_{V_R})$ respectively.}\label{Fig:Case- IIA Lambdab}
\end{figure}

We derive the involvement of two different unprimed coefficients on the branching ratio (top-left), forward-backward asymmetry (top-right), $R_{\Lambda_c}$(middle-left), tau (middle-right) and $\Lambda_c$  (bottom) longitudinal polarization asymmetry of $\Lambda_b \to \Lambda_c  \tau \bar \nu_\tau$ for case  IIA,  in Figure[\ref{Fig:Case- IIA Lambdab}]. The blue, green, cyan  and red color solid lines are representing the SM, $(C_{V_L},C_{S_L})$, $ (C_{S_L},C_{V_R})$ and ($C_{V_R},C_{S_R}) $ respectively. In the branching fraction plot, all the coefficients are giving significant deviation from the SM prediction but the $(C_{V_L},C_{S_L}) $ is showing maximum disparity. The forward-backward asymmetry representations show a reasonable deviation for all the coefficients but $(C_{V_L},C_{S_L})$ gives a significant difference to the zero crossing point. In $R_{\Lambda_c}$ representation, $(C_{V_L},C_{S_L})$ shows marginal deviation whereas $ (C_{S_L},C_{V_R})$ and $(C_{V_R},C_{S_R} )$ give a quite good variance. In case of $ P_L^{\tau} $,  $(C_{V_L},C_{S_L})$ gives a slight deviation whereas  $(C_{S_L},C_{V_R})$ and $(C_{V_R},C_{S_R} )$ show a observable deviation from SM.  In case of $ P_L^{\Lambda_c} $, there is no deviation for $(C_{V_L},C_{S_L})$ whereas  $(C_{ S_L},C_{V_R})$ and $(V_R,S_R) $ show significant deviation from the SM. Calculated central and $1\sigma$ uncertainty values  for branching fraction, forward- backward asymmetry, $R_{\Lambda_c}$, $ P_L^{\tau} $ and $P_L^{\Lambda_c}$ of  $\Lambda_b \to \Lambda_c  \tau \bar \nu_\tau$ process in the presence of two unprimed Wilson coefficients for case IIA are given in  Table \ref{Tab:CIIA-Lambdab}\,.
\begin{table}[htb]
\scriptsize
\caption{Calculated values of branching ratio and angular observables of  $\Lambda_b \to \Lambda_c \tau \bar \nu_\tau$ process  in the presence of new complex Wilson coefficients (case IIA ).}\label{Tab:CIIA-Lambdab}
\begin{center}
\begin{tabular}{|c|c|c|c|c|c|}
\hline
 ~Observables~&~Values for $(C_{V_R},C_{S_R})$~&~Values for $(C_{V_L},C_{S_L})$~&~Values for $(C_{S_L},C_{V_R})$ ~\\
\hline
~ Br$(\Lambda_b \to \Lambda_c  \tau \bar \nu_\tau)$~&~$0.064 \pm 0.006$~&~$0.084 \pm 0.004$~&~$0.058 \pm 0.007$~\\
\hline
~ Br$(\Lambda_b \to \Lambda_c  \mu \bar \nu_\tau)$~&~$0.197 \pm 0.003$~&~$0.247 \pm 0.008$~&~$0.186 \pm 0.006$~\\
\hline
~$ A_{FB}^{\Lambda_c}$~&~$0.135\pm 0.065$~&~$0.078\pm 0.008$~&~$0.112\pm 0.026$~\\
\hline
~$ R_{\Lambda_c}$~&~$0.324\pm 0.081$~&~$0.340\pm 0.047$~&~$0.311\pm 0.028$~\\
\hline
~$ P_L^{\Lambda_c} $~&~$0.292\pm 0.056$~&~$-0.797\pm 0.280$~&~$0.261\pm 0.059$~\\
\hline
~$ P_L^{\tau} $~&~$0.049\pm 0.006$~&~$-0.248\pm 0.348$~&~$0.090\pm 0.008$~\\
\hline
\end{tabular}
\end{center}
\end{table}

\subsubsection{B: Presence of $C^{'}_i ~\&~  C^{'}_j$  coefficients}  
\begin{figure}[htb]
\includegraphics[scale=0.3]{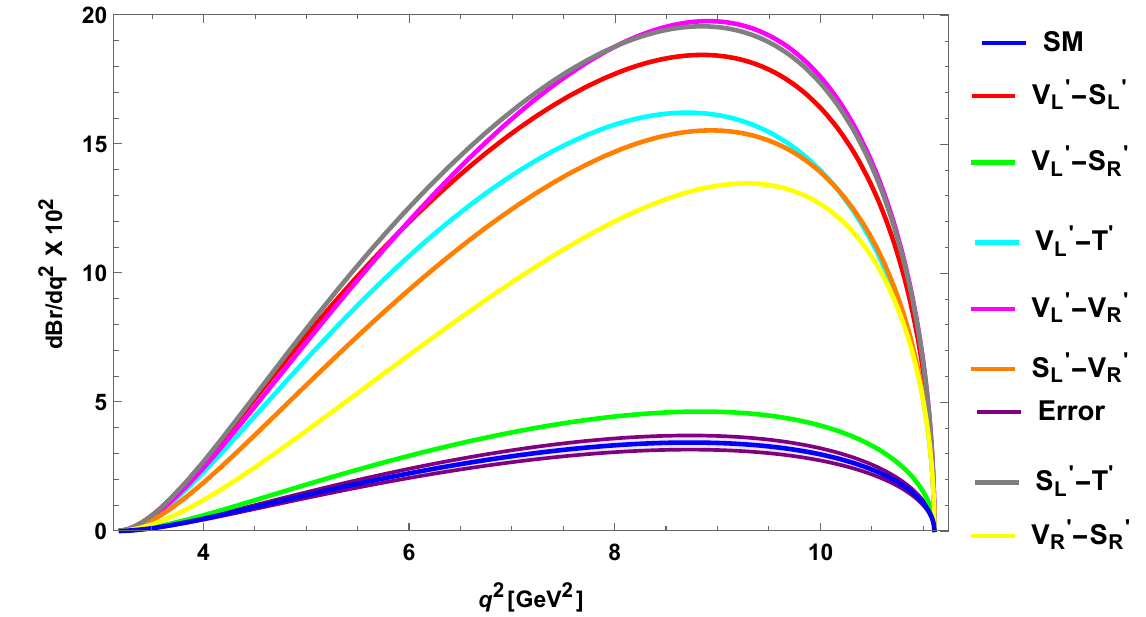}
\quad
\includegraphics[scale=0.3]{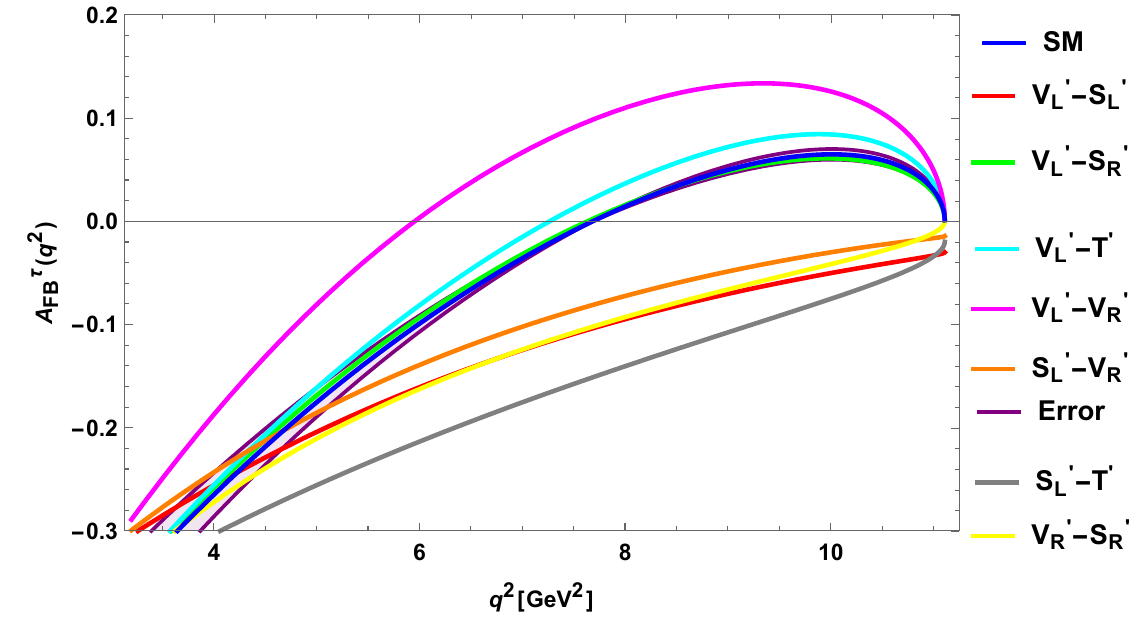}
\quad
\includegraphics[scale=0.3]{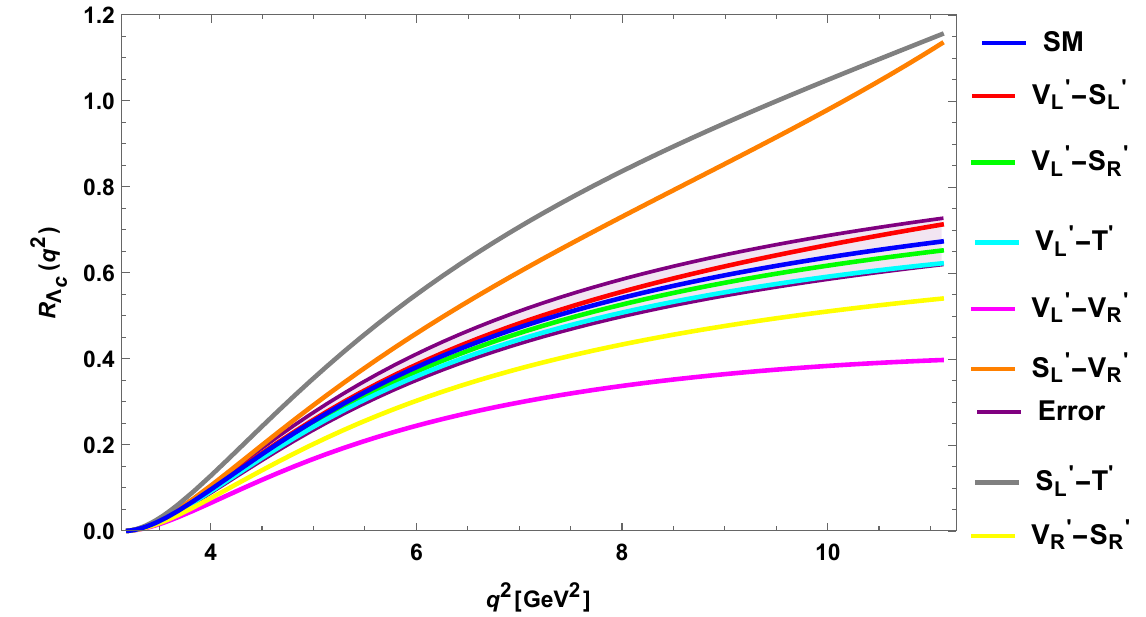}
\quad
\includegraphics[scale=0.3]{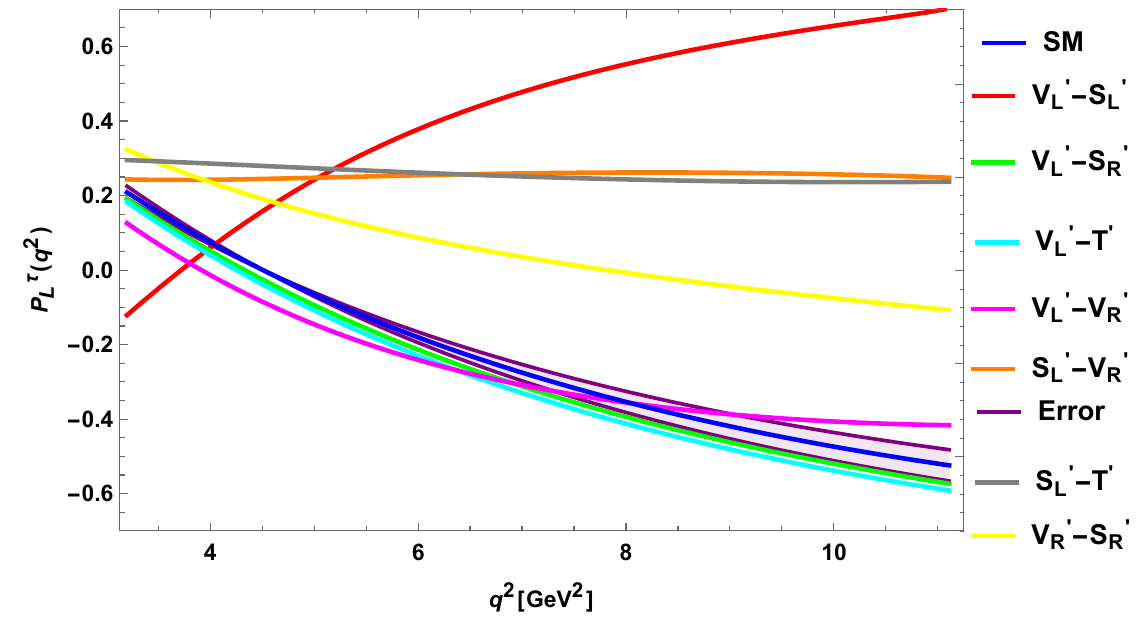}
\quad
\includegraphics[scale=0.3]{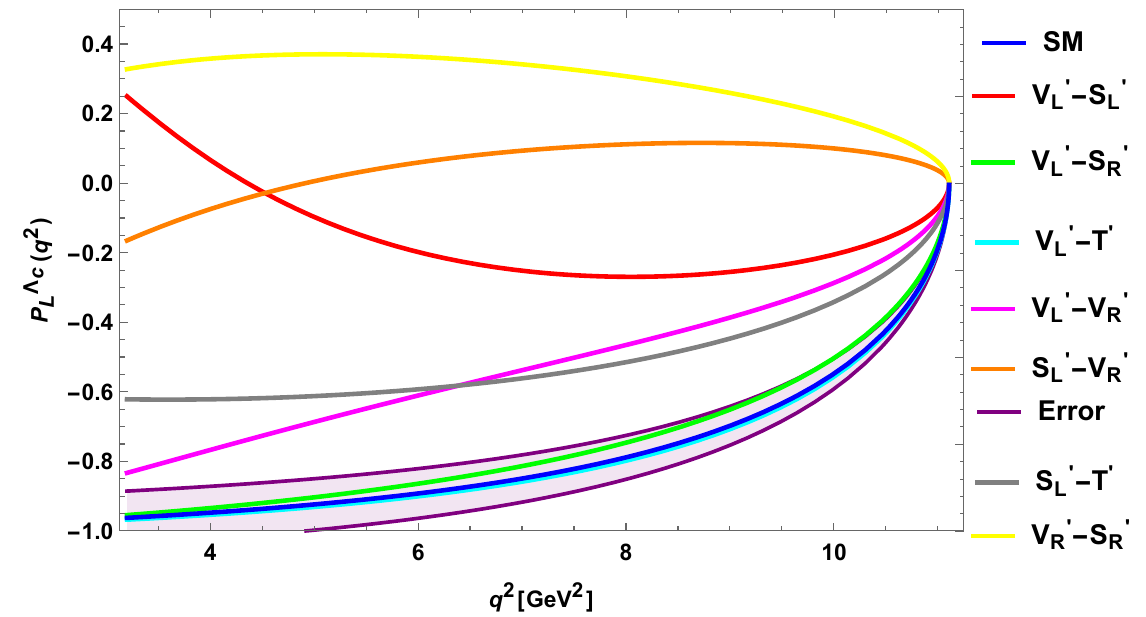}
\caption{ The branching ratio (top-left), forward-backward asymmetry (top-right), $R_{\Lambda_c}$ (middle-left), tau (middle-right) and $\Lambda_c$  (bottom) longitudinal polarization asymmetry of $\Lambda_b \to \Lambda_c  \tau \bar \nu_\tau$   for case  IIB, where the plot legends represent diffrent combinations of the new primed coefficients.}\label{Fig:Case- IIB Lambdab}
\end{figure}

Here we show the implications of two different primed coefficients on the branching ratio (top-left), forward-backward asymmetry (top-right), $R_{\Lambda_c}$ (middle-left), longitudinal polarization asymmetries of tau (middle-right) and $\Lambda_c$  (bottom) of $\Lambda_b \to \Lambda_c  \tau \bar \nu_\tau$   for case  IIB, in Figure[\ref{Fig:Case- IIB Lambdab}]. The blue, red, green, cyan blue, magenta, orange, gray and yellow color solid lines are representing the SM, ($C_{V_L}',C_{S_L}'$), $(C_{V_L}',C_{S_R}')$, $(C_{V_L}',C_{T}')$, $(C_{V_L}',C_{V_R}')$, $(C_{S_L}',C_{V_R}')$, $(C_{S_L}',C_T')$ and $(C_{V_R}',C_{S_R}')$ respectively, where as  Purple band represents the $1\sigma$ uncertainties for SM. In the branching fraction representation, all the coefficients are showing significant deviations from the SM prediction but the $(C_{V_L}',C_{S_R}')$ is giving minimum deviation. $(C_{V_L}',C_{V_R}')$ and $(C_{S_L}',C_T')$ show similar results which is the maximum deviation. In the forward-backward asymmetry plot $(C_{V_L}',C_{V_R}')$ shows maximum deviation where as $(V_L',S_R')$ expresses negligible deviation from SM value. All other constraints indicate some significant variance, but among those $(C_{V_L}',C_{V_R}')$ and $(C_{V_L}',C_T')$ predict  substantial deviations and also in the zero crossing point. In $R_{\Lambda_c}$, the $(C_{V_L}',C_{S_R}')$ conveys slight deviation whereas all other coefficients show significant deviation from SM prediction.  In $ P_L^{\tau} $ representation, each coefficients show a standard deviation from SM whereas  $(C_{V_L}',C_{S_R}')$ and  $(C_{V_L}',C_T')$ show some similar marginal discrepancies.  In $ P_L^{\Lambda_c} $ representation, each coefficients show profound deviation from SM whereas  $(C_{V_L}',C_{S_R}')$ show  slight discrepancy and  $(C_{V_L}',C_T')$ deviation is very negligible. Envisaged results for branching ratio, forward- backward asymmetry, $R_{\Lambda_c}$ and $ P_L^{\tau} $ of   $\Lambda_b \to \Lambda_c \tau \bar \nu_\tau$ process in the presence of new  Wilson coefficients for case IIB are given in the Table \ref{Tab:CIIB-Lambdab}\,.

\begin{table}[htb]
\tiny
\caption{Calculated values of branching ratio and angular observables of  $\Lambda_b \to \Lambda_c \tau \bar \nu_\tau$ process in the presence of new complex Wilson coefficients (case IIB ).}\label{Tab:CIIB-Lambdab}
\begin{center}
\begin{tabular}{|c|c|c|c|c|c|c|c|}
\hline
 ~Observables~&~ $(C_{V_L}',C_{S_L}')$~&~ $(C_{V_L}',C_{S_R}')$~&~$(C_{V_L}',C_T')$ ~&~$(C_{V_L}',C_{V_R}')$~&~$(C_{S_L}',C_{V_R}')$~&~ $(C_{S_L}',C_T')$~&~ $(C_{V_R}',C_{S_R}')$~\\
\hline
~ Br$(\Lambda_b \to \Lambda_c  \tau \bar \nu_\tau)$~&~$0.095\pm 0.001$~&~$0.023\pm 0.002$~&~$0.083\pm 0.008$~&~$0.099\pm 0.007$~&~$0.077\pm 0.003$~&~$ 0.100\pm 0.005$~&~$0.063\pm 0.007$~\\
\hline
~ Br$(\Lambda_b \to \Lambda_c  \mu \bar \nu_\tau)$~&~$0.271\pm 0.003$~&~$0.067\pm 0.002$~&~$0.252\pm 0.009$~&~$0.420\pm 0.004$~&~$0.170\pm 0.003$~&~ $ 0.199\pm 0.006 $~&~$ 0.197\pm 0.003 $~\\
\hline

~$A_{FB}^{\Lambda_c}$~&~$0.151\pm 0.019$~&~$0.083\pm 0.007$~&~$0.075 \pm 0.004$~&~$0.010 \pm 0.005$~&~$0.126\pm 0.048$~&~ $0.191\pm 0.019 $~&~ $ 0.134 \pm 0.011$~\\
\hline
~$ R_{\Lambda_c}$~&~$0.351\pm 0.040$~&~$0.349\pm 0.027$~&~$0.330\pm 0.078$~&~$0.236\pm 0.032$~&~$0.456\pm 0.025$~&~$0.503\pm 0.018$~&~$0.324\pm 0.082$~\\
\hline

~$ P_L^{\Lambda_c} $~&~$-0.163\pm 0.321$~&~$-0.761\pm 0.834$~&~$-0.804\pm 0.391$~&~$-0.536\pm 0.528$~&~$0.046\pm 0.004$~&~$ -0.508\pm 0.637 $~&~$0.291\pm 0.011$~\\
\hline
~$ P_L^{\tau} $~&~$0.445\pm 0.013$~&~$-0.247\pm 0.235$~&~$-0.256\pm 0.612$~&~$-0.240\pm 0.327$~&~$0.253\pm 0.013$~&~ $ 0.257\pm 0.011$~&~$ 0.050\pm 0.003 $~\\
\hline
\end{tabular}
\end{center}
\end{table}

\subsubsection{C: Presence of $C^{''}_i ~\&~ C^{''}_j $ coefficients}
\begin{figure}[htb]
\includegraphics[scale=0.3]{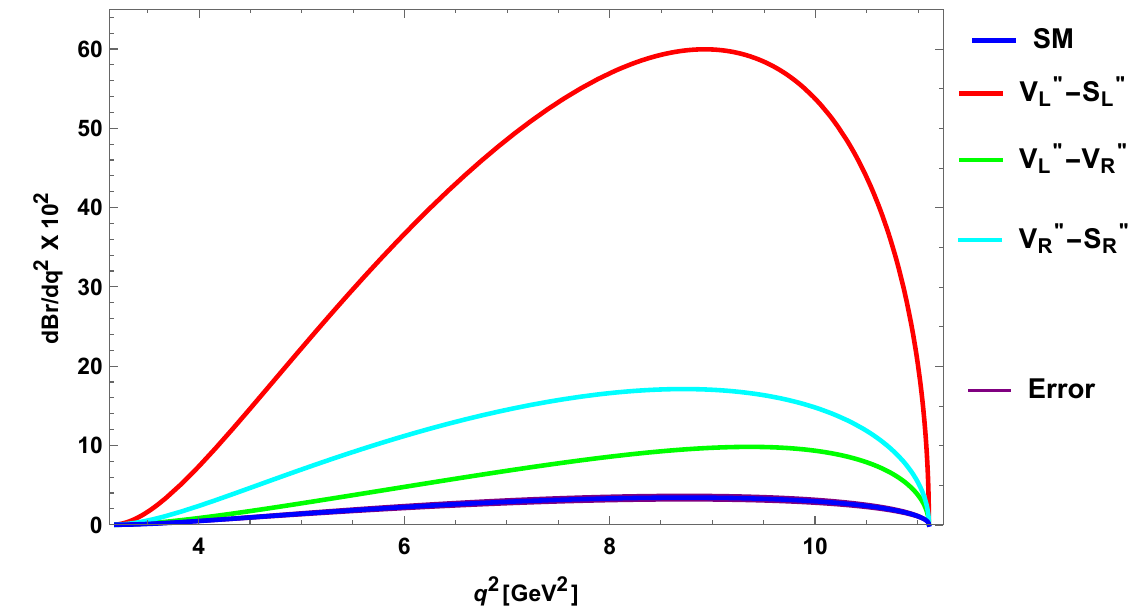}
\quad
\includegraphics[scale=0.3]{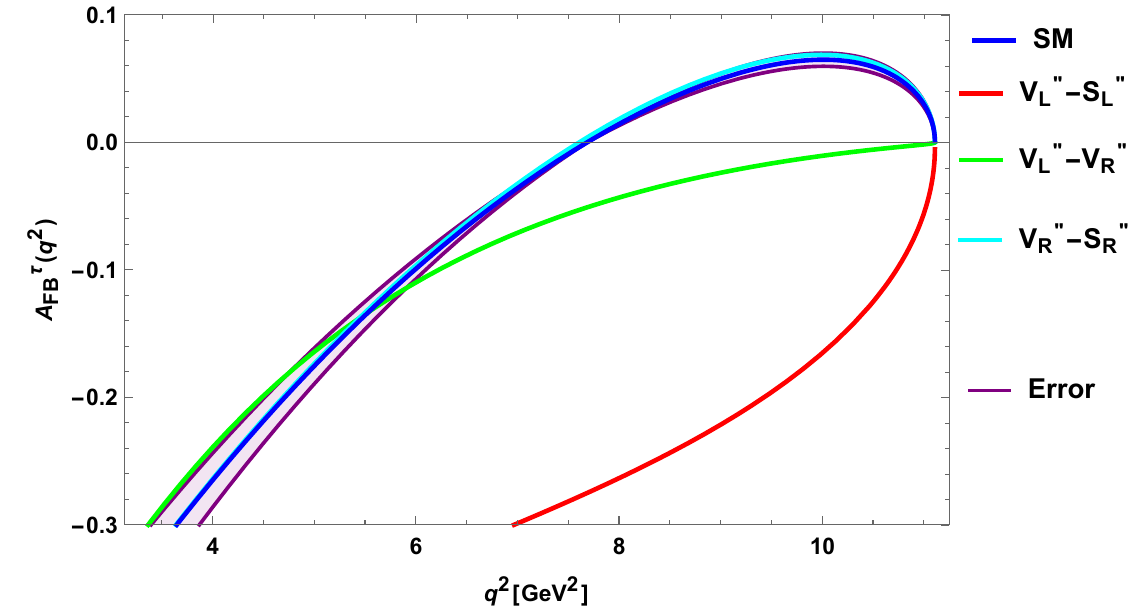}
\quad
\includegraphics[scale=0.3]{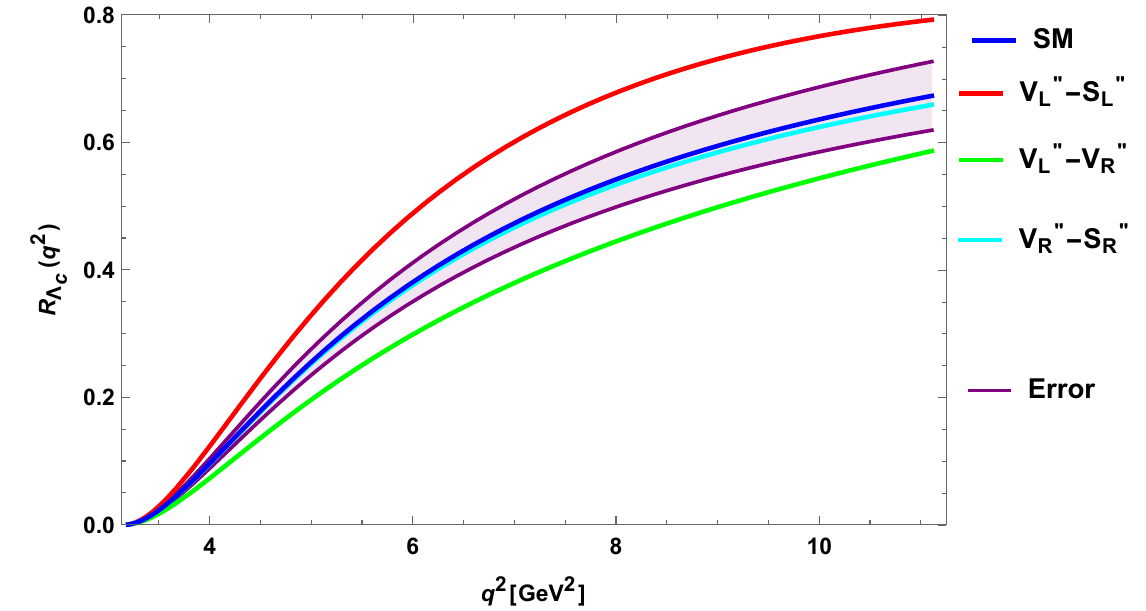}
\quad
\includegraphics[scale=0.3]{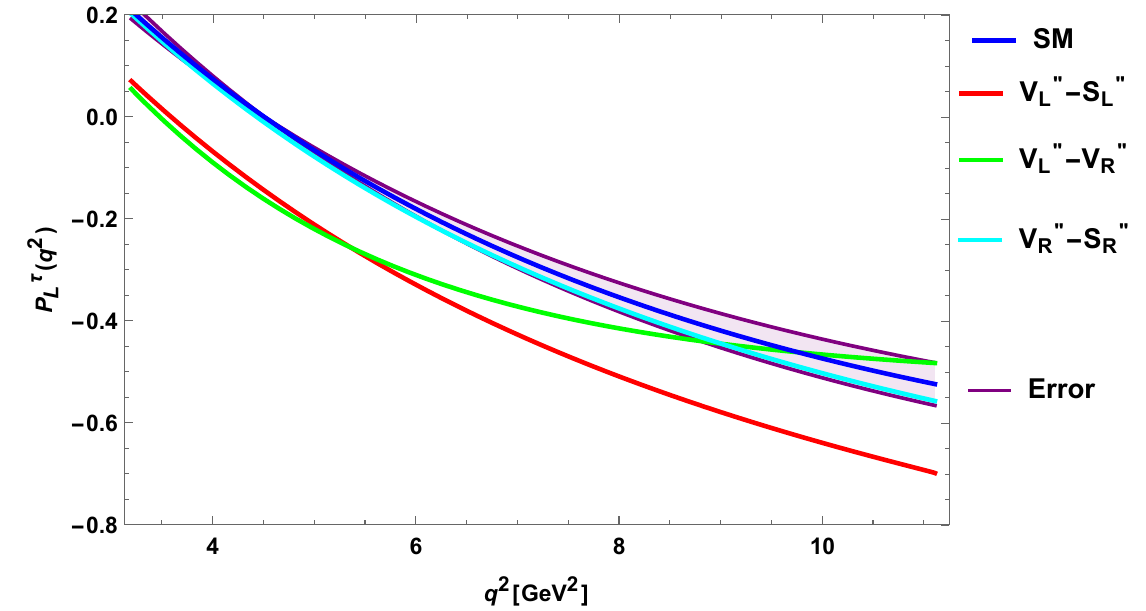}
\quad
\includegraphics[scale=0.3]{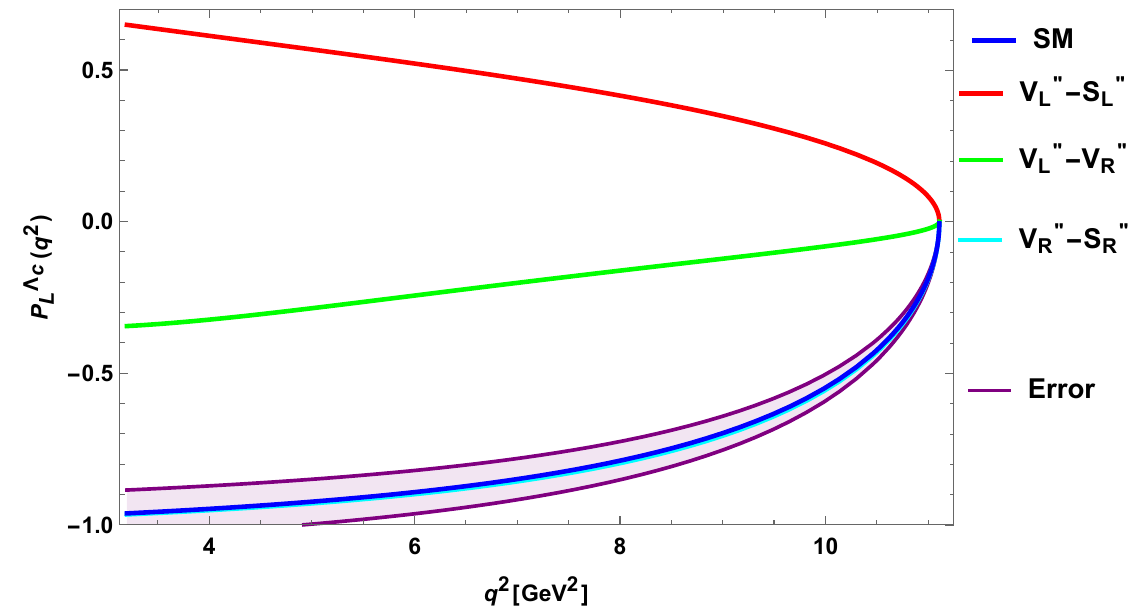}
\caption{ The branching ratio (top-left), forward-backward asymmetry (top-right), $R_{\Lambda_c}$ (middle-left), tau (middle-right) and $\Lambda_c$  (bottom) longitudinal polarization asymmetry of $\Lambda_b \to \Lambda_c  \tau \bar \nu_\tau$   for case  IIC, where the legends represent different combinations of the new double-primed coefficients.}\label{Fig:Case- IIC Lambdab}
\end{figure}

Investigating the case IIC, we consider the relations of two distinct double primed  coefficients on the branching ratio (top-left), forward-backward asymmetry (top-right), $R_{\Lambda_c}$ (middle-left), tau (middle-right) and $\Lambda_c$  (bottom) longitudinal polarization asymmetry of $\Lambda_b \to \Lambda_c  \tau \bar \nu_\tau$   for case  IIC, in Figure[\ref{Fig:Case- IIC Lambdab}]. The blue, red, green and cyan blue color lines are representing the SM, $(C_{V_L}'',C_{S_L}'')$, $ (C_{V_L}'',C_{V_R}'')$ and $(C_{V_R}'',C_{S_R}'')$ respectively, where as  Purple band represents the $1\sigma$ uncertainties for SM. In the graph of branching fraction, all the coefficients are showing significant deviations from the SM prediction but the $(C_{V_L}'',C_{S_L}'')$ shows the maximum discrepancy. In the forward-backward asymmetry, they show a standard deviation for all the coefficients but $(C_{V_R}'',C_{S_R}'')$ gives  a little deviation which is negligible. In $R_{\Lambda_c}$ representation $(C_{V_R}'',C_{S_R}'')$ shows modest deviation where as $(C_{V_L}'',C_{V_R}'')$ and $(C_{V_L}'',C_{S_L}'')$ give a profound fluctuation from SM prediction. In case of $ P_L^{\tau}$,  all the coefficients show standard deviations from SM where as $(C_{V_R}'',C_{S_R}'')$ gives marginal deviation. Anticipated values for branching fractions, forward- backward asymmetry, $R_{\Lambda_c}$, $\Lambda_c$ longitudinal polarization asymmetry and $ P_L^{\tau} $ of  $\Lambda_b \to \Lambda_c \tau \bar \nu_\tau$ process in the presence of new complex Wilson coefficients for case IIC are given in the Table \ref{Tab:CIIC-Lambdab}\,. 
 \begin{table}[htb]
\scriptsize
\caption{Calculated values of branching ratio and angular observables of  $\Lambda_b \to \Lambda_c \tau \bar \nu_\tau$ process  in the presence of new complex Wilson coefficients (case IIC ).}\label{Tab:CIIC-Lambdab}
\begin{center}
\begin{tabular}{|c|c|c|c|c|c|}
\hline
 ~Observables~&~Values for $(C_{V_L}'',C_{S_L}'')$~&~Values for $(C_{V_L}'',C_{V_R}'')$~&~Values for $(C_{V_R}'',C_{S_R}'')$ ~\\
\hline
~ Br$(\Lambda_b \to \Lambda_c  \tau \bar \nu_\tau)$~&~$0.302\pm 0.065$~&~$0.045\pm 0.002$~&~$0.088\pm 0.009$~\\
\hline
~ Br$(\Lambda_b \to \Lambda_c  \mu \bar \nu_\tau)$~&~$0.656\pm 0.074$~&~$0.140\pm 0.012$~&~$0.253\pm 0.015$~\\
\hline

~$ A_{FB}^{\Lambda_c}$~&~$0.290\pm 0.026$~&~$0.093\pm 0.009$~&~$0.087\pm 0.002$~\\
\hline
~$ R_{\Lambda_c}$~&~$0.461 \pm 0.041$~&~$0.327\pm 0.014$~&~$0.347\pm 0.072$~\\
\hline

~$ P_L^{\Lambda_c} $~&~$0.445\pm 0.034$~&~$-0.190\pm 0.327$~&~$-0.803\pm 0.724$~\\
\hline
~$ P_L^{\tau} $~&~$-0.381\pm 0.232$~&~$-0.331\pm 0.165$~&~$-0.224\pm 0.113$~\\
\hline
\end{tabular}
\end{center}
\end{table}
 
  \section{Numerical Analysis For $ B_c^+ \to \eta_c \tau^+ \nu_\tau$ decay processes. }
Including all the new physics operators, the  differential decay rate of $B_c^+ \to  \eta_c \tau^+ \nu_\tau$ processes, where $\eta_c$ is the pseudo-scalar mesons,  with respect to $q^2$ is given by \cite{Sakaki:2014sea8}
\bea
\frac{d\Gamma(B_c^+ \to  \eta_c \tau^+ \nu_\tau)}{dq^2} &=& {  |V_{cb}|^2 G_F^2 \over 192 M_B^3 \pi^3} q^2 \sqrt{\lambda(q^2)} \left( 1 - {m_\tau^2 \over q^2} \right)^2  \nn \\   && \times \Bigg \lbrace \Big | 1 + C^{\rm{eff}}_{V_L} + C^{\rm{eff}}_{V_R} \Big |^2 
\left[ \left( 1 + {m_\tau^2 \over 2q^2} \right) H_{0}^{2} + {3 \over 2}{m_\tau^2 \over q^2}  H_{t}^{2} \right] \nn \\ && + {3 \over 2} \left |C^{\rm{eff}}_{S_L} + C^{\rm{eff}}_{S_R} \right |^2 \, H_S^{2} + 8 \left |C^{\rm{eff}}_{T} \right |^2 \left(1+ \frac{2m_\tau^2}{q^2} \right) H_T^2 \nn \\ && +3{\rm Re}\left[ ( 1 + C^{\rm{eff}}_{V_L} +C^{\rm{eff}}_{V_R} ) ({C^{\rm{eff}}_{S_L}}^* + {C^{\rm{eff}}_{S_R}}^* ) \right] {m_l \over \sqrt{q^2}} \, H_S H_{t}  \nn \\ && -12{\rm Re}\left[ \left( 1 +C^{\rm{eff}}_{V_L} +C^{\rm{eff}}_{V_R} \right) {C^{\rm{eff}}_{T}}^* \right] \frac{m_\tau}{\sqrt{q^2}} H_T H_0  \Bigg \rbrace,  \label{br-exp}
\eea
where $
\lambda_\eta =\lambda (M_{B_c}^2, M_{\eta_c}^2, q^2)$, 
$M_{B_c}~(M_{\eta_c})$ is the mass of the $B_c~(\eta_c)$ meson, $m_\tau $ is the charged lepton mass and $H_{0,t,S,T}$ are the  helicity amplitudes which include the form factors $(F_{0,1,T})$ \cite{Sakaki:2014sea8,Murphy:2018sqg}. The expressions for helicity amplitudes and form factors are given in Appendix B in details.

\subsection{Case I}
\begin{figure}[htb]
\includegraphics[scale=0.30]{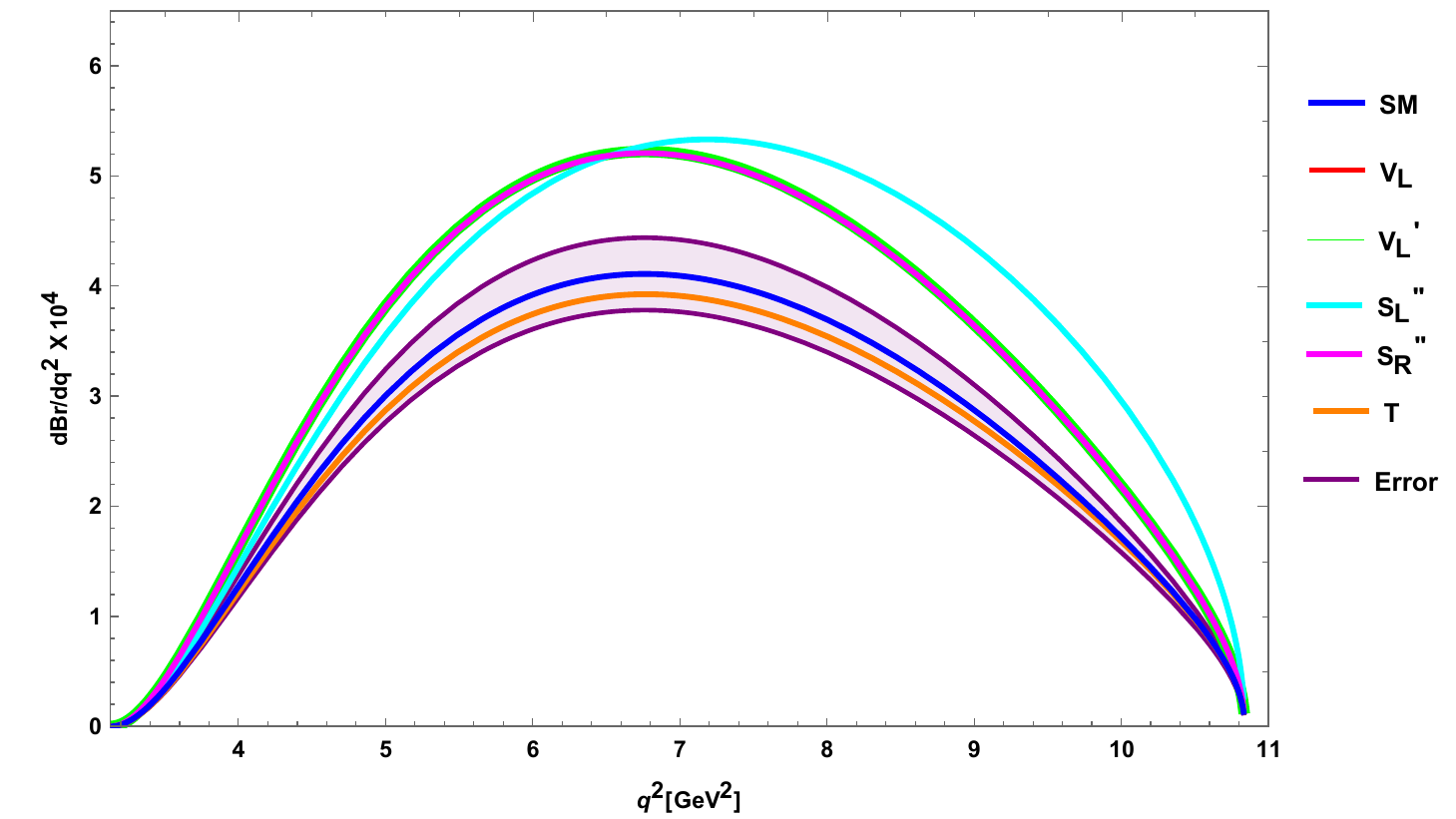}
\quad
\includegraphics[scale=0.30]{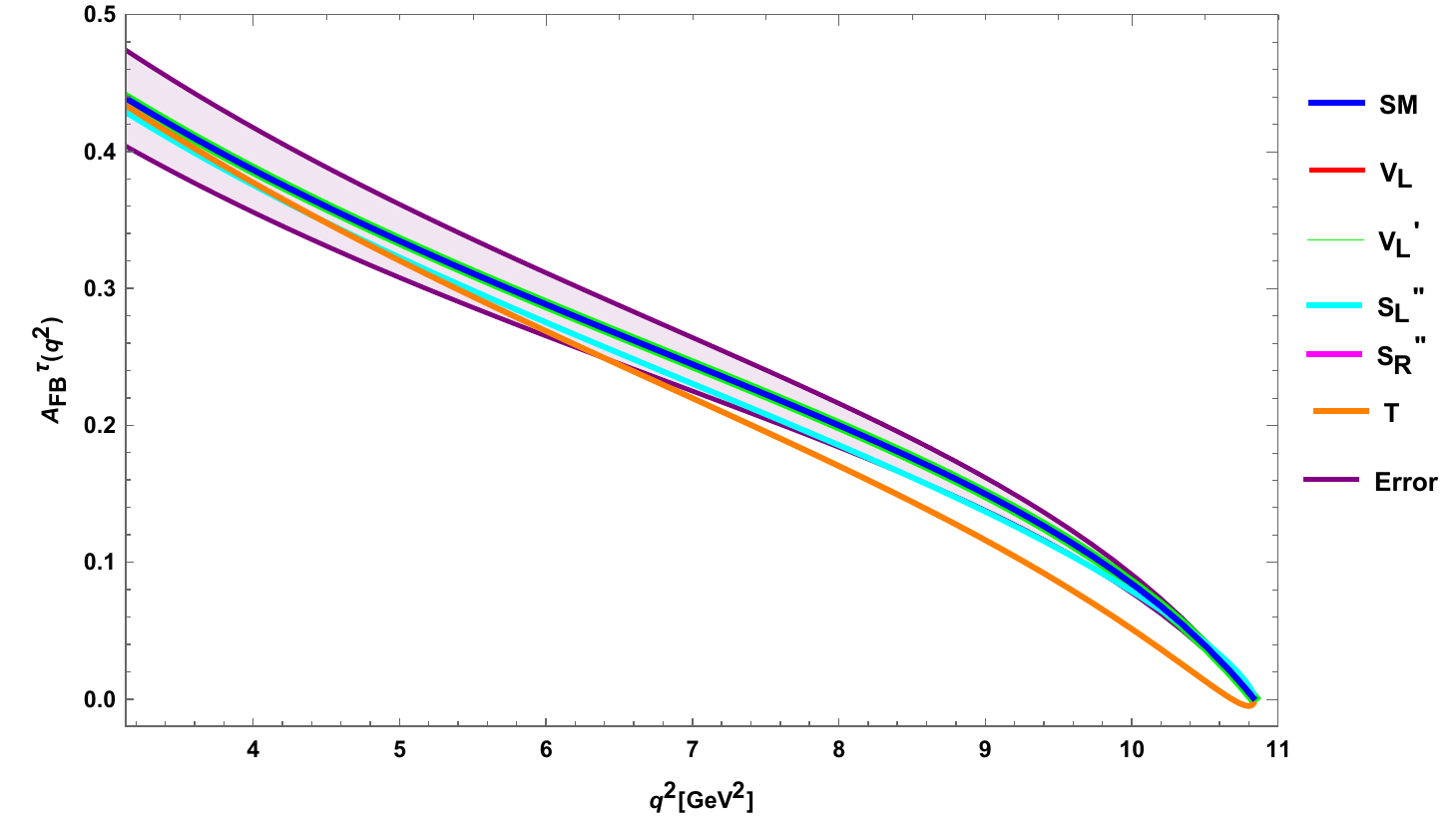}
\quad
\includegraphics[scale=0.30]{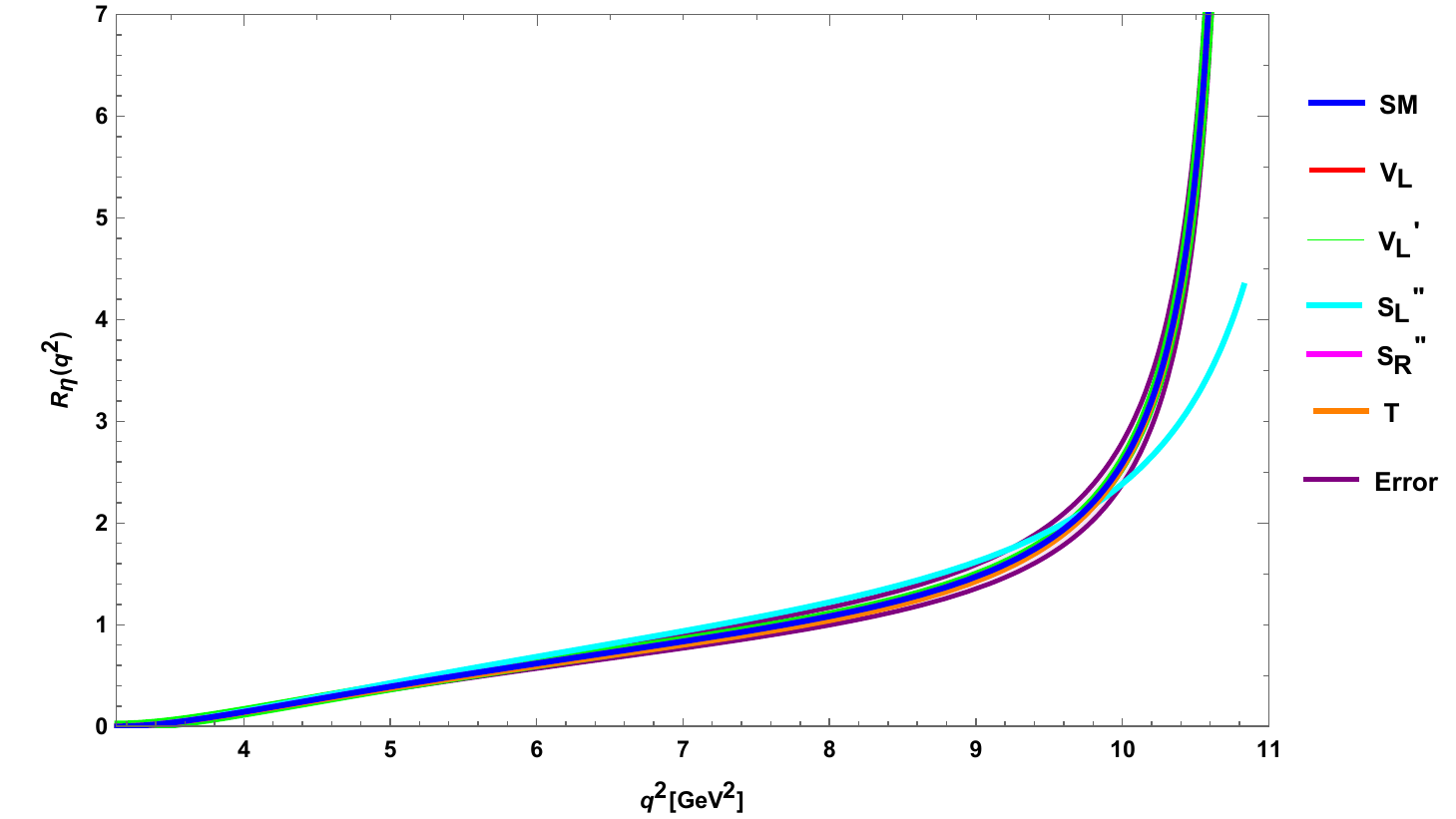}
\quad
\includegraphics[scale=0.30]{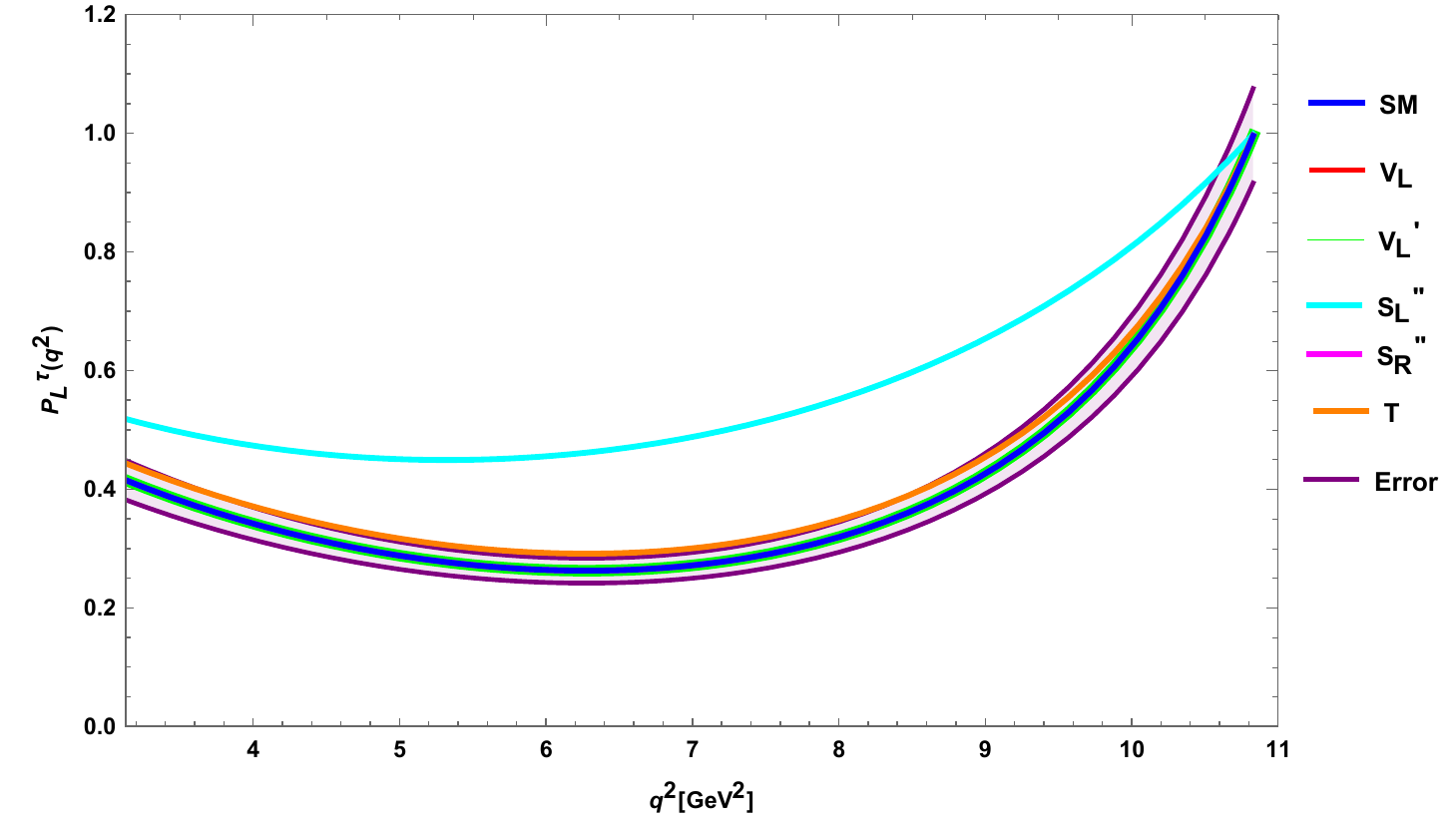}
 \caption{The branching ratio ( left top), forward-backward asymmetry (right top), $R_{\eta_c}$ (left bottom) and tau longitudinal polarization asymmetry (right bottom) of $B_c^+ \to \eta_c \tau^+ \nu_\tau$  for case I.}\label{Fig:Case-I Eta}
\end{figure}

Considering the presence of a single Wilson coefficient at a time, whose predicted best-fit values are presented in Table  \ref{Tab:Best-fit}\,, we show the branching ratio ( left top), forward-backward asymmetry (right top), $R_{\eta_c}$ (left bottom) and tau longitudinal polarization asymmetry (right bottom) of $B_c^+ \to \eta_c \tau^+ \nu_\tau$ decays for case I, in Figure[\ref{Fig:Case-I Eta}]. In these plots blue lines represent the SM predictions, Purple band represents the $1\sigma$ uncertainties for SM,and red, green, cyan blue, pink and orange lines are obtained by using the  $ C_{V_L}, C_{V_L}', C_{S_L}'',C_{S_R}''$ and $C_T$ coefficients respectively. In branching ratio plot, we observed that $ C_{S_L}''$ contribution gives significant deviation from the SM whereas $T$ gives nominal contribution as compared to other coefficients. Here we can see that due to the degeneracy of $ C_{V_L}, C_{V_L}' $ and $C_{S_R}''$ coefficients the plots are coincide with one line and show reasonable deviations from the SM prediction. From the forward-backward asymmetry graphical representation, we conclude that $C_T$ gives reasonable deviation from SM but $C_{S_L}''$ gives some marginal deviation, whereas  $ C_{V_L}, C_{V_L}'$ and $C_{S_R}''$ are not at all deviating. The effect of various new physics contributions on lepton non universality parameter $ R_{\eta_c}$ is almost negligible, while $C_{S_L}''$ gives a marginal deviation from SM prediction. In case of $ P_L^{\tau} $ asymmetry $C_{S_L}''$ gives profound deviation as compared to other coefficients and coefficient $C_T$ provides a minimal deviation from SM. Calculated central and $1\sigma$ uncertainty values of branching ratio and angular observable of  $B_c^+ \to \eta_c \tau^+ \nu_\tau$ process in the SM and in the presence of single new  Wilson coefficients for case I is given in the Table \ref{Tab:CI-eta}\,.
\begin{table}[htb]
\scriptsize
\caption{Calculated values of branching ratio and angular observables of  $B_c^+ \to \eta_c \tau^+ \nu_\tau$ process in the SM and in the presence of new complex Wilson coefficients (caseI ).}\label{Tab:CI-eta}
\begin{center}
\begin{tabular}{|c|c|c|c|c|c|c|}
\hline
 ~Observables~&~Values for SM~&~Values for $C_{V_L}$~&~Values for $C_{V_L}'$~&~Values for $C_{S_L}''$~&~Values for $C_{S_R}''$~&~Values for $C_T$~\\
\hline
~ Br$(B_c^+ \to \eta_c \tau^+ \nu_\tau)$~&~$0.0020 \pm 0.001$~&~$0.0026\pm 0.001$~&~$0.0026\pm 0.001$~&~$0.0028\pm 0.001$~&~$0.0026\pm 0.001$~&~$0.0019\pm 0.001$~\\
\hline
~ Br$(B_c^+ \to \eta_c \mu^+ \nu_\mu)$~&~$0.0072 \pm 0.002$~&~$0.0092\pm 0.003$~&~$0.0092\pm 0.003$~&~$0.0078\pm 0.002$~&~$0.0092\pm 0.004$~&~$0.0071\pm 0.002$~\\
\hline

~$ A_{FB}^{\eta}$~&~$1.850 \pm 0.028$~&~$1.850\pm 0.021$~&~$1.850\pm 0.021$~&~$1.765\pm 0.024$~&~$1.850\pm 0.021$~&~$1.678\pm 0.027$~\\
\hline
~$ R_{\eta}$~&~$0.284\pm 0.002$~&~$0.284\pm 0.004$~&~$0.284\pm 0.001$~&~$0.353\pm 0.006$~&~$0.284\pm 0.001$~&~$0.275\pm 0.006$~\\

\hline
~$ P_L^{\tau} $~&~$0.342\pm 0.009$~&~$0.342\pm 0.006$~&~$0.342\pm 0.006$~&~$0.554\pm 0.002$~&~$0.342\pm 0.006$~&~$0.371\pm 0.004$~\\
\hline
\end{tabular}
\end{center}
\end{table}

\subsection{Case II}
\subsubsection{A: Presence of $C_i ~\&~ C_j $ coefficients}
\begin{figure}[htb]
\includegraphics[scale=0.32]{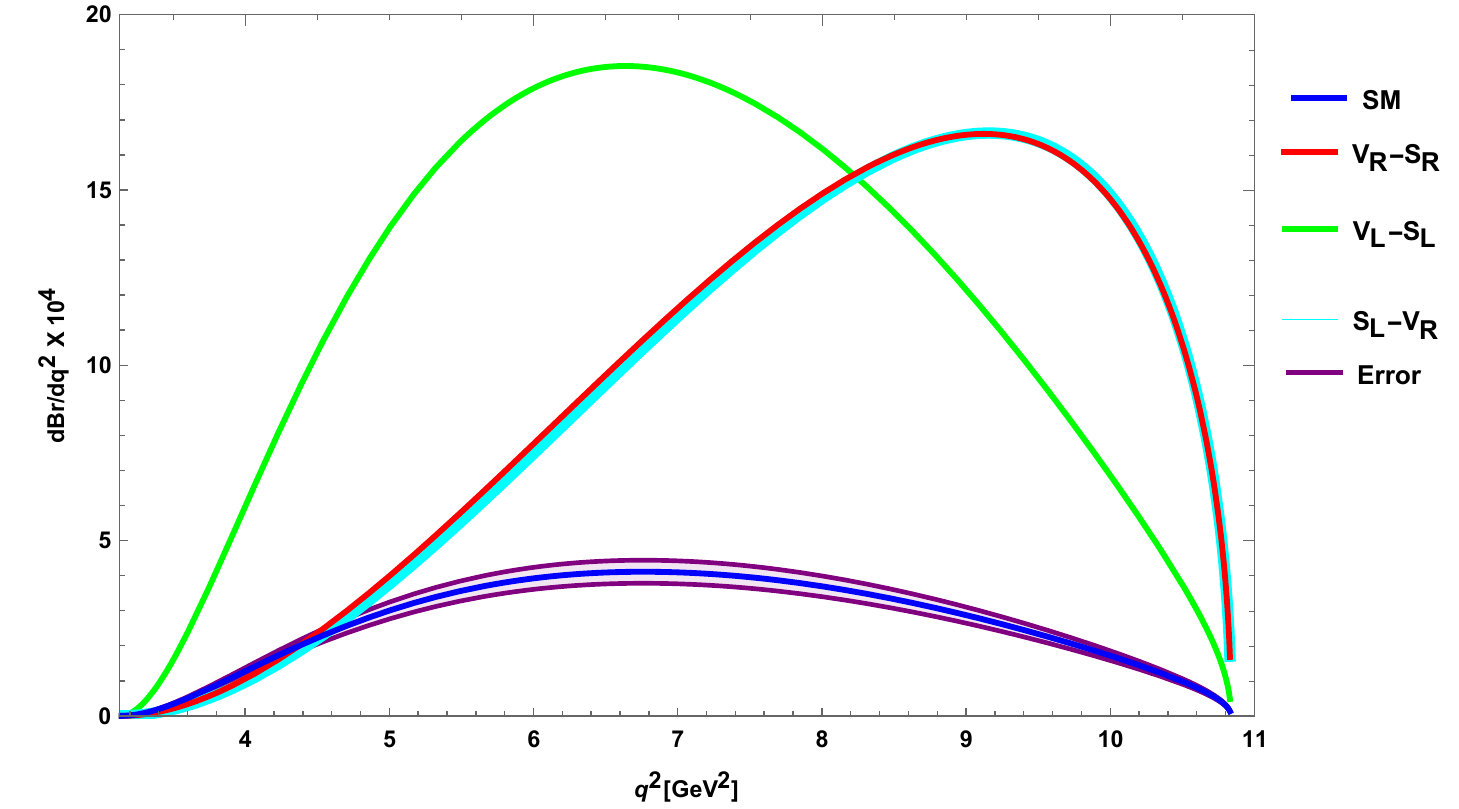}
\quad
\includegraphics[scale=0.32]{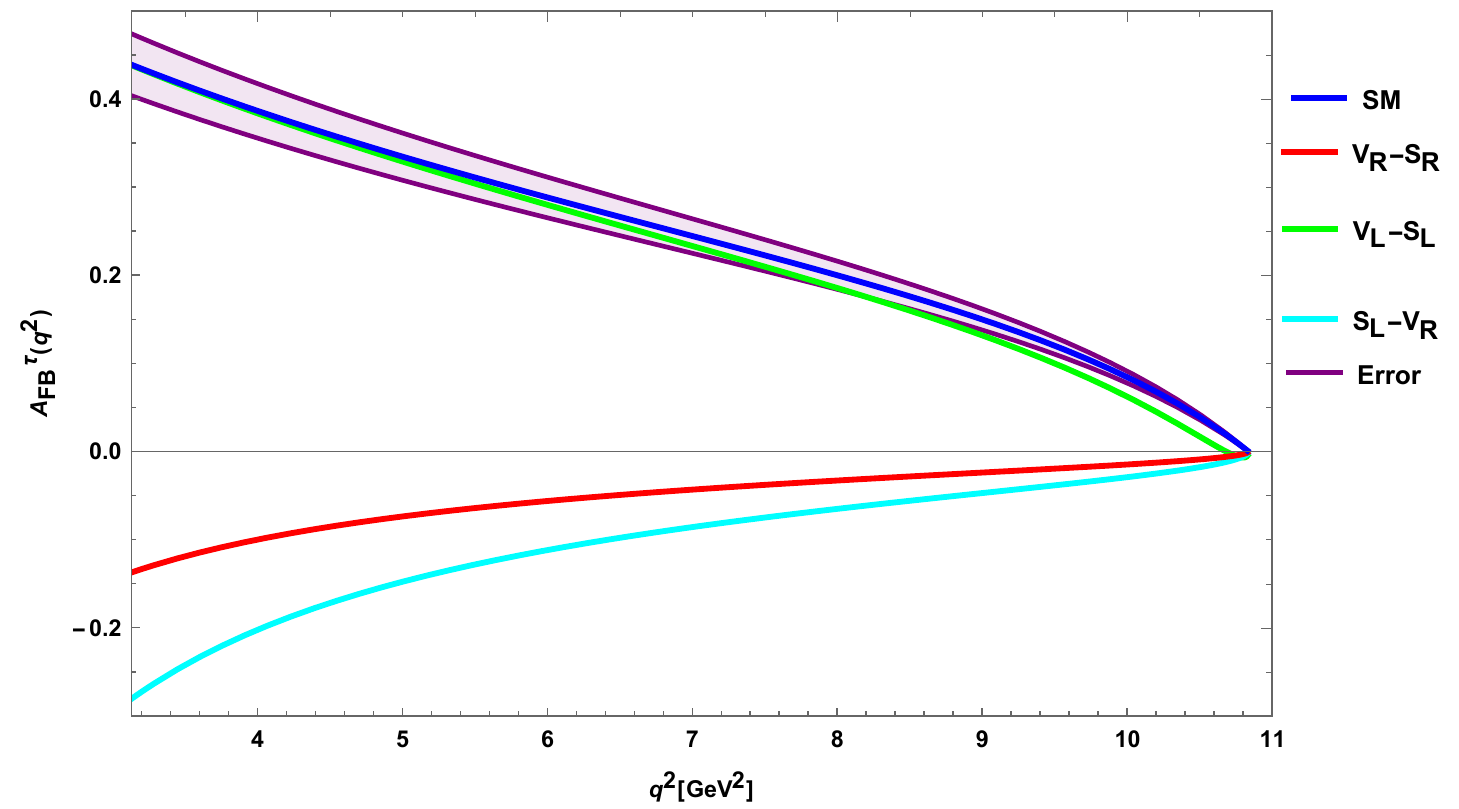}
\quad
\includegraphics[scale=0.32]{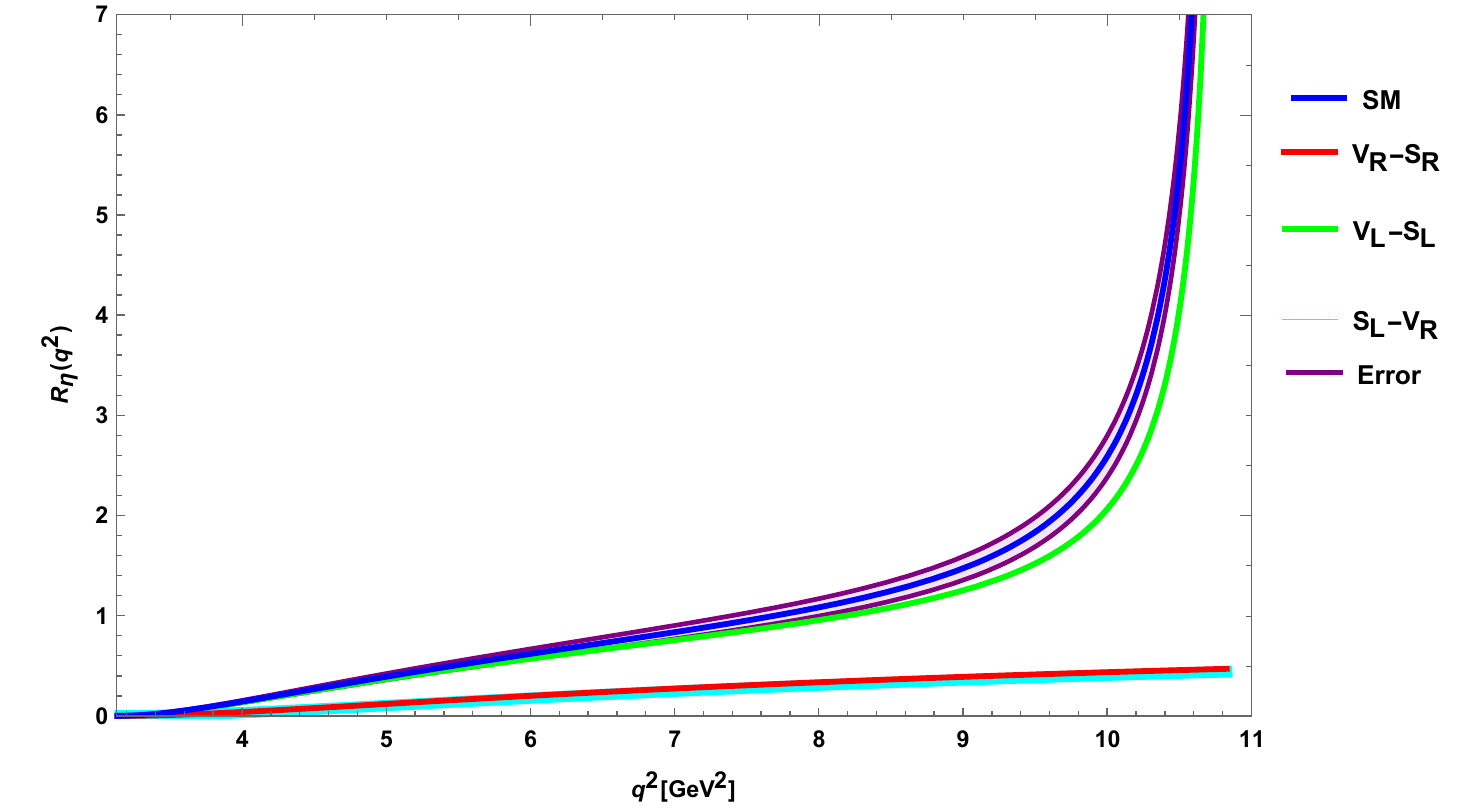}
\quad
\includegraphics[scale=0.32]{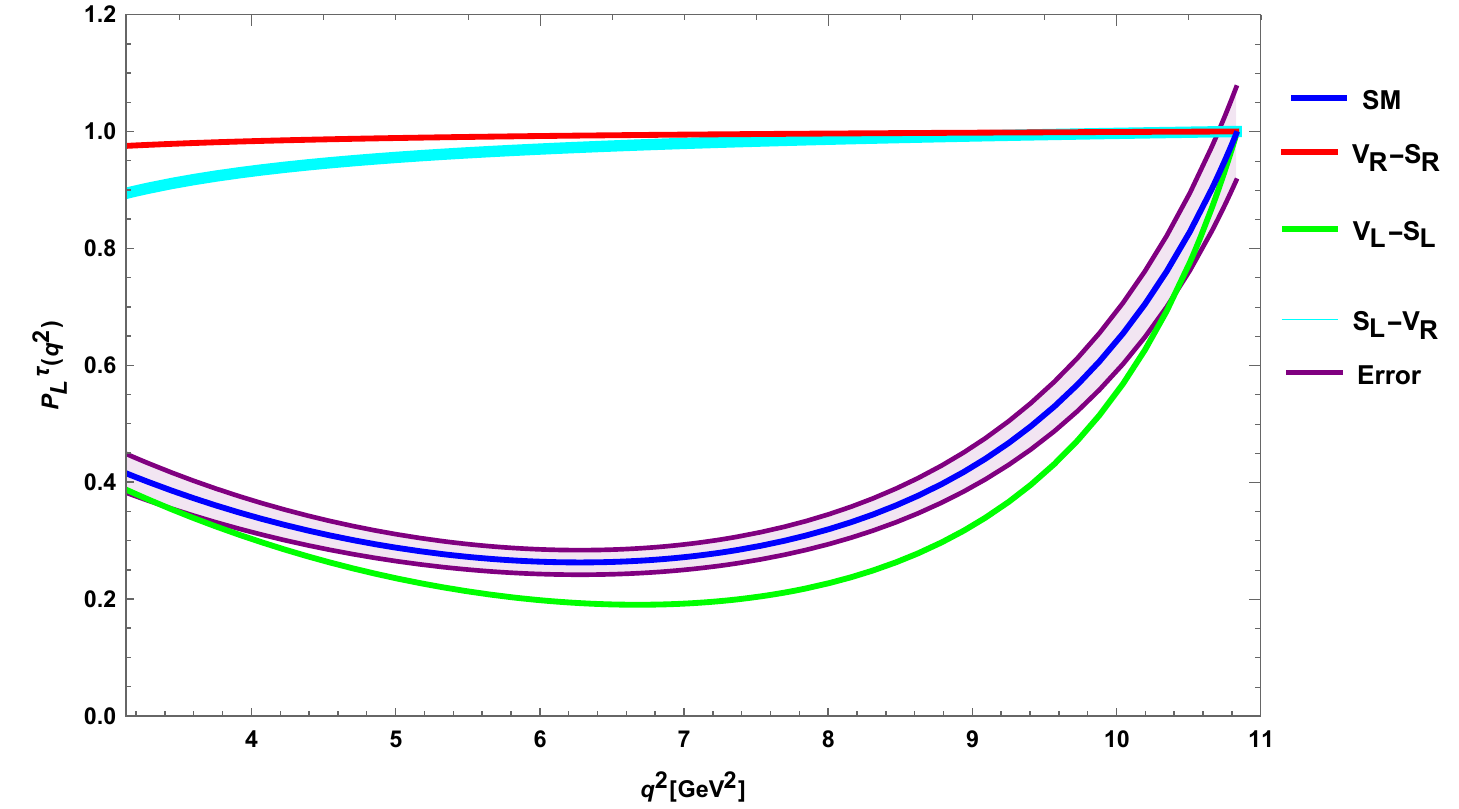}

\caption{ The branching ratio (top-left), forward-backward asymmetry (top-right), $R_{\eta_c}$ (bottom-left) and tau longitudinal polarization asymmetry (bottom-right) of $B_c^+ \to \eta_c \tau^+ \nu_\tau$  for case  IIA.}\label{Fig:Case- IIA eta}
\end{figure}

Here we discuss the implications of two different unprimed type of coefficients on the branching ratio ( top-left), forward-backward asymmetry (top-right), $R_{\eta_c}$ (bottom-left) and tau longitudinal polarization asymmetry (bottom-right) of $B_c^+ \to \eta_c \tau^+ \nu_\tau$  for case  IIA, in Figure[\ref{Fig:Case- IIA eta}]. The blue, green, cyan blue and red color solid lines are representing the SM, $(C_{V_L}, C_{S_L})$, $( C_{S_L}, C_{V_R})$ and $(C_{V_R},C_{S_R}) $ respectively, where as  Purple band represents the $1\sigma$ uncertainties for SM. In the branching fraction representation, all the coefficients are showing significant deviation from the SM prediction but the  $(C_{ S_L},V_R)$ and $(C_{V_R},C_{S_R}) $ are giving similar results. The forward-backward asymmetry plot shows noticeable  deviation for all the coefficients but $(C_{V_L}, C_{S_L})$ gives a marginal result in the positive axis whereas  $(C_{S_L},C_{V_R})$ and $C_{V_R},C_{S_R}) $ show the discrepancies in the negative axis. In $R_{\eta_c}$ graphical representation $(C_{V_L},C_{S_L})$ shows slight deviation whereas $(C_{ S_L},C_{V_R})$ and $(C_{V_R},C_{S_R}) $ give a profound deviation from SM.  In case of $ P_L^{\tau} $ asymmetry $(C_{V_L},C_{S_L})$ gives an acceptable deviation whereas  $ (C_{S_L},C_{V_R})$ and $(C_{V_R}, C_{S_R}) $ show a maximum deviation from SM. Predicted values for branching fraction, forward- backward asymmetry, $R_{\eta_c}$ and $ P_L^{\tau} $ of  $B_c^+ \to \eta_c \tau^+ \nu_\tau$ process with $1\sigma$ uncertainty in the presence of new complex Wilson coefficients for case IIA is given in the Table \ref{Tab:CIIA-eta}\,.
 
\begin{table}[htb]
\scriptsize
\caption{Calculated values of branching ratio and angular observables of  $B_c^+ \to \eta_c \tau^+ \nu_\tau$ process in the presence of new complex Wilson coefficients (case IIA ).}\label{Tab:CIIA-eta}
\begin{center}
\begin{tabular}{|c|c|c|c|c|c|}
\hline
 ~Observables~&~Values for $(C_{V_R}, C_{S_R})$~&~Values for $(C_{V_L},C_{S_L})$~&~Values for $(C_{S_L},C_{V_R})$ ~\\
\hline
~ Br$(B_c^+ \to \eta_c \tau^+ \nu_\tau)$~&~$0.0072\pm 0.004$~&~$0.0091\pm 0.006$~&~$0.0071\pm 0.003$~\\
\hline
~ Br$(B_c^+ \to \eta_c \mu^+ \nu_\mu)$~&~$0.0309\pm 0.008$~&~$0.0359\pm 0.004$~&~$0.0336\pm 0.009$~\\
\hline

~$ A_{FB}^{\eta}$~&~$-0.394\pm 0.165$~&~$1.761\pm 0.034$~&~$-0.791\pm 0.239$~\\
\hline
~$ R_{\eta}$~&~$0.234\pm 0.008$~&~$0.255\pm 0.003$~&~$0.213\pm 0.005$~\\
\hline
~$ P_L^{\tau} $~&~$0.996\pm 0.021$~&~$0.262\pm 0.018$~&~$0.985\pm 0.023$~\\
\hline
\end{tabular}
\end{center}
\end{table}

\subsubsection{B:  Presence of $C^{'}_i ~\&~ C^{'}_j $ coefficients}
\begin{figure}[htb]
\includegraphics[scale=0.28]{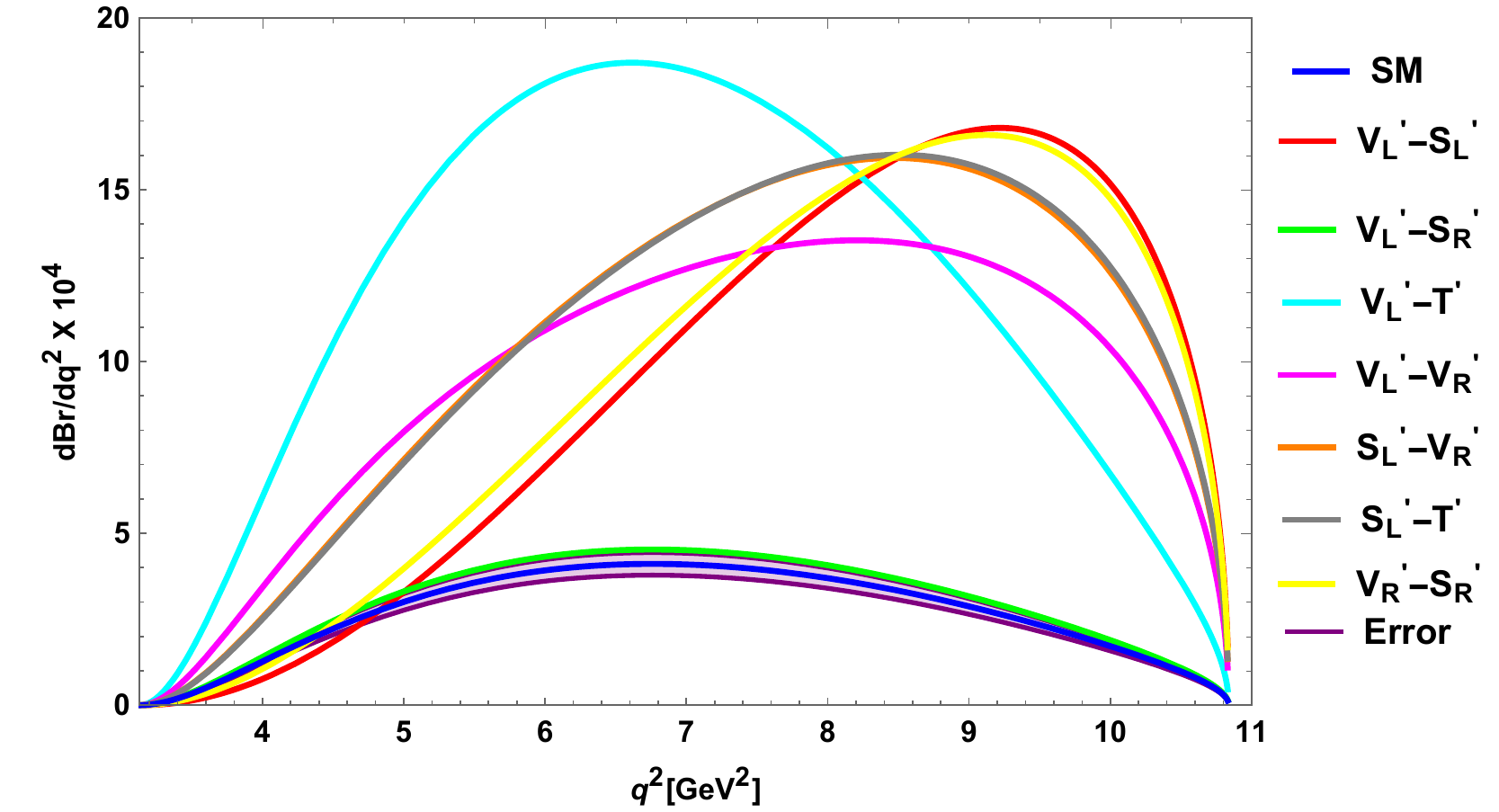}
\quad
\includegraphics[scale=0.28]{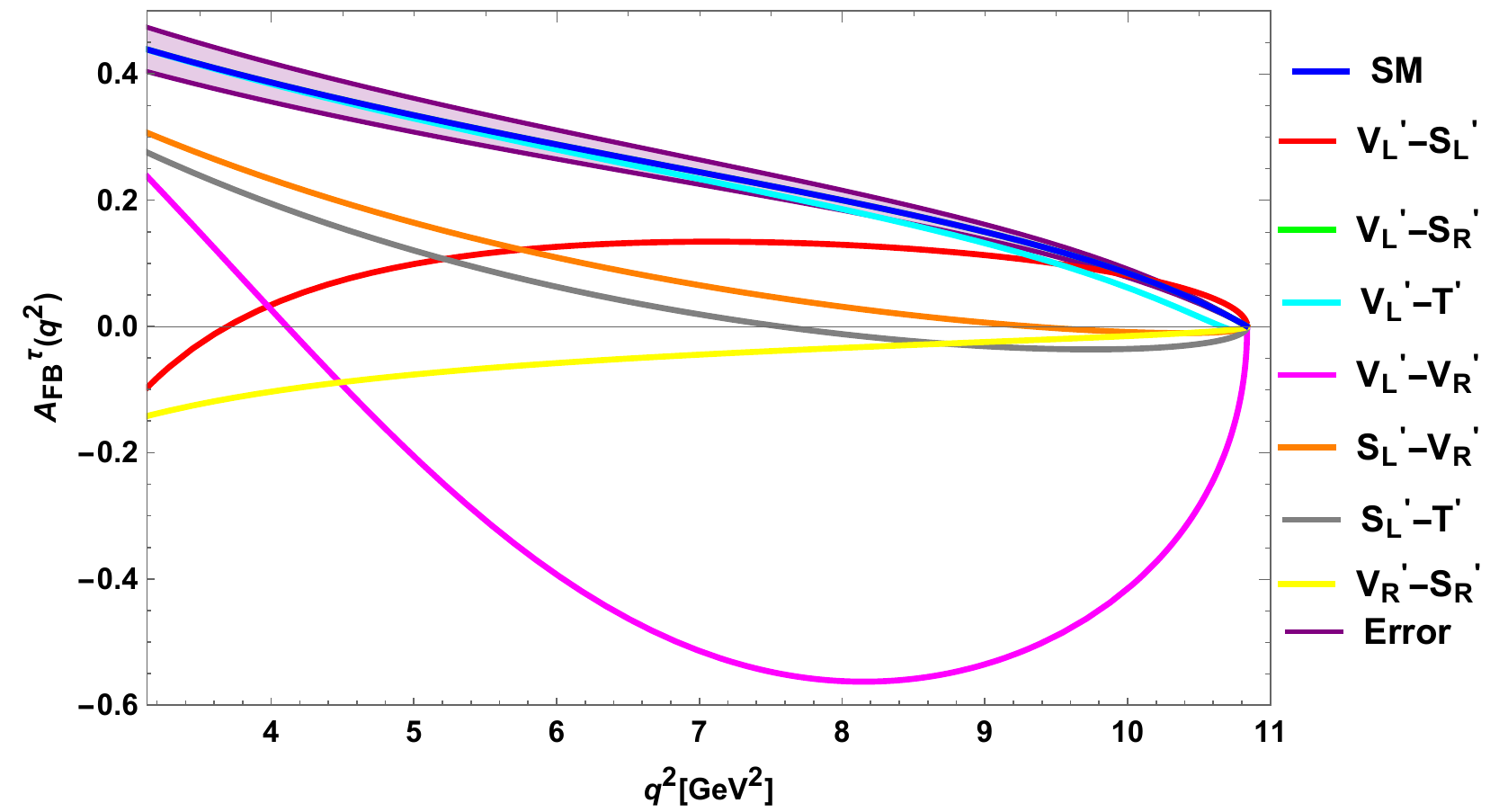}
\quad
\includegraphics[scale=0.28]{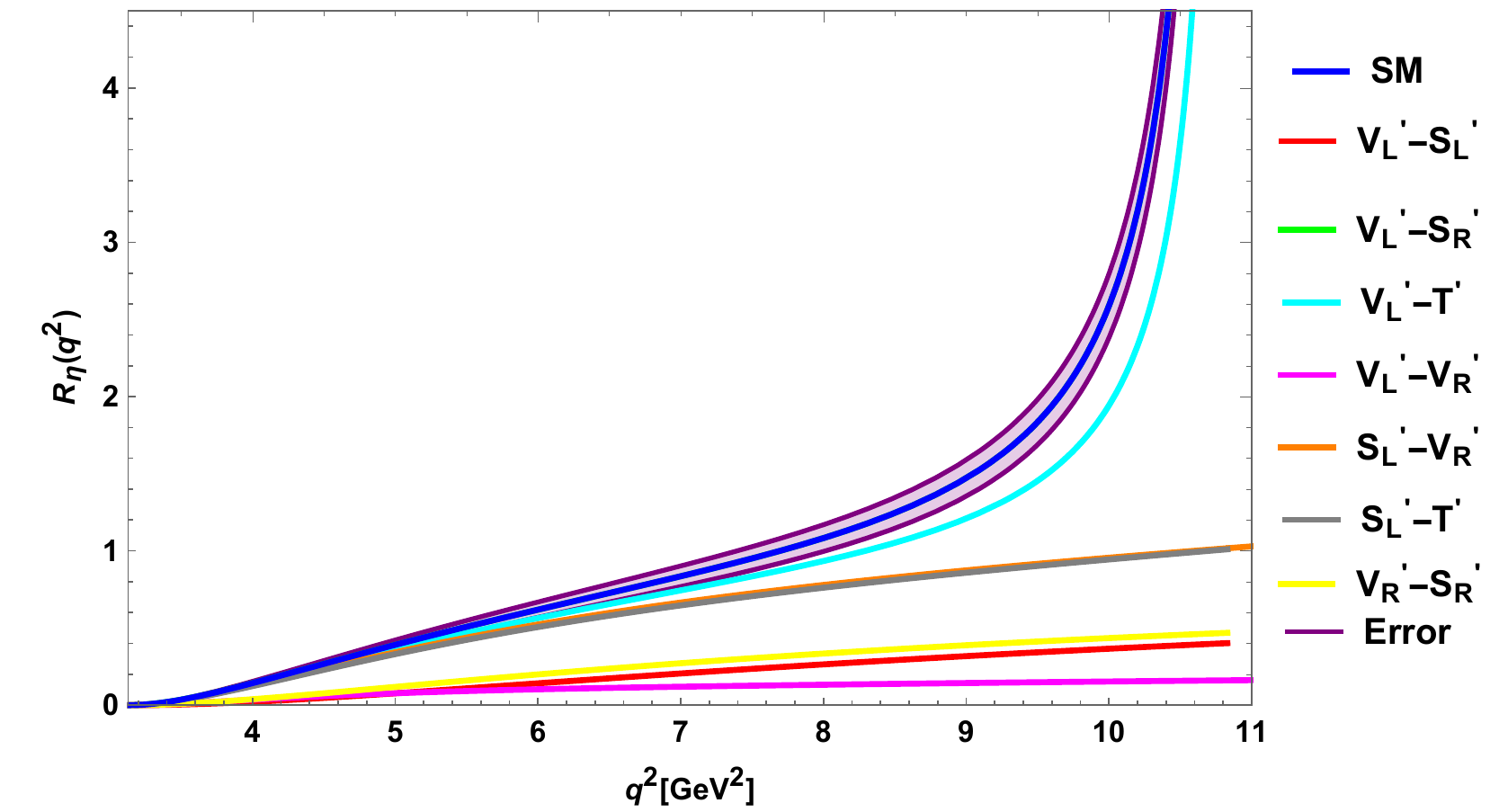}
\quad
\includegraphics[scale=0.28]{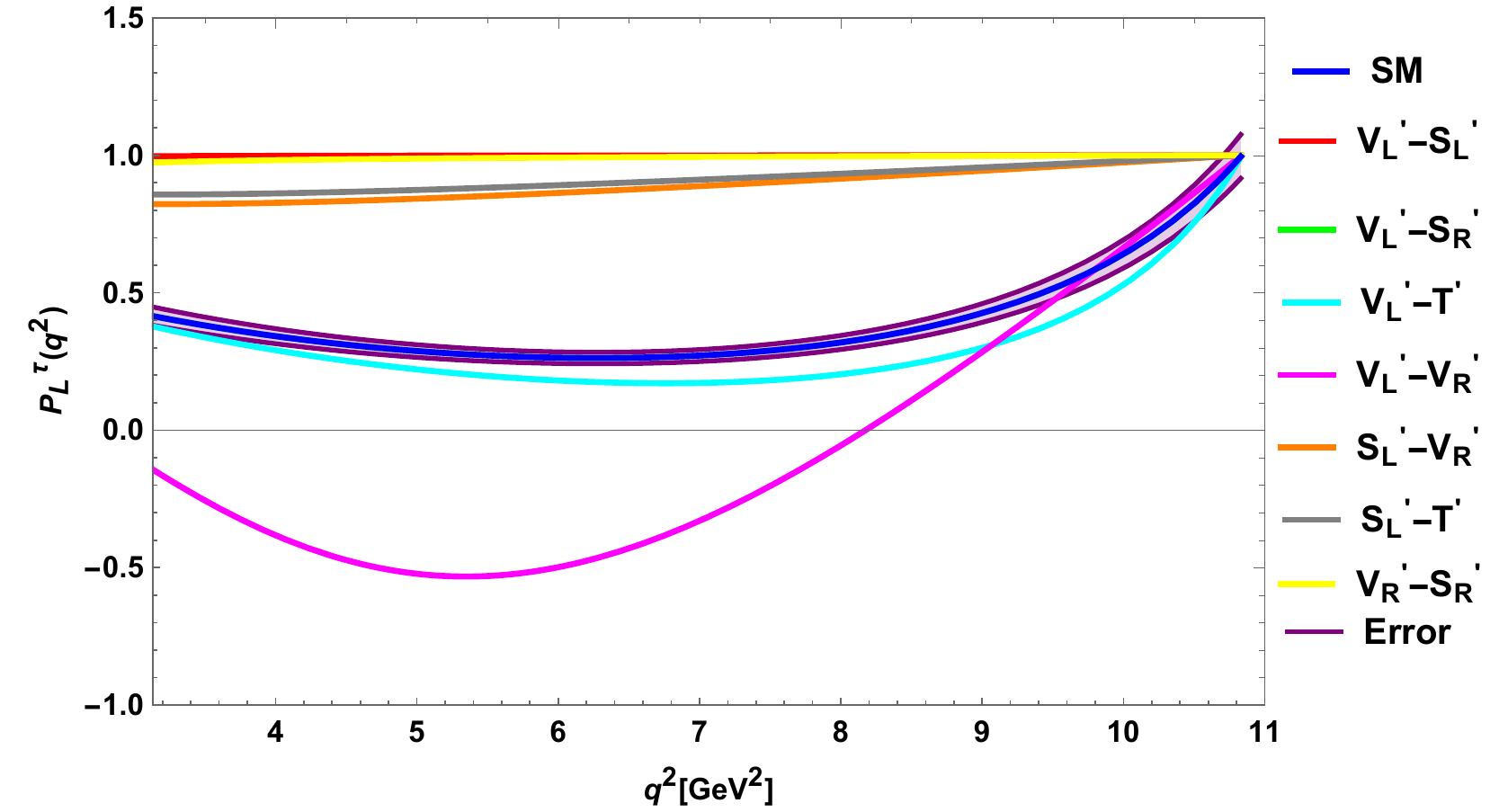}

\caption{The branching ratio (top left), forward-backward asymmetry (top right), $R_{\eta_c}$ (bottom left) and tau longitudinal polarization asymmetry (bottom right) of $B_c^+ \to \eta_c \tau^+ \nu_\tau$  for case  IIB.}\label{Fig:Case- IIB eta}
\end{figure}

Considering the case IIB, here we show the implications of two different primed coefficients on the branching ratio (top-left), forward-backward asymmetry (top-right), $R_{\eta_c}$ (bottom-left) and tau longitudinal polarization asymmetry (bottom-right) of $B_c^+ \to \eta_c \tau^+ \nu_\tau$, in Figure[\ref{Fig:Case- IIB eta}]. The blue, red, green, cyan blue, magenta, orange, gray and yellow color solid lines are representing the SM, $(C_{V_L}',C_{S_L}')$, $(C_{V_L}',C_{S_R}')$, $(C_{V_L}',C_T')$, $(C_{V_L}',C_{V_R}')$, $(C_{S_L}',C_{V_R}')$, $(C_{S_L}',C_T')$ and $(C_{V_R}',C_{S_R}')$ respectively. In the branching fraction representation, all the coefficients are showing significant deviations from the SM prediction but the $(C_{V_L}',C_{S_R}')$ is giving slight deviation. $(C_{V_L}',C_{S_L}')$ and $(C_{V_R}',C_{S_R}')$ show similar results whereas $(C_{S_L}',C_{V_R}')$ and $(C_{S_L}',C_T')$ give same outcomes. In the forward-backward asymmetry plot $(C_{V_L}',C_{V_R}')$ shows maximum deviation whereas $(C_{V_L}',C_T')$ expresses a little deviation from SM values. All other constraints indicate some significant variance, but among those $(C_{S_L}',C_{V_R}')$, $(C_{V_L}',C_{S_L}')$ and $(C_{S_L}',C_T')$ predict some sizeable deviations with regard to zero crossing point. In $R_{\eta_c}$ plot, the $(C_{V_L}',C_{S_R}')$ convey no deviation although $(C_{V_L}',C_T')$ gives a marginal deviation. All other coefficients show significant deviation from SM which is a good indication for the presence of new physics.  In $ P_L^{\tau} $ representation, $(C_{V_L}',C_T')$ gives a minimum deviation whereas all other show a standard deviation from SM. Envisaged results for branching ratio, forward-backward asymmetry, $R_{\eta_c}$ and $ P_L^{\tau} $ of  $B_c^+ \to \eta_c \tau^+ \nu_\tau$ process in the presence of new  Wilson coefficients for case IIB are given in the Table \ref{Tab:CIIB-eta}\,.

\begin{table}[htb]
\tiny
\caption{Calculated values of branching ratio and angular observables of  $B_c^+ \to \eta_c \tau^+ \nu_\tau$ process in the presence of new complex Wilson coefficients (case IIB ).}\label{Tab:CIIB-eta}
\begin{center}
\begin{tabular}{|c|c|c|c|c|c|c|c|}
\hline
 ~Observables~&~ $(C_{V_L}',C_{S_L}')$~&~ $(C_{V_L}',C_{S_R}')$~&~$(C_{V_L}',C_T')$ ~&~$(C_{V_L}',C_{V_R}')$~&~$(C_{S_L}',C_{V_R}')$~&~ $(C_{S_L}',C_T')$~&~ $(C_{V_R}',C_{S_R}')$~\\
\hline
~ Br$(B_c^+ \to \eta_c \tau^+ \nu_\tau)$~&~$0.0070\pm 0.003$~&~$0.0090\pm 0.002$~&~$0.0092\pm 0.004$~&~$0.0073\pm 0.002$~&~$0.0080\pm 0.001$~&~ $ 0.0080\pm 0.001 $~&~ $ 0.0072 \pm 0.001$~\\
\hline
~ Br$(B_c^+ \to \eta_c \mu^+ \nu_\mu)$~&~$0.0391\pm 0.001$~&~$0.0316\pm 0.004$~&~$0.0367\pm 0.003$~&~$0.095\pm 0.009$~&~$0.0196\pm 0.005$~&~ $ 0.0196\pm 0.002 $~&~ $ 0.0301\pm 0.007 $~\\
\hline

~$ A_{FB}^{\eta}$~&~$0.698\pm 0.011$~&~$1.850\pm 0.019$~&~$1.761\pm 0.028$~&~$-2.570\pm 0.121$~&~$0.707\pm 0.080$~&~ $0.410\pm 0.012 $~&~ $-0.407\pm 0.390$~\\
\hline
~$ R_{\eta}$~&~$0.179\pm 0.002$~&~$0.284\pm 0.008$~&~$0.250\pm 0.006$~&~$0.076\pm 0.004$~&~$0.415\pm 0.001$~&~$0.408\pm 0.005$~&~ $ 0.234 \pm 0.007$~\\
\hline
~$ P_L^{\tau} $~&~$0.999\pm 0.003$~&~$0.342\pm 0.007$~&~$0.242\pm 0.004$~&~$-0.052\pm 0.032$~&~$0.910\pm 0.008$~&~ $ 0.929\pm 0.001$~&~ $ 0.995\pm 0.001 $~\\
\hline
\end{tabular}
\end{center}
\end{table}

\subsubsection{C: Presence of  $C^{''}_i ~\&~ C^{''}_j $ coefficients}
\begin{figure}[htb]
\includegraphics[scale=0.28]{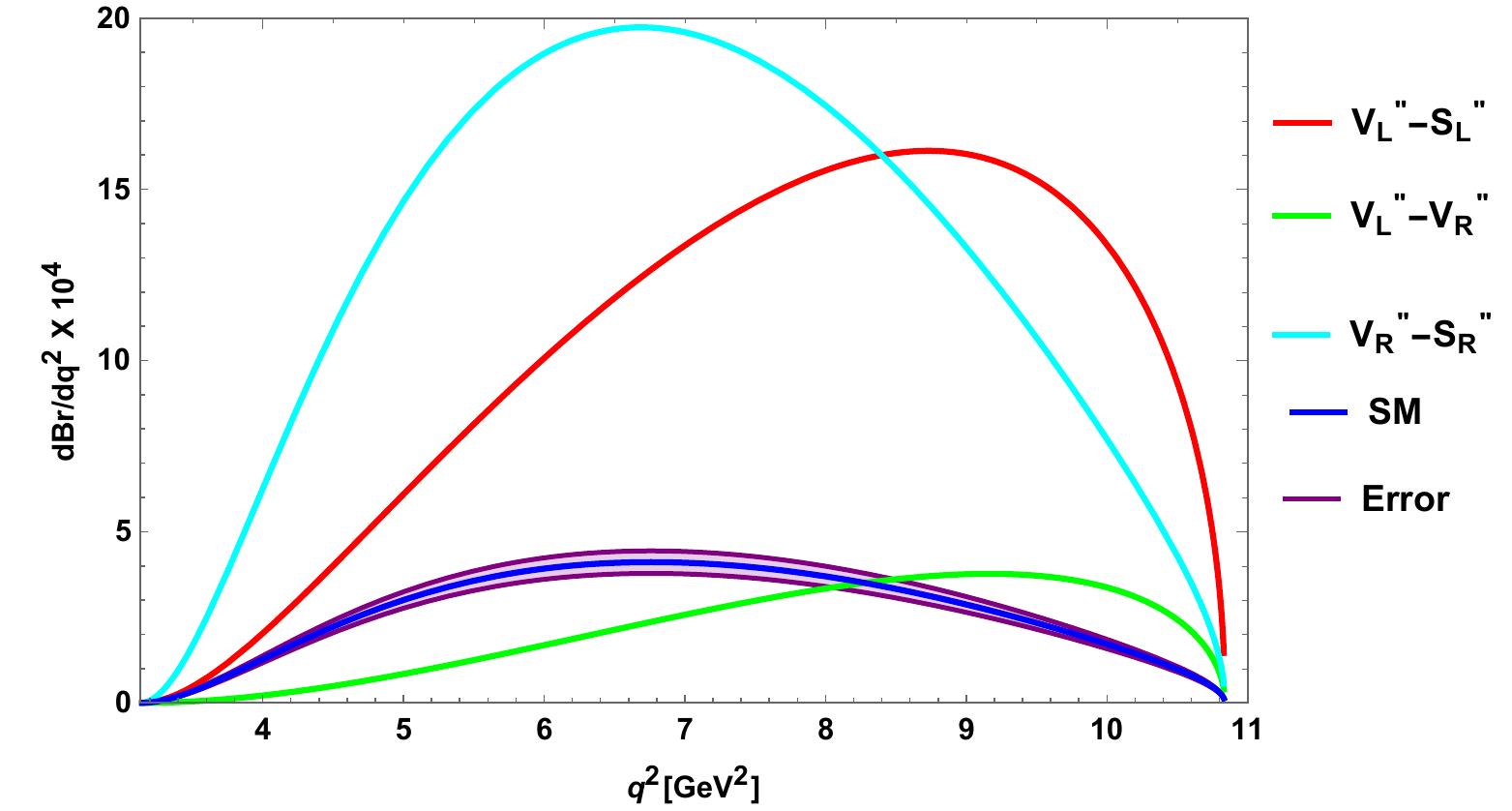}
\quad
\includegraphics[scale=0.28]{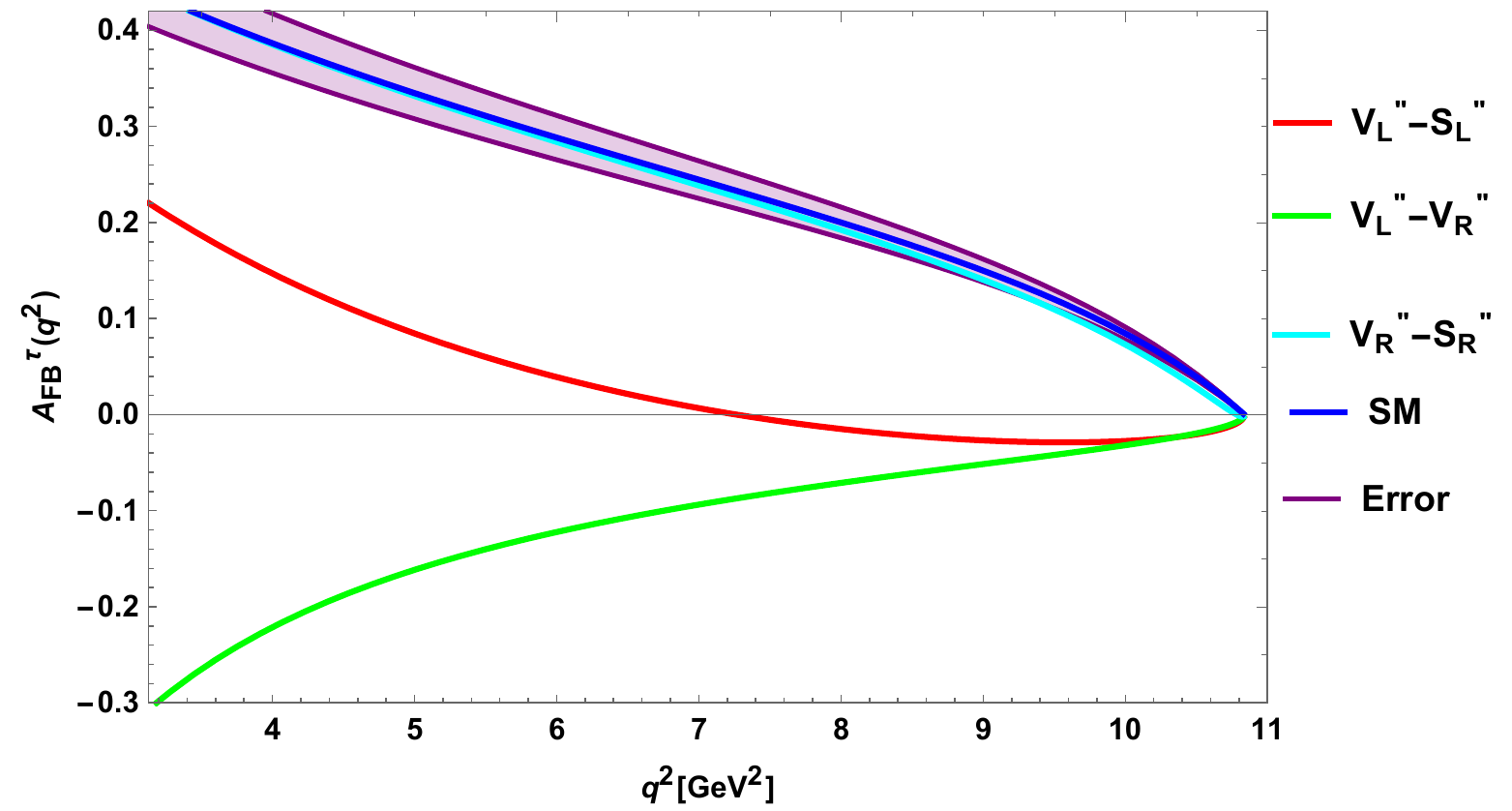}
\quad
\includegraphics[scale=0.28]{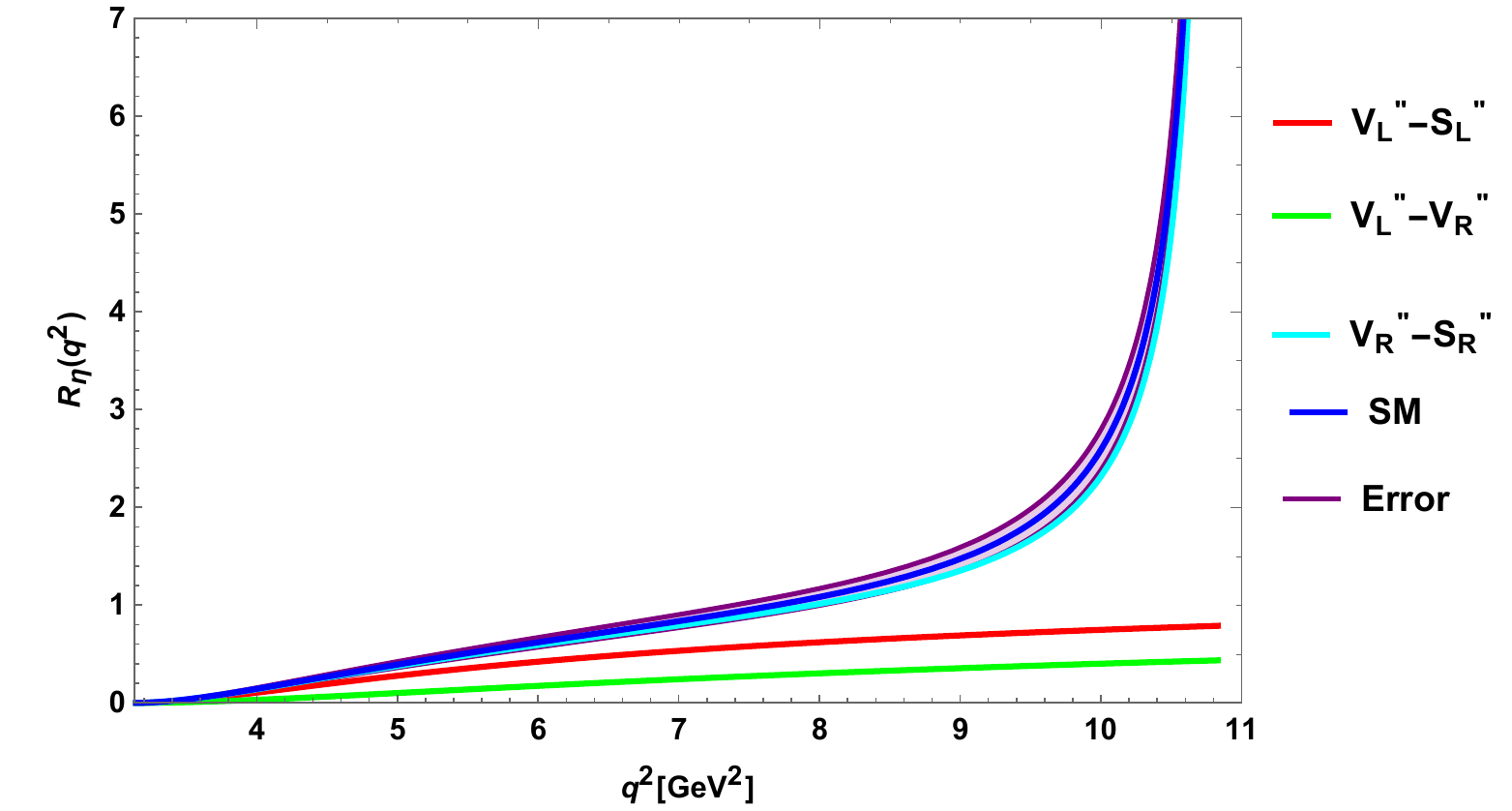}
\quad
\includegraphics[scale=0.28]{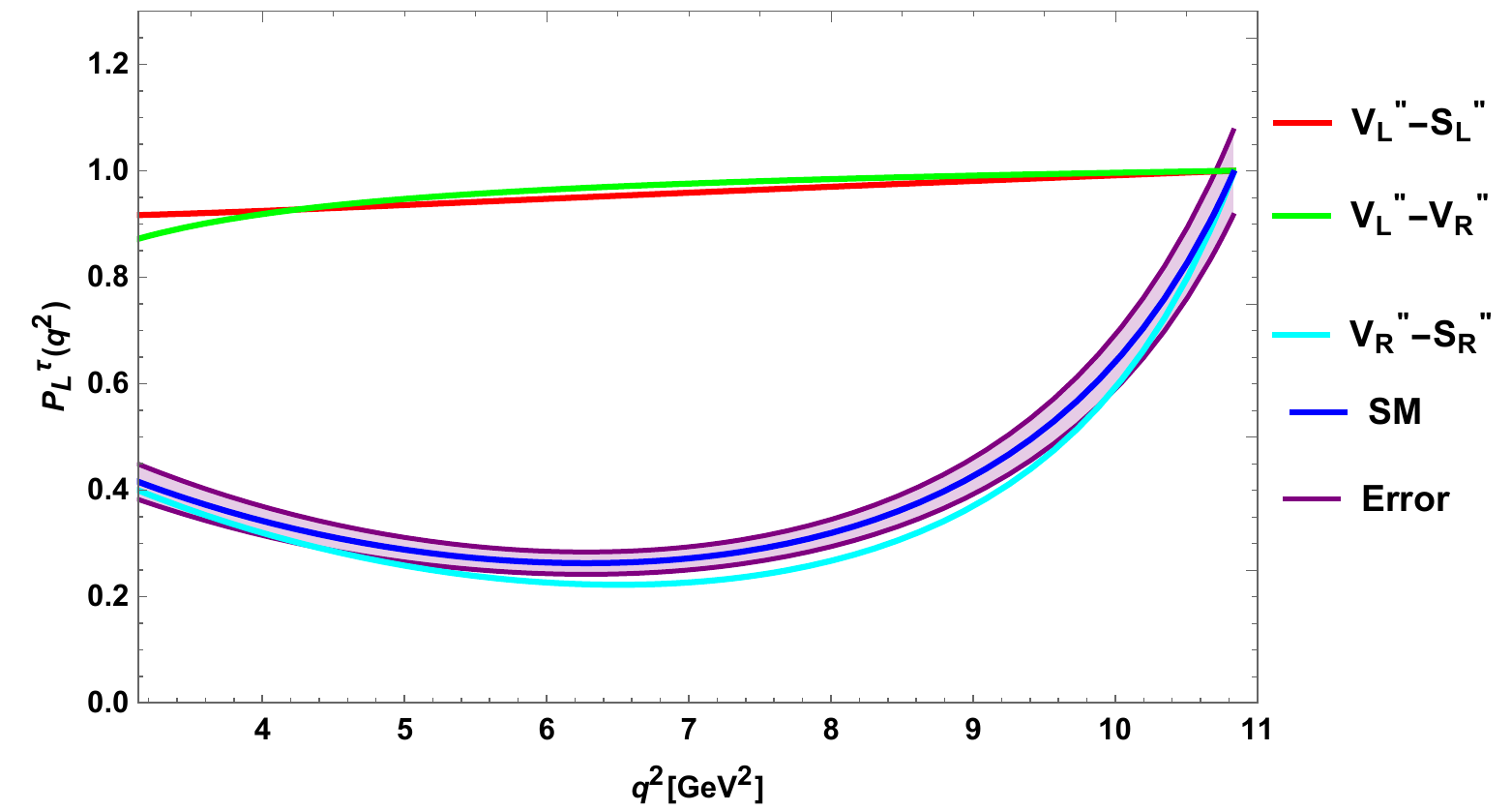}

\caption{ The branching ratio (top left), forward-backward asymmetry (top right), $R_{\eta_c}$ (bottom left) and tau longitudinal polarization asymmetry (bottom right) of $B_c^+ \to \eta_c \tau^+ \nu_\tau$  IIC.}\label{Fig:Case- IIC eta}
\end{figure}

Coming to the case IIC, we talk about the connections of two different double primed  coefficients on the branching ratio (top left), forward-backward asymmetry (top right), $R_{\eta_c}$ (bottom left) and tau longitudinal polarization asymmetry (bottom right) of $B_c^+ \to \eta_c \tau^+ \nu_\tau$, in Figure[\ref{Fig:Case- IIC eta}]. The blue, red, green and cyan blue color solid lines are representing the SM, $(C_{V_L}'',C_{S_L}'')$, $(C_{ V_L}'',C_{V_R}'')$ and $(C_{V_R}'',C_{S_R}'')$ respectively, where as Purple band represents the $1\sigma$ uncertainties for SM,. In the branching fraction, all the coefficients are showing significant deviation from the SM prediction. In the forward-backward asymmetry, $(C_{V_R}'',C_{S_R}'')$ shows a marginal deviation whereas coefficients $(C_{V_L}'', C_{V_R}'')$ gives maximum discrepancy and $(C_{V_L}'',C_{S_L}'')$ gives notable result to the zero crossing point. In $R_{\eta_c}$ graphical representation $(C_{V_R}'',C_{S_R}'')$ shows mild deviation whereas $(C_{V_L}'',C_{V_R}'')$ and $(C_{V_L}'',C_{S_L}'')$ give a profound fluctuation from SM.  In case of $ P_L^{\tau}$ representation,  all the coefficients show significant deviations from SM whereas $(C_{V_R}'',C_{S_R}'')$ gives minimum deviation as compared to others. Anticipated values for branching fractions, forward- backward asymmetry, $R_{\eta_c}$ and $ P_L^{\tau} $ of  $B_c^+ \to \eta_c \tau^+ \nu_\tau$ process in the presence of new complex Wilson coefficients for case IIC is given in the Table \ref{Tab:IIC-eta}\,.

 \begin{table}[htb]
\scriptsize
\caption{Calculated values of branching ratio and angular observables of   $B_c^+ \to \eta_c \tau^+ \nu_\tau$  process in the presence of new complex Wilson coefficients (case IIC ).}\label{Tab:IIC-eta}
\begin{center}
\begin{tabular}{|c|c|c|c|c|c|}
\hline
 ~Observables~&~Values for $(C_{V_L}'',C_{S_L}'')$~&~Values for $(C_{V_L}'',C_{V_R}'')$~&~Values for $(C_{V_R}'',C_{S_R}'')$ ~\\
\hline
~ Br$( B_c^+ \to \eta_c \tau^+ \nu_\tau )$~&~$0.0016\pm 0.001$~&~$0.0078\pm 0.001$~&~$0.0098\pm 0.002$~\\
\hline
~ Br$( B_c^+ \to \eta_c \mu^+ \nu_\mu)$~&~$0.0077\pm 0.004$~&~$0.0202\pm 0.028$~&~$0.0367\pm 0.030$~\\
\hline

~$ A_{FB}^{\eta}$~&~$0.282\pm 0.061$~&~$-0.864\pm 0.012$~&~$1.803\pm 0.061$~\\
\hline
~$ R_{\eta}$~&~$0.209\pm 0.006$~&~$0.385\pm 0.009$~&~$0.267\pm 0.005$~\\
\hline
~$ P_L^{\tau} $~&~$0.967\pm 0.007$~&~$0.982\pm 0.001$~&~$0.297\pm 0.005$~\\
\hline
\end{tabular}
\end{center}
\end{table}

\section{Numerical Analysis For $ B \to D^{**} \tau \bar \nu_\tau$   decay processes }

In this section, we inspect $B \to  D^{**} \tau \bar \nu_\tau$ decays,
where $D^{**} = \{D^*_0, D_1^*, D_1, D_2^*\}$ signifies the $1P$ orbitally excited states and these are also known as the four lightest excited charmed mesons in the quark model. Assuming no right-handed neutrinos, we include contributions from all feasible four-fermion operators which are serving for the extension of the form factors even going beyond the leading order in the heavy quark expansion.  Phenomenologically considering the $B \to \dss\ell\bar\nu$ decays exactly, we notice that certain form factor combinations are suppressed by the light lepton mass which cannot be constrained by $B \to \dss l \nu$ but in the semitauonic rates it remains unsuppressed \cite{Bernlochner:2016bci}. For the comprehensively improvement in the upcoming predictions for $B\to \dss \tau\bar\nu$, we  use heavy quark effective theory (HQET), which will yield complementary sensitivity to new physics. For a good theoretical control of these modes will help to boost the determinations of the CKM elements $|V_{cb}|$ and $|V_{ub}|$, both from exclusive and inclusive $B$ decays \cite{HFT:2014bci}. The isospin averaged widths  and masses of the four lightest excited $D$ meson states are shown in  Table \ref{tab:charm}\,. 
\begin{table}[htb]
\begin{tabular}{|c|c|c|c|c|}
\hline
Particle  &    $s_l^{\pi_l}$ &  $J^P$  &  $\Gamma$ (MeV)   & $m$ (MeV)\\
\hline
$\dSs$ &  $\frac12^+$  &  $0^+$  &   $236$   & $2349$ \\
$\dVs$ &  $\frac12^+$  &  $1^+$  &  $384$  & $2427$   \\
\hline
$\dV$ &  $\frac32^+$  &  $1^+$  &   $31$ & $2421$  \\
$\dTs$ &  $\frac32^+$  &  $2^+$  &   $47$ & $2461$  \\
\hline
\end{tabular}
\caption{Isospin averaged widths  and masses of the lightest charm mesons, rounded to 1\,MeV.~}
\label{tab:charm}
\end{table}
In the differential decay rates $\theta$ is connected to the charged lepton energy and elucidated as the angle betwixt the charged lepton and the charmed meson in the center of momentum frame. The $1+\cos^2\theta$ and $\cos\theta$ terms determine the helicity $\lambda=\pm1$ rates while the $\sin^2\theta$ terms are the helicity zero rates \cite{Isgur:1991wq}. The decay rates for $|\lambda|=1$ disappear for massless leptons at maximal recoil, $w_{\rm max} = (1+r^2-\rt)/(2r)$, as implied by the $(1-2rw+r^2-\rt)$ factors. In the double differential rates, the expressions $r=m_{D^{**}}/m_B$ for each $D^{**}$ state, and $\Gamma_0 = {G_F^2\,|V_{cb}|^2\,m_B^5 /(192\pi^3)}$ are useful. Here we defining the branching fraction only for the SM and the basis of new physics of vector operator. So we are considering the contributions from $ C_{V_L}, C_{V_L}'$ in case I, $(C_{V_L}',C_{V_R}')$ in case IIB and $(C_{V_L}'',C_{V_R}'')$ in case IIC. The comprehensive expressions for form factors used in the decay rates are given in the Appendix C \cite{Bernlochner:2016bci}.
\subsection{$B\to D_0^*\ell\bar\nu$}
\begin{figure}[htb]
\includegraphics[scale=0.4]{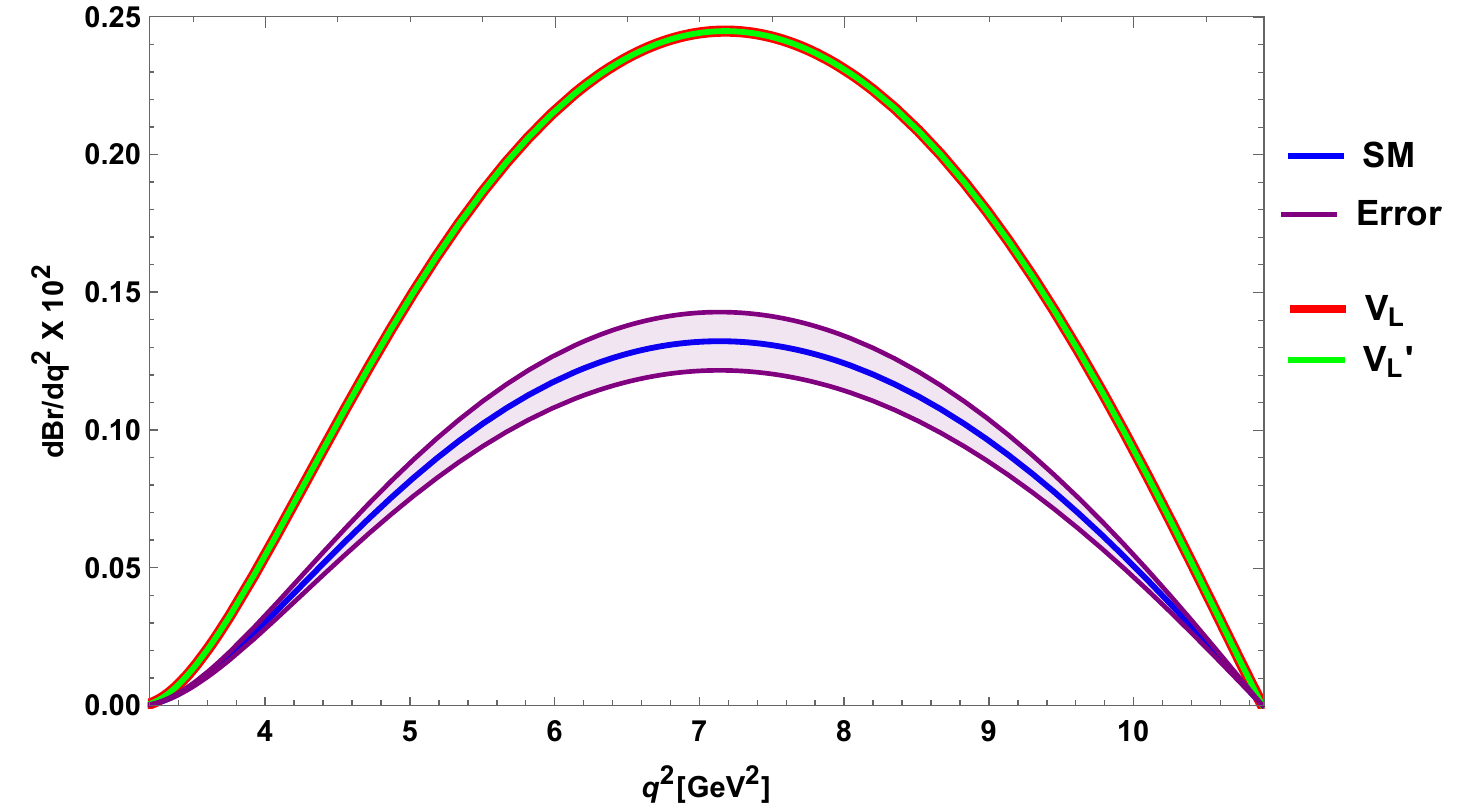}
\quad
\includegraphics[scale=0.4]{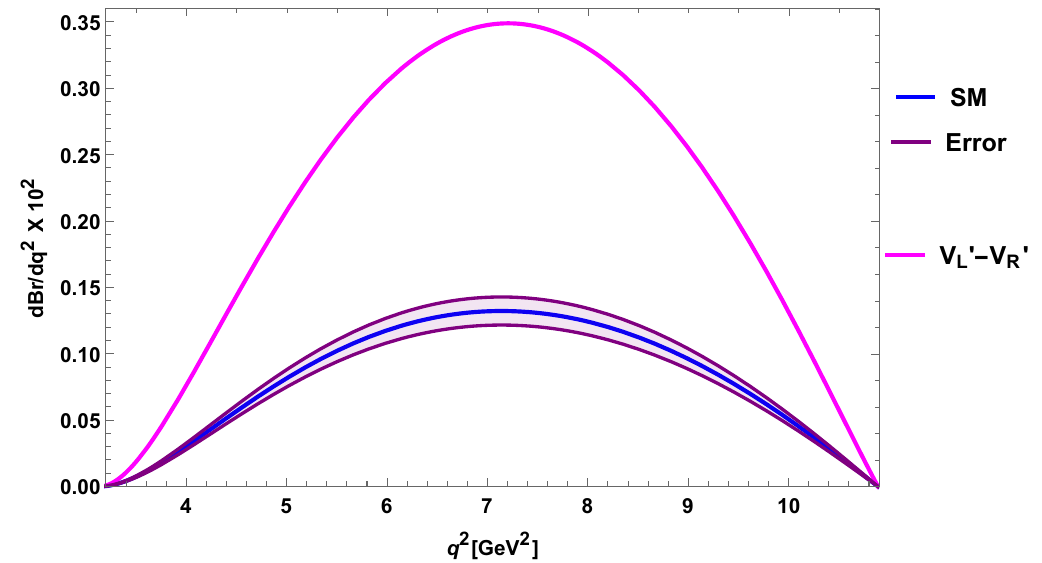}
\quad
\includegraphics[scale=0.4]{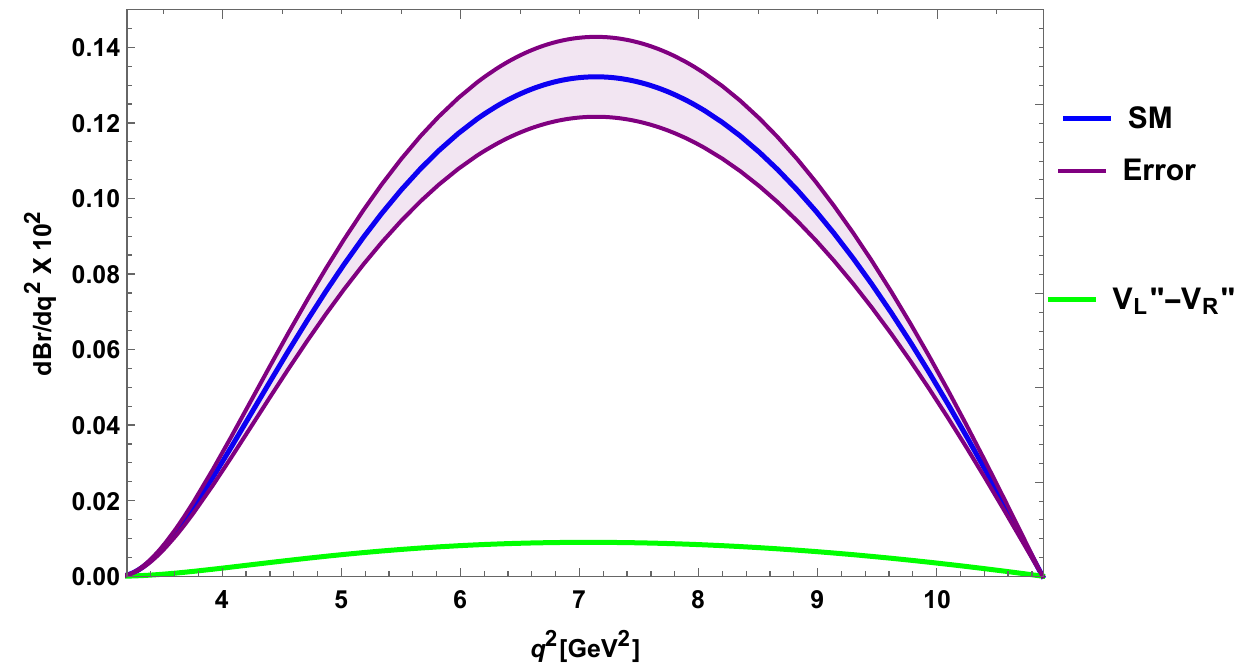}
\caption{The branching ratio for case I (top), case IIB (bottom-left) and case IIC (bottom-right) of $ B \to {{D_0}^*} \tau \bar \nu_\tau$ decay.}\label{Fig:Case-I D0s}
\end{figure}

For the double differential rates in the SM for the $s_l^\pi = \frac12^+$
states $B\to D_0^*\ell\bar\nu$, we find
\begin{align}\label{D0rate}
\frac{\d\Gamma_{D_0^*}}{ \d \cos\theta  \d w\,} & = 
  3 \Gamma_0\, r^3 \big(1-2 r w+r^2-\rl\big)^2 \sqrt{w^2-1} \nn\\*
& \bigg\{\!\sin^2\theta\, \frac{ (w^2-1) [g_+(1+r) - g_-(1-r)]^2\, 
  + \rl [ g_-^2 (w-1)+ g_+^2(w+1)]}{(1-2 rw+r^2)^2} \nn\\*
& + (1+\cos^2\theta)\, \rl\, \frac{
  \big[g_-^2 (w-1)+g_+^2 (w+1)\big] \big(w-2r+r^2 w\big)
  - 2 g_- g_+ (1-r^2) (w^2-1)}{(1-2 r w+r^2)^3} \nn\\*
&- 2 \cos\theta\, \rl\, \sqrt{w^2-1}\, \frac{[g_+ (1+r) - g_- (1-r)]\,
  [(w-1) g_- (1+r) -  (w+1)g_+ (1-r)]}{(1-2 r w+r^2)^3} \bigg\} \,.
\end{align}
Here we show the graphical representation of the branching ratio for case I (top left), case IIB (top right) and case IIC (bottom) of $ B \to {{D_0}^*} \tau \bar \nu_\tau$ decay channel in the Figure [\ref{Fig:Case-I D0s}]. In these plots, blue line represents the SM prediction, Purple band represents the $1\sigma$ uncertainties for SM, red, green, pink and green lines are obtained by using the  $ C_{V_L}, C_{V_L}'$ in case I, $(C_{V_L}',C_{V_R}')$ in case IIB and $(C_{V_L}'',C_{V_R}'')$ in case IIC respectively. We observed that, $ C_{V_L}, C_{V_L}'$ give the same  deviations from SM predictions due to their degeneracy. But in case of  IIB, we  found that the discrepancy lies above the SM value, while it is lying below for case IIC. Predicted values of branching ratio is displayed for  $B\to D_0^*\ell\bar\nu$ process in the SM and in the presence of new Wilson coefficients  in the Table \ref{Tab:Br}\,.
\subsection{$B\to D_1^*\ell\bar\nu$}
\begin{figure}[htb]
\includegraphics[scale=0.4]{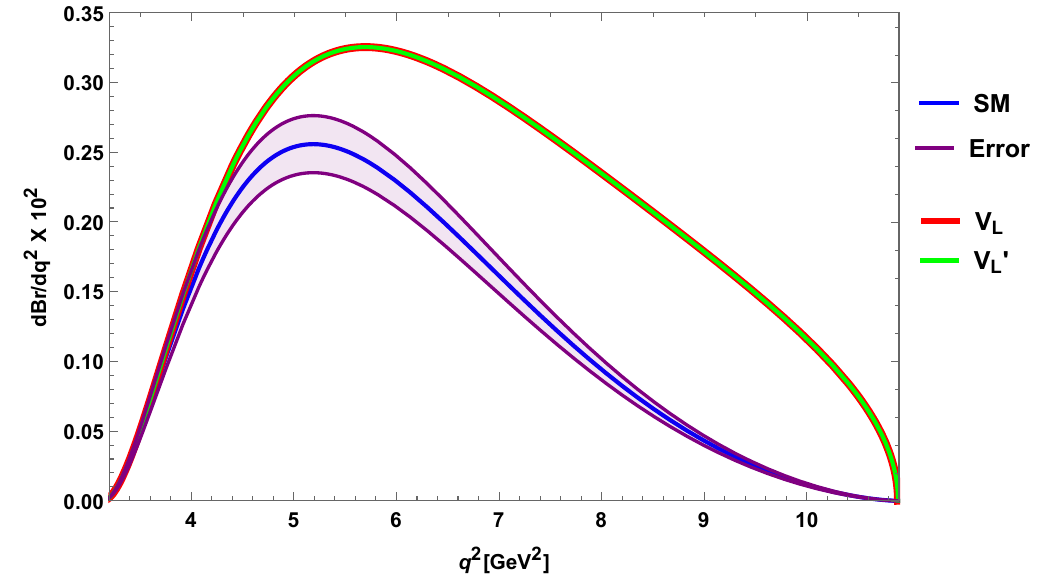}
\quad
\includegraphics[scale=0.4]{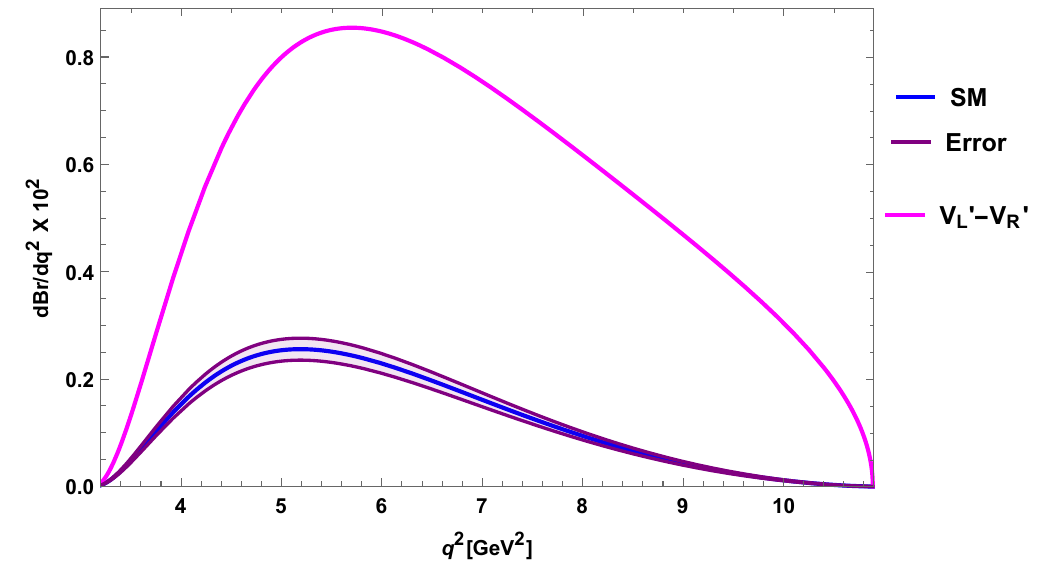}
\quad
\includegraphics[scale=0.4]{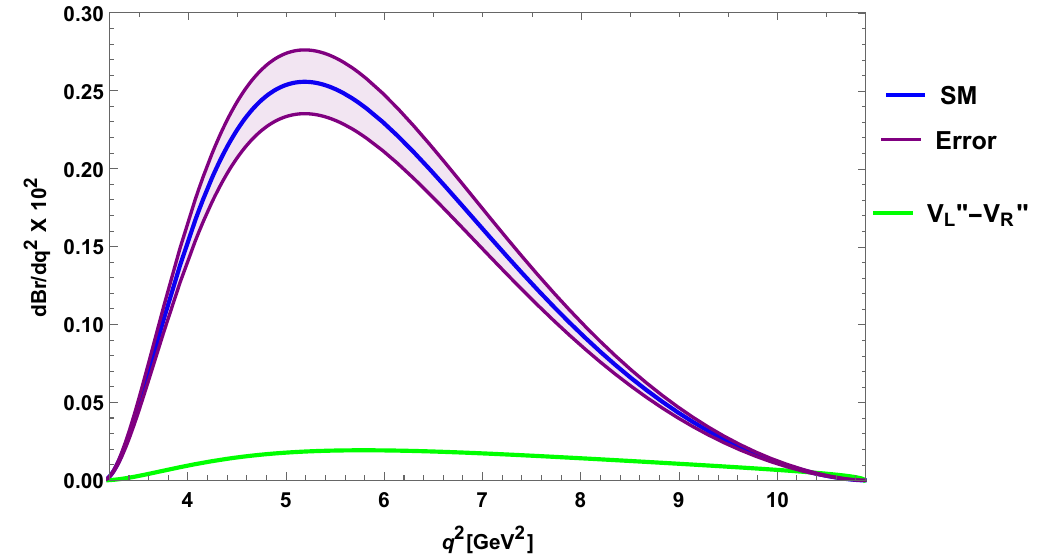}
\caption{The branching ratio for case I (top-left), case IIB (top-right) and case IIC (bottom) of $ B \to {{D_1}^*} \tau \bar \nu_\tau$ decay.}\label{Fig:Case-I D1s}
\end{figure}

For the double differential rates in the SM for the $s_l^\pi = \frac12^+$
state $B\to D_1^*\ell\bar\nu$, we find
\begin{align}\label{D1srate}
& \frac{\d\Gamma_{D_1^*}}{ \d\cos\theta \d w\,} =
  3 \Gamma_0\, r^3 \big(1-2 r w+r^2-\rl\big)^2  \sqrt{w^2-1}\,  \nn\\*
& \Bigg\{\! \sin^2\theta\, \bigg[
   \rl \frac{g_{V_1}^2 + (w^2-1) \big(2g_A^2+g_{V_2}^2+g_{V_3}^2 + 2g_{V_1}g_{V_2}+ 2wg_{V_2}g_{V_3}\big)}{2 (1-2rw+r^2)^2}+ \nn\\*
&\qquad\quad \frac{\big[g_{V_1} (w-r) + 
   (g_{V_3} + r g_{V_2})(w^2-1)\big]^2}{(1-2rw+r^2)^2} 
  \bigg] + (1+\cos^2\theta)\, \bigg[ \frac{g_{V_1}^2 +g_A^2(w^2-1)}{1-2r w+r^2}
 +\nn\\*
&\qquad \rl  \frac{ [g_{V_1}^2 + (w^2-1)g_{V_3}^2](2w^2-1+r^2-2rw)}
  {2 (1-2 rw+ r^2)^3}  + \rl (w^2-1) \frac{2g_{V_1} g_{V_2}(1-r^2) + 4 g_{V_1}g_{V_3}(w-r)}{  2 (1-2 rw+ r^2)^3} + \nn\\*
&\qquad\rl (w^2-1)\frac{ g_{V_2}^2(1-2rw-r^2+2r^2w^2) + 2g_{V_2}g_{V_3} (w-2r+r^2w)}{2 (1-2 rw+ r^2)^3} \bigg] \nn\\*
&\qquad - 2 \cos\theta\,\sqrt{w^2-1}\, \bigg[ \frac{2 g_A g_{V_1}}{1-2 r w+r^2} - \rl \frac{\big[g_{V_1} (w-r) + (g_{V_3} + r g_{V_2}) (w^2-1)\big]}{(1-2r w+r^2)^3} \times \nn\\*
&\qquad~~~~~~~~~~~~~~~~~~~~~~~~~~~~~~~~~~~~~~~~~~~~~\frac{[g_{V_1} +g_{V_2}(1-rw) +g_{V_3}(w-r)]}{(1-2r w+r^2)^3}\bigg]\! \Bigg\}\,.
\end{align}
Here we show the graphical representation of the branching ratio for case I (top left), case IIB (top right) and case IIC (bottom) of $ B \to {{D_1}^*} \tau \bar \nu_\tau$ decay channel in the Figure [\ref{Fig:Case-I D1s}].  In these plots, blue line represents the SM prediction, Purple band represents the $1\sigma$ uncertainties for SM,and red, green, pink and green lines are obtained by using the  $C_{V_L}, C_{V_L}'$ in case I, $(C_{V_L}',C_{V_R}')$ in case IIB and $(C_{V_L}'',C_{V_R}'')$ in case IIC respectively. We observed similar behaviour as in the case of $B\to D_0^*\ell\bar\nu$ process due to the effects of these new coefficients.
 Predicted values of branching ratio is displayed for  $B\to D_1^*\ell\bar\nu$ process in the SM and in the presence of new Wilson coefficients in the Table \ref{Tab:Br}\,.
\subsection{\textbf{$ B \to {D_1} \tau \bar \nu_\tau$}} 
\begin{figure}[htb]
\includegraphics[scale=0.4]{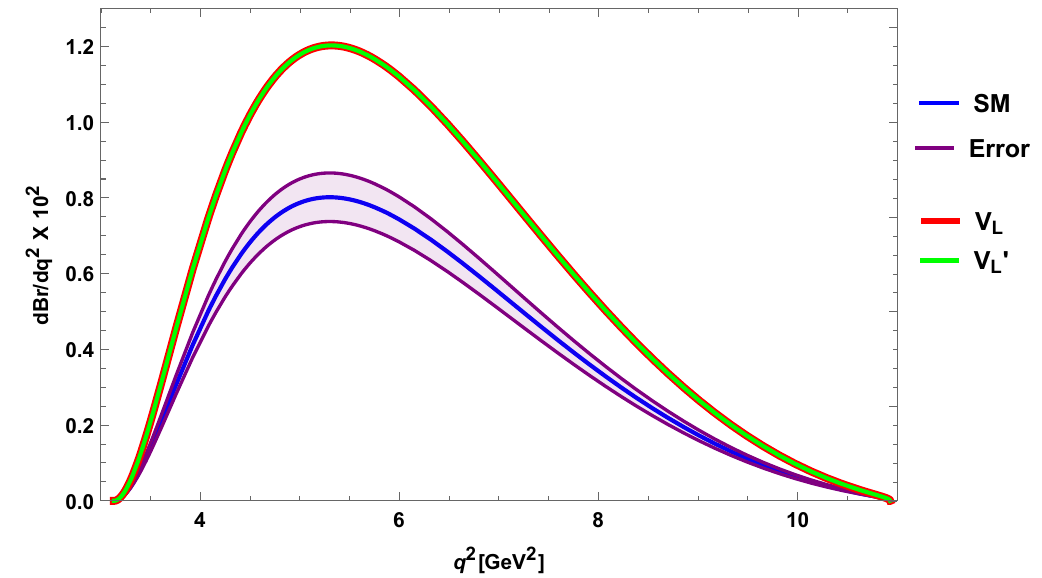}
\quad
\includegraphics[scale=0.4]{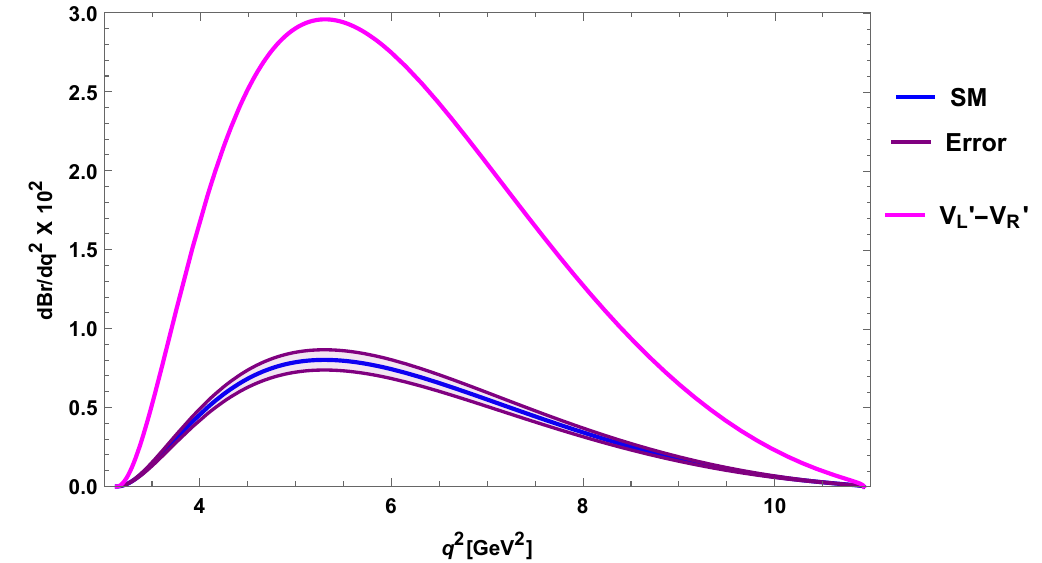}
\quad
\includegraphics[scale=0.4]{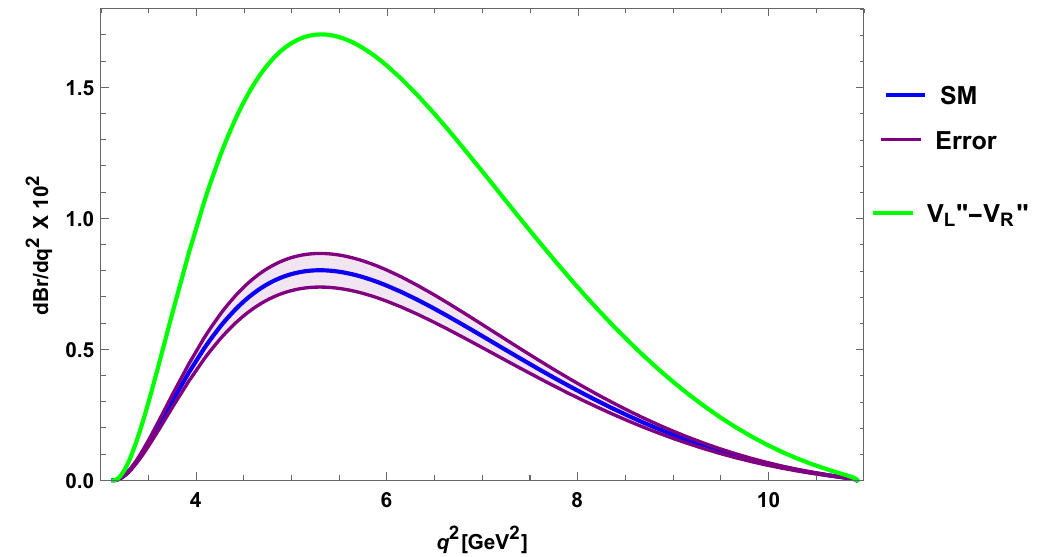}
\caption{The branching ratio for case I (top-left), case IIB (top-right) and case IIC (bottom) of $ B \to {D_1} \tau \bar \nu_\tau$ decay.}\label{Fig:Case-I D1}
\end{figure}

For the double differential rates in the SM for the $s_l^\pi = \frac32^+$
states, we obtain
\begin{widetext}
\begin{align}
\tiny
& \frac{\d\Gamma_{D_1}}{ \d\cos\theta \d w\,} =
  3 \Gamma_0\, r^3 \sqrt{w^2-1}\, \big(1-2 r w+r^2-\rl\big)^2 \nn\\*
& ~~~~~~~~~~~\Bigg\{\! \sin^2\theta\, \bigg[
   \rl \frac{f_{V_1}^2 + \big(2f_A^2+f_{V_2}^2+f_{V_3}^2 + 2f_{V_1}f_{V_2}+ 2wf_{V_2}f_{V_3}\big) (w^2-1)}{2 (1-2rw+r^2)^2}+  \nn\\*
&~~~~~~~~~~~~\frac{\big[f_{V_1} (w-r) + 
   (f_{V_3} + r f_{V_2})(w^2-1)\big]^2}{(1-2rw+r^2)^2} 
  \bigg] + (1+\cos^2\theta)\, \bigg[ \frac{f_{V_1}^2 +f_A^2(w^2-1)}{1-2r w+r^2}  \nn\\*
&\qquad + \rl \frac{ [f_{V_1}^2 + (w^2-1)f_{V_3}^2](2w^2-1+r^2-2rw)}
  {2 (1-2 rw+ r^2)^3} + \rl (w^2-1)  \nn\\*
&\qquad\quad \frac{2f_{V_1} f_{V_2}(1-r^2) + 4 f_{V_1}f_{V_3}(w-r)   
  + f_{V_2}^2(1-2rw-r^2+2r^2w^2) 
  + 2f_{V_2}f_{V_3} (w-2r+r^2w)}{2 (1-2 rw+ r^2)^3} \bigg] \nn\\*
&\qquad - 2 \cos\theta\,\sqrt{w^2-1}\, \bigg[
  \frac{2 f_A f_{V_1}}{1-2 r w+r^2} \nn\\*
&\qquad
 ~~~~~~~~~~~~~~~ - \rl \frac{\big[f_{V_1} (w-r) + (f_{V_3} + r f_{V_2}) (w^2-1)\big] 
  [f_{V_1} +f_{V_2}(1-rw) +f_{V_3}(w-r)]}{(1-2r w+r^2)^3}\bigg]\! \Bigg\}\,.
\end{align}\label{D1rate}
\end{widetext} 
 Here we show the graphical representations of branching ratio for case I (top left), case IIB (top right) and case IIC (bottom) of $ B \to {D_1} \tau \bar \nu_\tau$ decay channel in  Figure \ref{Fig:Case-I D1}.  In these plots blue line represents the SM prediction, Purple band represents the $1\sigma$ uncertainties for SM,and red, green, pink and green lines are obtained by using the  $ C_{V_L}, C_{V_L}'$ in case I, $(C_{V_L}',C_{V_R}')$ in case IIB and $(C_{V_L}'',C_{V_R}'')$ in case IIC respectively. We observed that $ C_{V_L}, C_{V_L}'$ give the same standard deviations from SM predictions due to their degeneracy. In case of case IIB and  IIC, we can found that the discrepancy sprawl above the SM value. Predicted values of branching ratio is displayed for  $B\to D_1\ell\bar\nu$ process in the SM and in the presence of new Wilson coefficients  in the Table \ref{Tab:Br}\,.
\subsection{\textbf{$ B \to {D_2^*} \tau \bar \nu_\tau$}} 
\begin{figure}[htb]
\includegraphics[scale=0.4]{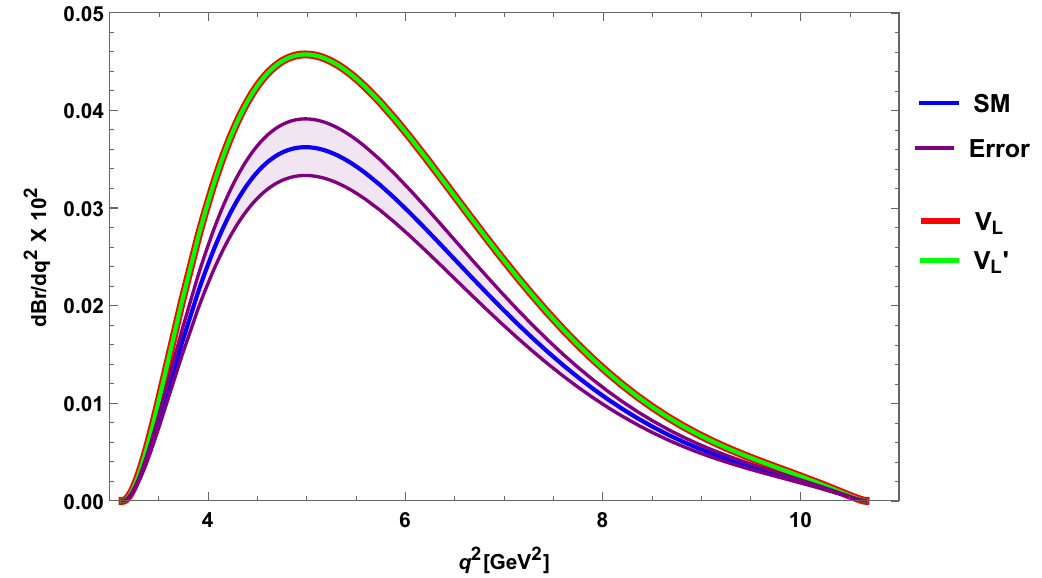}
\quad
\includegraphics[scale=0.4]{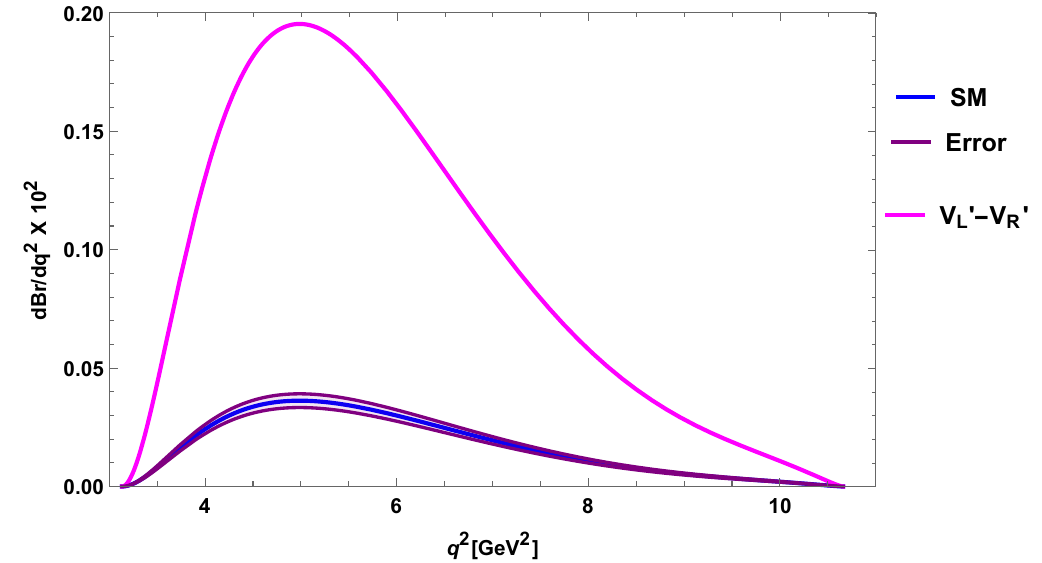}
\quad
\includegraphics[scale=0.4]{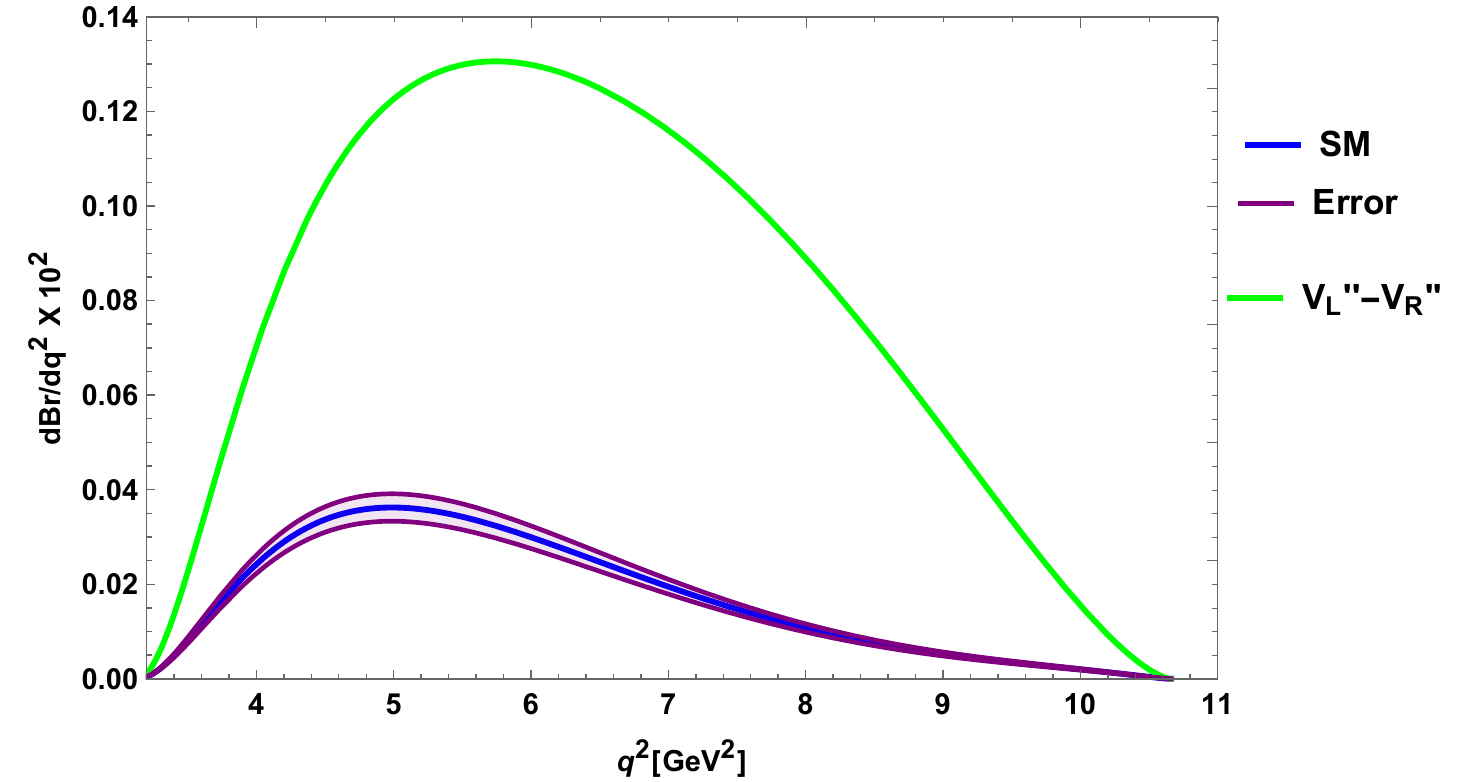}

\caption{The branching ratio for case I (top-left), case IIB (top-right) and case IIC (bottom) of $ B \to {D_2^*} \tau \bar \nu_\tau$ decay.}\label{Fig:Case-I D2s}
\end{figure}

For the double differential rates in the SM for the $s_l^\pi = \frac32^+$
states we obtain
For $B\to D_2^*\ell\bar\nu$ we find
\begin{align}\label{D2srate}
& \frac{\d\Gamma_{D_2^*}}{ \d \cos\theta \d w\,} =
  \Gamma_0\, r^3 \, \big(1+r^2-\rl-2 r w\big)^2 (w^2-1)^{3/2} \! \Bigg\{\! \sin^2\theta\, \bigg[
  \frac{2\big[(w-r)k_{A_1} + (w^2-1) (k_{A_3} + r k_{A_2})\big]^2}{(1-2rw+r^2)^2} \nn\\*
& + \rl \frac{3k_{A_1}^2 + (w^2-1)\big(3 k_V^2 +2k_{A_2}^2 +2k_{A_3}^2 + 4k_{A_1}k_{A_2} + 4w k_{A_2}k_{A_3}\big) }{2 (1-2 r w+r^2)^2} \bigg] + (1+\cos^2\theta)\, \bigg[ 
  \frac32\, \frac{k_{A_1}^2 + k_V^2 (w^2-1)}{1-2 r w+r^2}  \nn\\*
&\qquad + \rl \frac{ [k_{A_1}^2 + (w^2-1)k_{A_3}^2](2w^2-1+r^2-2rw)}{(1-2 rw+r^2)^3} + \rl (w^2-1)  \nn\\*
&\qquad\quad \frac{2k_{A_1} k_{A_2}(1-r^2) + 4 k_{A_1}k_{A_3}(w-r)   
   + k_{A_2}^2(1-2rw-r^2+2r^2w^2) 
   + 2k_{A_2}k_{A_3} (w-2r+r^2w)}{(1-2 rw+ r^2)^3} \bigg] \nn\\*
&\qquad - 2\cos\theta\,\sqrt{w^2-1}\,
  \bigg[ \frac{3 k_V k_{A_1}}{1-2 r w+r^2}
  - 2 \rl  \nn\\*
&\qquad \frac{\big[k_{A_1} (w-r)+(k_{A_3} + r k_{A_2}) (w^2-1)\big] 
  [k_{A_1} + k_{A_2}(1-rw)+ k_{A_3}(w-r)]}{(1-2r w+r^2)^3} \bigg]\!\Bigg\}.
\end{align}
 Here we show the graphical representations of branching ratio for case I (top left), case IIB (top right) and case IIC (bottom) of $ B \to{D_2^*} \tau \bar \nu_\tau$ decay channel in  Figure \ref{Fig:Case-I D2s}.  In these plots blue line represents the SM predictions, Purple band represents the $1\sigma$ uncertainties for SM,and red, green, pink and green lines are obtained by using the  $ C_{V_L}, C_{V_L}'$ in case I, $(C_{V_L}',C_{V_R}')$ in case IIB and $(C_{V_L}'',C_{V_R}'')$ in case IIC respectively. Here also we observed  similar behaviour as in the case of  $ B \to {D_1} \tau \bar \nu_\tau$ due to the effect of these new couplings.
  Predicted values of branching ratio is displayed for  $B\to{D_2^*}\ell\bar\nu$ process in the SM and in the presence of new Wilson coefficients  in the Table \ref{Tab:Br}\,.
\begin{table}[htb]
\scriptsize
\caption{Calculated values of branching ratio of $B\to \{D^*_0, D_1^*, D_1, D_2^*\} \tau \bar\nu$ process in the SM and in the presence of new complex Wilson coefficients.}\label{Tab:Br}
\begin{tabular}{|c|c|c|c|c|c|c|}
\hline
~~&~  SM ~ &~Case I ~&~Case I ~&~ Case IIB~&~ Case IIC~\\
\hline
 ~Observables~&~Values for SM~&~Values for $C_{V_L}$~&~Values for $C_{V_L}'$~&~Values for $(C_{V_L}',C_{V_R}')$~&~Values for $(C_{V_L}'',C_{V_R}'')$ ~\\
\hline
~ Br($B\to D_0^*\tau\bar\nu$)~&~$0.0063 \pm 0.001$~&~$0.0116  \pm 0.004$~&~$0.0116  \pm 0.008$~&~$0.0165 \pm 0.002$~& $0.00041  \pm 0.001$\\
\hline
~ Br($B\to D_1^*\tau\bar\nu$)~&~$0.0022 \pm 0.001$~&~$0.0028 \pm 0.001$~&~$0.0028 \pm 0.001$~&~$0.0126 \pm 0.009$~& $0.00023 \pm 0.008$\\
\hline
~Br($B\to D_1 \tau \bar\nu$)~&~$0.0343 \pm 0.004$~&~$0.0423 \pm 0.006$~& $0.0423 \pm 0.006$&~$0.1900 \pm 0.034$~&~$0.1212 \pm 0.028$~\\
\hline
~ Br($B\to D_2^* \tau \bar\nu$)~&~$0.0012 \pm 0.001$~&~$0.0016 \pm 0.001$~&~$0.0016 \pm 0.001$~&~$0.0072 \pm 0.003$~& $0.0060 \pm 0.002$\\
\hline
\end{tabular}

\end{table}
\section{Conclusion}
In this paper, we have investigated the pertinent semileptonic  $\Lambda_b$ and $B$ meson decay processes concerning the $ b \rightarrow c \tau \bar{\nu_\tau}$ transition in a model independent approach. Using the generalized effective Hamiltonian in the presence of new physics coefficients, it brings forth the addition of scalar, vector and tensor contributions to the SM effects. Here we performed a $\chi^2$ fitting, considering either one NP operator or a combination of two similar operators (for example [$O_{V_L}$, $O_{V_R}$], [$O_{S_L}$, $O_{S_R}$], [$O^{'}_{V_L}$, $O^{'}_{V_R}$] and [$O^{''}_{S_L}$, $O^{''}_{S_R}$]) at a time while making the fit to the experimental observables like $R_D$, $R_{D^*}$, $R_{J/\Psi}$,  $P_{\tau}$ and the upper limit on Br($B_c^+ \to \tau^+ \nu_\tau$) as functions of the various Wilson coefficients. Measurements of various flavor ratios, such as $R_D$, $R_{D^\star}$, $P_{\tau}^{D^\star}$, $R_{J/\psi}$, and $R_{\Lambda_c}$, are associated with processes like $B\to D^{(\star)}\tau\nu$, $B_c\to J/\psi\tau\nu$, and $\Lambda_b \to \Lambda_c \tau\nu$. These measurements reveal discrepancies compared to the predictions of the Standard Model. If these disparities persist in future experiments, it would offer clear and indisputable evidence of physics that goes beyond what the Standard Model can explain. Using the best-fit values of the total of $18$ constraints extracting from the fitting, we evaluated the branching ratios, forward-backward asymmetries, lepton non-universality parameter, lepton and hadron polarization asymmetries of $\Lambda_b \to \Lambda_c \tau \bar \nu_\tau$, $ B_c^+ \to \eta_c \tau^+   \nu_\tau$,  and $ B \to D^{**} \tau \bar \nu_\tau$   (where $D^{**} = \{D^*_0, D_1^*, D_1, D_2^*\}$ in the full $q^2$ (in ${\rm GeV}^2$) limit. Mostly, we have investigated  the sensitivity of the new couplings towards the different angular observables of $ b \rightarrow c \tau \bar{\nu}_\tau$ transition, where  we have taken into consideration  two different cases. First case includes the presence of only one new coefficient whereas second case consists of two new coefficients at a time, i.e., unprimed, primed and double primed respectively. In $\Lambda_b \to \Lambda_c \tau \bar \nu_\tau$ decay channel, we observed that in case I due to the degeneracy of $ C_{V_L}, C_{V_L}' $ and $C_{S_R}''$ they provide the same results and show no deviation from the SM for every angular observable.  For other coefficients, branching ratio, forward-backward asymmetry and LNU parameter give  appreciable deviation. In case IIA, all the couplings are showing significant deviation whereas for other observable $(C_{V_L},C_{S_L})$ shows very negligible deviation. In case IIB, all the seven constraints are revealing an outstanding discrepancies from SM for branching fraction, but for all other observable $(C_{V_L}',C_{S_R}')$ and  $(C_{V_L}',C_T')$ show some marginal deviations. In case IIC, we can sense that  $(C_{V_L}'',C_{S_L}'')$ shows the maximum deviation for all the observable. Except branching ratio  $(C_{V_R}'',C_{S_R}'')$ shows no deviation from SM predictions. For $ B_c^+ \to \eta_c \tau^+  \nu_\tau$, we observed that in case I $C_{S_L}''$ gives the maximum deviation where as $C_T$ shows a marginal deviation from SM prediction. Other constraints except $C_{S_L}''$ and $C_T$ show no deviation in other angular observable. From case IIA we can derive that all the couplings show some significant deviation in case of branching fraction. For forward-backward asymmetry $(C_{V_L},C_{S_L})$ display an outstanding deviation and it gives idea about the zero crossing point. In case of LNU parameter  $(C_{V_L},C_{S_L})$ shows slight deviation but all the other coupling show some standard deviation. In lepton polarization asymmetry  $(C_{V_L},C_{S_L})$ gives an unique deviation compared to other couplings. In case IIB, the deviation coming from $(C_{V_L}',C_{S_R}')$ is very negligible. But all other couplings show outstanding discrepancies from the SM prediction for all the angular observable. From case IIB, we conclude that $(C_{V_R}'',C_{S_R}'')$ and $(C_{V_L}'',C_{S_L}'')$ show a good deviation in branching ratio, forward-backward and lepton polarization  asymmetry. But in LNU parameter $(C_{V_R}'',C_{S_R}'')$ gives very negligible divergence. In $ B \to D^{**} \tau \bar \nu_\tau$ decay channels we manifest that all the vector couplings show extreme deviation as compared to other decay modes.\\
~~~If we consider some model dependent scenario,  $2HDM$ model certainly, these findings and their correlations suggest that the High-Luminosity Large Hadron Collider (HL-LHC) has the capacity not only to examine but also potentially rule out the presence of a charged Higgs boson in the lower mass range. Additionally, the HL-LHC can conduct thorough investigations into the unusual patterns observed in flavor-related measurements linked to transitions like $b \to c \tau\nu$., covering a wide range of aspects. This, in turn, can yield valuable insights into the fundamental physics mechanisms responsible for these flavor anomalies. The HL-LHC's ability to explore these phenomena represents a significant opportunity to enhance our comprehension of flavor physics and potentially uncover novel physics that goes beyond the confines of the Standard Model. Coming to an end we can say that we investigated the branching fraction and angular observable of the above decay processes for one and two coefficients at a time for full $q^2$ range. From these above data we have remarked the sensitivity of new couplings on angular observable which will furnish a clear proposal on the fabric of new physics. 
\acknowledgements
AB would like to acknowledge DST INSPIRE program for financial support. RM acknowledges the support from University of Hyderabad through IoE project grant No. RC1-20-012. The computational work done at CMSD, University of Hyderabad is duly acknowledged.
\appendix
\section{Helicity Amplitude and Form Factor Relations For $\Lambda_b \to \Lambda_c  \tau \bar \nu_\tau$ decay}
The helicity amplitudes in terms of the various form factors and the NP couplings are given as\cite{Shivashankara:2015cta}:
\bea
H_{\frac{1}{2}\,0}^V &=& \left(1+C^{\rm{eff}}_{V_L}+C^{\rm{eff}}_{V_R} \right)\,\frac{\sqrt{Q_-}}{\sqrt{q^2}}\,\Big[(M_{\Lambda_b + \Lambda _c})\,f_1(q^2) - q^2\,f_2(q^2)\Big]\,, \nonumber \\
H_{\frac{1}{2}\,0}^A &=& \left(1+C^{\rm{eff}}_{V_L}-C^{\rm{eff}}_{V_R}\right)\,\frac{\sqrt{Q_+}}{\sqrt{q^2}}\,\Big[(M_{\Lambda_b - \Lambda _c})\,g_1(q^2) + q^2\,g_2(q^2)\Big]\,, \nonumber \\
H_{\frac{1}{2}\,+}^V &=& \left(1+C^{\rm{eff}}_{V_L}+C^{\rm{eff}}_{V_R}\right)\,\sqrt{2\,Q_-}\,\Big[-f_1(q^2) + (M_{\Lambda_b + \Lambda _c})\,f_2(q^2)\Big]\,, \nonumber \\
H_{\frac{1}{2}\,+}^A &=& \left(1+C^{\rm{eff}}_{V_L}-C^{\rm{eff}}_{V_R}\right)\,\sqrt{2\,Q_+}\,\Big[-g_1(q^2) - (M_{\Lambda_b - \Lambda _c})\,g_2(q^2)\Big]\,, \nonumber \\
H_{\frac{1}{2}\,t}^V &=& \left(1+C^{\rm{eff}}_{V_L}+C^{\rm{eff}}_{V_R}\right)\,\frac{\sqrt{Q_+}}{\sqrt{q^2}}\,\Big[(M_{\Lambda_b - \Lambda _c})\,f_1(q^2) + q^2\,f_3(q^2)\Big]\,, \nonumber \\
H_{\frac{1}{2}\,t}^A &=& \left(1+C^{\rm{eff}}_{V_L}-C^{\rm{eff}}_{V_R}\right)\,\frac{\sqrt{Q_-}}{\sqrt{q^2}}\,\Big[(M_{\Lambda_b + \Lambda _c})\,g_1(q^2) - q^2\,g_3(q^2)\Big]\,, \nn \\
H_{\frac{1}{2}\,0}^{S} &=& \left(C^{\rm{eff}}_{S_L}+C^{\rm{eff}}_{S_R}\right)\,\frac{\sqrt{Q_+}}{m_b - m_{q}}\,\Big[(M_{\Lambda_b - \Lambda _c})\,f_1(q^2) + q^2\,f_3(q^2)\Big]\,, \nonumber \\
H_{\frac{1}{2}\,0}^{P} &=&\left(C^{\rm{eff}}_{S_L}-C^{\rm{eff}}_{S_R}\right)\,\frac{\sqrt{Q_-}}{m_b + m_{q}}\,\Big[(M_{\Lambda_b + \Lambda _c})\,g_1(q^2) - q^2\,g_3(q^2)\Big]\,,\nn \\
H_{\frac{1}{2},+,0}^T&=&-C^{\rm{eff}}_{T}\sqrt{\frac{2}{q^2}}\left(f_T\sqrt{Q_+}(M_{\Lambda_b - \Lambda _c})+g_T\sqrt{Q_-}(M_{\Lambda_b + \Lambda _c})\right)\,,\nn \\
H_{\frac{1}{2},+,-}^T&=&-C^{\rm{eff}}_{T}\left(f_T\sqrt{Q_+}+g_T\sqrt{Q_-}\right)\,,\nn \\
H_{\frac{1}{2},+,t}^T&=&C^{\rm{eff}}_{T}\Big[-\sqrt{\frac{2}{q^2}}\left(f_T\sqrt{Q_-}(M_{\Lambda_b + \Lambda _c}) +g_T\sqrt{Q_+}(M_{\Lambda_b - \Lambda _c})\right)\nn \\ &&
+\sqrt{2q^2}\left(f_T^V\sqrt{Q_-}-g_T^V\sqrt{Q_+}\right)\Big]\,,\nn \\
H_{\frac{1}{2},0,t}^T&=&C^{\rm{eff}}_{T}\Big [-f_T\sqrt{Q_-}-g_T\sqrt{Q_+}+f_T^V\sqrt{Q_-}(M_{\Lambda_b + \Lambda _c})-g_T^V\sqrt{Q_+}(M_{\Lambda_b - \Lambda _c})\nn \\&&
+f_T^S\sqrt{Q_-}Q_++g_T^S\sqrt{Q_+}Q_-\Bigg ]\,,\nn \\
H_{-\frac{1}{2},+,-}^T&=&C^{\rm{eff}}_{T}\Big [f_T\sqrt{Q_+}-g_T\sqrt{Q_-}\Big]\,,\nn\\
H_{-\frac{1}{2},0,-}^T&=&C^{\rm{eff}}_{T}\Big[\sqrt{\frac{2}{q^2}}\left(f_T\sqrt{Q_+}(M_{\Lambda_b - \Lambda _c})-g_T\sqrt{Q_-}(M_{\Lambda_b + \Lambda _c})\right)\Big]\,,\nn
\eea
\bea
H_{-\frac{1}{2},-,t}^T&=&C^{\rm{eff}}_{T} \Big [-\sqrt{\frac{2}{q^2}}\left(f_T\sqrt{Q_-}(M_{\Lambda_b + \Lambda _c})-g_T\sqrt{Q_+}(M_{\Lambda_b - \Lambda _c})\right)\nn \\&&
+\sqrt{2q^2}\left(f_T^V\sqrt{Q_-}+g_T^V\sqrt{Q_+}\right)\Big ]\,,\nn\\
H_{-\frac{1}{2},0,t}^T&=&C^{\rm{eff}}_{T}\Big [-f_T\sqrt{Q_-}+g_T\sqrt{Q_+}+f_T^V\sqrt{Q_-}(M_{\Lambda_b + \Lambda _c})
+g_T^V\sqrt{Q_+}(M_{\Lambda_b - \Lambda _c})\nn \\&&+f_T^S\sqrt{Q_-}Q_+-g_T^S\sqrt{Q_+}Q_- \Big ]\,.
\eea
where $M_{\Lambda_b \pm \Lambda _c}= M_{\Lambda_b} \pm M_{\Lambda_c} $, $ Q_\pm =\left(M_{\Lambda_b} \pm M_{\Lambda_c}\right)^2-q^2$ and $f_i^{(a)}, g_i^{(b)} ~(i=1,2,3, T~ \&~ a,b=V,S)$ are the various form factors.
The relation betwen various form factors are given as \cite{Feldmann:2011xf,Chen:2001zc}
\bea
 f_0 &=&  \frac{q^2}{M_{\Lambda_b}-M_{\Lambda_c}} \, f_3 +f_1 \,,\qquad  f_+ = -\frac{q^2}{M_{\Lambda_b}+M_{\Lambda_c}} \, f_2 +f_1  \,,
\qquad  f_\perp = -(M_{\Lambda_b}+M_{\Lambda_c}) \, f_2+f_1 \,,\nn \\
g_0 &=& -\frac{q^2}{M_{\Lambda_b}+M_{\Lambda_c}} \, g_3 + g_1 \,, \qquad  g_+ = \frac{q^2}{M_{\Lambda_b}-M_{\Lambda_c}} \, g_2+g_1   \,,\qquad
 g_\perp = (M_{\Lambda_b}-M_{\Lambda_c}) \, g_2 +g_1 
\,,\nn \\
 h_+ &=& - \frac{M_{\Lambda_b}+M_{\Lambda_c}}{q^2} \, f_1^T +f_2^T \,,
\qquad
 h_\perp = - \frac{1}{M_{\Lambda_b}+M_{\Lambda_c}} \, f_1^T +f_2^T 
\,,\nn \\
 \tilde h_+ &=&  \frac{M_{\Lambda_b}-M_{\Lambda_c}}{q^2} \, g_1^T +g_2^T  \,,
\qquad
 \tilde h_\perp =  \frac{1}{M_{\Lambda_b}-M_{\Lambda_c}} \, g_1^T+g_2^T \,,
\eea
with 
\bea
f_2^T&=&f_T-f_T^S q^2,\qquad f_1^T=-\frac{q^2}{M_{\Lambda_b}-M_{\Lambda_c}} f_3^T, \qquad f_1^T=\left(f_T^V+f_T^S(M_{\Lambda_b}-M_{\Lambda_c}) \right)q^2, \nn \\
g_2^T&=&g_T-g_T^S q^2,\qquad g_1^T=\frac{q^2}{M_{\Lambda_b}+M_{\Lambda_c}} g_3^T , \qquad g_1^T=\left(g_T^V+g_T^S(M_{\Lambda_b}+M_{\Lambda_c}) \right)q^2\,.
\eea
\section{Form Factor Expansion for $  B_c^+ \to \eta_c \tau^+ \nu_\tau$ decay }
 The expressions for helicity amplitudes and the relation between the form factors are \cite{Sakaki:2014sea8,Murphy:2018sqg}.
\bea
  H_0(q^2) &=& \sqrt{\lambda_{\eta_c}(q^2) \over q^2} F_1(q^2) \,, \,\qquad H_t(q^2) = {M_{B_c}^2 - M_{\eta_c}^2 \over \sqrt{q^2}} F_0(q^2) \,, \\    
 H_S(q^2)  &=& {M_{B_c}^2 - M_{\eta_c}^2 \over m_b-m_c} F_0(q^2) \,, \, \qquad H_T^s(q^2) = -{\sqrt{\lambda_{\eta_c}(q^2)} \over M_{B_c} + M_{\eta_c}}F_T(q^2) \,.\\
  F_1(q^2) &=& {1 \over 2\sqrt{M_{B_c} M_{\eta_c}}} \left[ (M_{B_c} + M_{\eta_c} ) h_+ - (M_{B_c} - M_{\eta_c}) h_- \right] \,, \\
   F_T(q^2) &=& { M_{B_c} + M_{\eta_c} \over 2\sqrt{M_{B_c} M_{\eta_c}} } h_T \,\\   
   F_0(q^2) &=& {1 \over 2\sqrt{M_{B_c} M_{\eta_c}}} \left[ { ( M_{B_c} + M_{\eta_c})^2 - q^2 \over M_{B_c} + M_{\eta_c} } \, h_+ \right. \nn \\&& \left. \quad - { ( M_{B_c} - M_{\eta_c} )^2 - q^2 \over M_{B_c} - M_{\eta_c}} \, h_- \right] \,.           
 \eea
 
\section{Form Factor Expansion for $ B \to D^{**} \tau \bar \nu_\tau$ decay}
\begin{itemize}
\item The form factors for $B \to D_0^*\, \ell\, \bar\nu$ are
\begin{eqnarray}\label{FFD0}
g_+ &=& \varepsilon_c \bigg[ 2(w-1)\zeta_1
  - 3\zeta\, {w\bar\Lambda^*-\bar\Lambda\over 1+w} \bigg] \nn\\*
&&{} - \varepsilon_b \bigg[ {\bar\Lambda^*(1+2w)-\bar\Lambda(2+w)\over 1+w}\,
  \zeta - 2(w-1)\,\zeta_1 \bigg] , \nn\\*
g_- &=&  \varepsilon_b\, \chi_b 
  + \zeta + \varepsilon_c \Big[ \chi_{\rm ke}+6\chi_1-2(1+w)\chi_2 \Big]  \,. 
\end{eqnarray}
\item The form factors for $B \to D_1^*\, \ell\, \bar\nu$ are
\begin{widetext}
\begin{eqnarray}\label{FFD1s}
g_A &=& \zeta - 
  \varepsilon_b\, \bigg[ {\bar\Lambda^*(1+2w)-\bar\Lambda(2+w)\over 1+w}\,
  \zeta - 2(w-1)\,\zeta_1 - \chi_b \bigg] + \varepsilon_c \bigg[\frac{w\bar\Lambda^*-\bar\Lambda}{1+w} \zeta
  + \chi_{\rm ke}-2\chi_1 \bigg] ,\nn\\
g_{V_1} &=& (w-1) \zeta + \varepsilon_c 
  \Big[(w\bar\Lambda^*-\bar\Lambda)\zeta + (w-1)(\chi_{\rm ke}-2\chi_1) \Big]\nn \\&&~~~~~~~~~~~~
  - \varepsilon_b \Big\{\! \big[\bar\Lambda^*(1+2w)-\bar\Lambda(2+w)\big] \zeta 
  - 2(w^2-1) \zeta_1 - (w-1)\chi_b \Big\} , \nn\\
g_{V_2} &=& 2\varepsilon_c\, (\zeta_1-\chi_2) \,, \nn\\*
g_{V_3} &=& - \zeta  + \varepsilon_b \bigg[ {\bar\Lambda^*(1+2w)-\bar\Lambda(2+w)\over 1+w}\,
  \zeta - 2(w-1)\,\zeta_1 - \chi_b \bigg]
 \nn \\&&~~~~~~~~~~~~ - \varepsilon_c \bigg[ {w\bar\Lambda^*-\bar\Lambda \over 1+w}\zeta 
  + 2\zeta_1 + \chi_{\rm ke} - 2\chi_1 +2\chi_2 \bigg]
  . 
\end{eqnarray}
\end{widetext}
\item The form factors for $B \to D_1\, \ell\, \bar\nu$ are
\begin{widetext}
\begin{eqnarray}\label{FFD1}
\sqrt6\, f_A &=&
  - \varepsilon_b \big\{ (w-1) \big[(\bar\Lambda'+\bar\Lambda)\tau 
  - (1+2w)\tau_1-\tau_2\big] + (1+w)\eta_b \big\} \nn\\*
&&{} - \varepsilon_c \big[ 4(w\bar\Lambda'-\bar\Lambda)\tau - 3(w-1) (\tau_1-\tau_2) 
  + (1+w) (\eta_{\rm ke}-2\eta_1-3\eta_3) \big] - (w+1)\tau  \,,\nn\\*
\sqrt6\, f_{V_1} &=& 
  - \varepsilon_b (w^2-1) \big[(\bar\Lambda'+\bar\Lambda)\tau 
  - (1+2w)\tau_1-\tau_2 + \eta_b \big] \nn\\*
&&{} - \varepsilon_c \big[ 4(1+w)(w\bar\Lambda'-\bar\Lambda)\tau
  - (w^2-1)(3\tau_1-3\tau_2-\eta_{\rm ke}+2\eta_1+3\eta_3) \big]+  (1-w^2)\tau \,, \nn\\
\sqrt6\, f_{V_2} &=& - 3\varepsilon_b \big[(\bar\Lambda'+\bar\Lambda)\tau 
  - (1+2w)\tau_1-\tau_2 + \eta_b\big] \nn\\* 
&&{} - \varepsilon_c \big[ (4w-1)\tau_1+5\tau_2 +3\eta_{\rm ke} +10\eta_1 
  + 4(w-1)\eta_2-5\eta_3 \big]  -3\tau \,, \nn\\*
\sqrt6\, f_{V_3} &=&  
  \varepsilon_b \big\{ (2+w) \big[(\bar\Lambda'+\bar\Lambda)\tau 
  - (1+2w)\tau_1-\tau_2\big] - (2-w)\eta_b \big\} \nn\\*
&& + \varepsilon_c \big[ 4(w\bar\Lambda'-\bar\Lambda)\tau + 
  (2+w)\tau_1 + (2+3w)\tau_2 \nn\\*
&& \quad +(w-2)\eta_{\rm ke} -2(6+w)\eta_1 -4(w-1)\eta_2 -(3w-2)\eta_3 \big]+(w-2)\tau  \,. 
\end{eqnarray}
\end{widetext}
\item The form factors for $B \to D_2^*\, \ell\, \bar\nu$ are
\begin{widetext}
\begin{eqnarray}\label{FFD2}
k_V &=& - \varepsilon_b \big[(\bar\Lambda'+\bar\Lambda)\tau 
  - (2w+1)\tau_1-\tau_2 + \eta_b\big] 
  - \varepsilon_c (\tau_1-\tau_2+\eta_{\rm ke}-2\eta_1+\eta_3) - \tau \,, \nn\\*
k_{A_1} &=& - \varepsilon_b \big\{ (w-1)
  \big[(\bar\Lambda'+\bar\Lambda)\tau - (2w+1)\tau_1-\tau_2\big] 
  + (1+w)\eta_b \big\} \nn\\*
&&{} - \varepsilon_c \big[ (w-1)(\tau_1-\tau_2)
  + (w+1)(\eta_{\rm ke}-2\eta_1+\eta_3) \big]  - (1+w)\tau\,, \nn\\
k_{A_2} &=& - 2\varepsilon_c (\tau_1+\eta_2) \,, \nn\\*
k_{A_3} &=& \tau- \varepsilon_c (\tau_1+\tau_2-\eta_{\rm ke}+2\eta_1-2\eta_2-\eta_3)\nn\\*
&&{} + \varepsilon_b \big[(\bar\Lambda'+\bar\Lambda)\tau 
  - (2w+1)\tau_1-\tau_2 + \eta_b\big] 
   \,.
\end{eqnarray}
\end{widetext}
\end{itemize}

\end{document}